\documentclass[subeqn]{article}
  \usepackage[T1]{fontenc}
  \usepackage{amsmath,amssymb,mathrsfs}
  \usepackage{xspace}
  \usepackage[left=3.6cm, right=3.6cm, top=3cm, bottom=3cm]{geometry}
  \usepackage{enumitem}
\usepackage[toc,page]{appendix}

\usepackage{hyperref}
\usepackage{tikz}

\usepackage[ansinew]{inputenc}
\usepackage{cases}
\usepackage{cases}
\usepackage{array}
\usepackage{mathenv}
\usepackage{graphicx,color}
\usepackage{eurosym}
\usepackage{amsmath,amsfonts,amssymb,amsthm}
\usepackage{wasysym}
\usepackage{mathrsfs}
\usepackage{oldstyle}
\usepackage[T1]{fontenc}
\usepackage{pb-diagram}
\usepackage[nottoc, notlof, notlot]{tocbibind}

  \newtheorem{theo}{Theorem}
  \newtheorem{pro}{Proposition}
   \newtheorem{lem}{Lemma}
  \theoremstyle{remark}
 \newtheorem{remark}{Remark}
  \newtheorem{notation}[remark]{Notation}
   
  \theoremstyle{definition}

\title{\textbf{Charge renormalisation in a mean-field approximation of QED}}
\author{Sok J\'er\'emy\\
 Ceremade, UMR 7534, Universit\'e Paris-Dauphine,\\
  Place du Mar\'echal de Lattre de Tassigny,\\
  75775 Paris Cedex 16, France.\\ \\
}%

\begin{document}
%\pageblanche0
\maketitle
%%%%%%%%%%%%%%%%%%%%%%%%%%%%%%
%\romanpagenumbers
%\input premieres.tex
%%%%%%%%%%%%%%%%%%%%%%%%%%%%%%

%%%%%%%%%%%%%%%%%%%%%%%%%%%%% nouvel environnement
 \newenvironment{dem}{
\trivlist \item[\hskip \labelsep{\bf\ Proof:}]}{\hfill \makebox[2em]{\hfill{\footnotesize $\Box$}}
\vspace{2ex}}

%%%%%%%%%%%%%%%%%%%%%%%%%%%%%% Formalisme MQ
\newcommand{\ket}[1]{\ensuremath{|#1\rangle}\xspace}
\newcommand{\bra}[1]{\ensuremath{\langle #1|}\xspace}
\newcommand{\psh}[2]{\ensuremath{\langle #1\,,\,#2\rangle}\xspace}
\newcommand{\upp}[1]{\ensuremath{^{(#1)}}\xspace}
%%%%%%%%%%%%%%%%%%%%%%%%%%%%%% Sommes

\newcommand{\ssum}{\ensuremath{\displaystyle\sum}}
\newcommand{\dint}{\ensuremath{\displaystyle\int}}
\newcommand{\diint}{\ensuremath{\displaystyle\iint}}
\newcommand{\diiint}{\ensuremath{\displaystyle\iiint}}
\newcommand{\diiiint}{\ensuremath{\displaystyle\iiiint}}
\newcommand{\diiintdens}{\ensuremath{\displaystyle\int\!\!\!\!\int\!\!\!\!\int}}
\newcommand{\diiiintdens}{\ensuremath{\displaystyle\int\!\!\!\!\int\!\!\!\!\int\!\!\!\!\int}}

%%%%%%%%%%%%%%%%%%%%%%%%%%%%%% Symboles
\newcommand{\D}{\ensuremath{\mathcal{D}^0}}
\newcommand{\ee}[1]{\ensuremath{E\left(#1\right)}}
\newcommand{\ef}[1]{\ensuremath{E_m\left(#1\right)}}
\newcommand{\ed}[1]{\ensuremath{\widetilde{E}\left(#1\right)}}
\newcommand{\PP}{\ensuremath{\mathcal{P}^0_-}}
\newcommand{\PPP}{\ensuremath{\mathcal{P}^0_+}}

\newcommand{\EE}{\ensuremath{\mathcal{E}_{\text{BDF}}^0}}

\newcommand{\g}{\ensuremath{\gamma}}
\newcommand{\G}{\ensuremath{\Gamma}}

\newcommand{\llo}{\ensuremath{\log(\Lambda)}}
\newcommand{\WW}{\ensuremath{b_{\Lambda}}}
\newcommand{\la}{\ensuremath{\lambda}}
\newcommand{\La}{\ensuremath{\Lambda}}

\newcommand{\hl}{\ensuremath{\mathfrak{H}_\Lambda}}

\newcommand{\rr}{\ensuremath{\mathfrak{R}}}
\newcommand{\ph}{\ensuremath{\varphi}}
\newcommand{\unp}{\ensuremath{\underline{\psi}}}
\newcommand{\lpsi}{\ensuremath{\psi_\lambda}}

\newcommand{\dd}{\ensuremath{\mathrm{d}}}
\newcommand{\eps}{\ensuremath{\varepsilon}}
\newcommand{\oo}[1]{\ensuremath{\omega_#1}}
\newcommand{\om}{\ensuremath{\omega}}

\newcommand{\tr}{\ensuremath{\mathrm{Tr}_{0}}}
\newcommand{\ttr}{\ensuremath{\mathrm{Tr}}}

\newcommand{\tigma}{\ensuremath{T_{\sigma}}}
\newcommand{\talpha}{\ensuremath{T_{\boldsymbol{\alpha}}}}
\newcommand{\malpha}{\ensuremath{\boldsymbol{\alpha}}}

\newcommand{\CC}{\ensuremath{\mathbb{C}^4}}
\newcommand{\RR}{\ensuremath{\mathbb{R}^3}}

% spe sokd 
\newcommand{\Ebf}[1]{\ensuremath{\overline{E}_{#1}}}
\newcommand{\Dbf}{\ensuremath{\mathbf{D}}}
\newcommand{\Dt}{\ensuremath{\widetilde{D}}}

\newcommand{\abs}{\ensuremath{\mathfrak{a}}}

\newcommand{\sbf}[1]{\ensuremath{\mathbf{s}_{#1}}}
\newcommand{\Tbf}{\ensuremath{\mathbf{T}}}
\newcommand{\fla}{\ensuremath{f_{\Lambda}}}
\newcommand{\Fla}{\ensuremath{F_{\Lambda}}}
\newcommand{\Flac}{\ensuremath{\check{F}_{\Lambda}}}
\newcommand{\gla}{\ensuremath{g_{\Lambda}}}

\newcommand{\kappab}[1]{\ensuremath{\boldsymbol{\kappa}_{#1}}}
\newcommand{\ash}{\ensuremath{\mathrm{arcsinh}}}
\newcommand{\eed}[1]{\ensuremath{\widetilde{E}_{#1}}}
\newcommand{\weak}{\ensuremath{\rightharpoonup}}
\newcommand{\sign}{\ensuremath{\tfrac{D_0}{|D_0|}}}

%spe nobind
\newcommand{\pvacno}{\ensuremath{\boldsymbol{\pi}_{\mathrm{vac}}}}
\newcommand{\rgg}{\ensuremath{\text{R}_g}}
\newcommand{\etabfc}[1]{\ensuremath{\boldsymbol{\eta}_{c\text{R}_g}^{(#1)}}}
\newcommand{\ala}{\ensuremath{a[\La]}}
%\newcommand{\rg}{\ensuremath{R_{\text{g}}}}

% spe Intro
\newcommand{\Pup}{\ensuremath{P_{\uparrow}}}
\newcommand{\Pdow}{\ensuremath{P_{\downarrow}}}

\newcommand{\YYY}{\ensuremath{\mathrm{Y}}}
\newcommand{\Cha}{\ensuremath{\mathrm{C}}}
\newcommand{\Isym}{\ensuremath{\mathrm{I}_{\mathrm{s}}}}
\newcommand{\Kbf}{\ensuremath{\mathbf{\mathrm{K}}}}
\newcommand{\Lbf}{\ensuremath{\mathbf{\mathrm{L}}}}
\newcommand{\Spbf}{\ensuremath{\mathbf{\mathrm{S}}}}
\newcommand{\Jbf}{\ensuremath{\mathbf{\mathrm{J}}}}
\newcommand{\PPh}{\ensuremath{\Phi_{\mathrm{SU}}}}

\newcommand{\hla}{\ensuremath{\mathfrak{h}_\Lambda}}
\newcommand{\pvac}{\ensuremath{\overline{\boldsymbol{\pi}}}}
\newcommand{\pvt}{\ensuremath{\boldsymbol{\pi}}}
\newcommand{\gvac}{\ensuremath{\gamma_{\mathrm{vac}}}}

%%%%%%%%%%%%%%%%%%%%%%%%%%%%%% Transformations sur symboles

\newcommand{\ov}[1]{\ensuremath{\overline{#1}}}
\newcommand{\un}[1]{\ensuremath{\underline{#1}}}
\newcommand{\ww}[1]{\ensuremath{\widehat{#1}}}
\newcommand{\wt}[1]{\ensuremath{\widetilde{#1}}}
\newcommand{\wh}[1]{\ensuremath{\widehat{#1}}}

\newcommand{\wit}[1]{\ensuremath{\widetilde{#1}}}

%%%%%%%%%%%%%%%%%%%%%%%%%%%%%% Normes

\newcommand{\nq}[1]{\ensuremath{\lVert#1\rVert_{\mathcal{Q}}}}
\newcommand{\nqu}[1]{\ensuremath{\lVert#1\rVert_{q_1}}}
\newcommand{\nqo}[1]{\ensuremath{\lVert#1\rVert_{q_0}}}
\newcommand{\nqf}[1]{\ensuremath{\lVert#1\rVert_{\mathcal{Q}_f}}}
\newcommand{\nc}[1]{\ensuremath{\lVert#1\rVert_{\mathfrak{C}}}}
\newcommand{\ncu}[1]{\ensuremath{\lVert#1\rVert_{\mathfrak{C}_1}}}
\newcommand{\ncf}[1]{\ensuremath{\lVert#1\rVert_{\mathfrak{C}_f}}}
\newcommand{\nqr}[1]{\ensuremath{\lVert#1\rVert_{\mathcal{R}}}}
\newcommand{\ns}[2]{\ensuremath{\lVert#2\rVert_{\mathfrak{S}_{#1}}}}
\newcommand{\nlp}[2]{\ensuremath{\lVert#2\rVert_{L^{#1}}}}
\newcommand{\nso}[2]{\ensuremath{\lVert#2\rVert_{H^{#1}}}}
\newcommand{\nhi}[1]{\ensuremath{\lVert#1\rVert_{E}}}
\newcommand{\nqq}[1]{\ensuremath{\lVert#1\rVert_{\text{Ex}}}}
\newcommand{\nqkin}[1]{\ensuremath{\lVert#1\rVert_{\text{Kin}}}}
\newcommand{\nb}[1]{\ensuremath{\lVert#1\rVert_{\mathcal{B}}}}
\newcommand{\ncc}[1]{\ensuremath{\lVert#1\rVert_{\mathcal{C}}}}
\newcommand{\nx}[1]{\ensuremath{\lVert#1\rVert_{\mathcal{X}}}}
\newcommand{\nxu}[1]{\ensuremath{\lVert#1\rVert_{\mathcal{X}_1}}}
\newcommand{\nxf}[1]{\ensuremath{\lVert#1\rVert_{\mathcal{X}_f}}}

\newcommand{\norm}[1]{\ensuremath{\lVert#1\rVert}}
\newcommand{\nqbf}[1]{\ensuremath{\lVert#1\rVert_{\mathbf{Q}}}}
\newcommand{\nqkino}[1]{\ensuremath{\lVert#1\rVert_{\mathrm{T}}}}
\newcommand{\nqbfg}[1]{\ensuremath{\lVert#1\rVert_{\mathbf{Q}_w}}}
\newcommand{\nqbfu}[1]{\ensuremath{\lVert#1\rVert_{\mathbf{Q}_1}}}
\newcommand{\ncg}[1]{\ensuremath{\lVert#1\rVert_{\mathfrak{C}_w}}}
\newcommand{\ncgp}[1]{\ensuremath{\lVert#1\rVert_{\mathfrak{C}'_w}}}
\newcommand{\nxg}[1]{\ensuremath{\lVert#1\rVert_{\mathcal{X}_w}}}

\abstract{We study the Bogoliubov-Dirac-Fock (BDF) model, a no-photon, mean-field approximation of quantum electrodynamics that allows to study relativistic electrons interacting with the vacuum. It is a variational model in which states are represented by Hilbert-Schmidt operators. We prove a charge renormalisation formula that holds close to the non-relativistic limit: the density of a ground state is shown to be integrable although such a state is known not to be trace-class. We prove that we can take the non-relativistic limit by keeping track of the vacuum polarisation. We get an altered Hartree-Fock model due to the screening effect.}

%We study the Bogoliubov-Dirac-Fock model that allows to consider relativistic electrons interacting with the vacuum i%n the presence of an external electrostatic field. It can be seen as a Hartree-Fock approximation of QED, where photons are neglected. A state is described by its one-body density matrix: an infinite rank, self-adjoint operator which is a compact perturbation of the negative spectral projector of the free Dirac operator. 

%sWe are interested in the properties of minimizers of the BDF-energy in the presence of an external field with charge density $\nu\ge 0$ in the regime $\alpha$, $\alpha \log(\Lambda)$ and $\alpha \nu$ (in some norms) small where $\alpha$ is the coupling constant and $\Lambda$ the ultraviolet cut-off. We prove that the density of such minimizer is $L^1$ and compute the effective charge of the system extending the results of \cite{gs} to the full BDF model. We also ensure the existence of minimizers under charge constraint $M\in\mathbb{N}^*$ provided that there holds $M-1< \int \nu$ close to the nonrelativistic limit $\alpha\to 0$ with $\alpha\llo$ \emph{fixed} to a small value. This constrats with the assumptions of \cite{at} where $\La$ is fixed. As a consequence, the nonrelativistic model we obtain in the limit keeps track of the charge renormalisation and is different from the Hartree-Fock model obtained in \cite{at}.}

\section{Introduction}

The relativistic quantum theory of electrons is based on the Dirac operator \cite{Th}: 

\noindent$mc^2\beta-\sum_{j=1}^3i\hbar c\alpha_j\cdot \partial_j$. Here $c$ is the speed of light, $m$ the mass of electron, $\hbar$ the Planck's constant,
\[
\beta:=\begin{pmatrix}\text{id}_{\mathbb{C}^2}& 0\\ 0& -\text{id}_{\mathbb{C}^2} \end{pmatrix},\ \alpha_j:=\begin{pmatrix}0& \sigma_j \\ \sigma_j& 0 \end{pmatrix}\in\text{End}(\mathbb{C}^4),
\]
where the $\sigma_j$'s are the Pauli matrices:
\begin{equation}
\begin{array}{lll}
\sigma_1=\begin{pmatrix} 0 & 1\\ 1 & 0 \end{pmatrix}, & \sigma_2=\begin{pmatrix} 0 & -i \\ i & 0 \end{pmatrix}, & \sigma_3=\begin{pmatrix} 1 & 0 \\ 0 & -1 \end{pmatrix}.
\end{array}
\end{equation} 

\noindent The Dirac operator is a self-adjoint operator acting on $\mathfrak{H}:=L^2(\mathbb{R}^3,\mathbb{C}^4)$ and whose domain is $H^1(\mathbb{R}^3,\mathbb{C}^4)$. In the one-particle theory, the energy of a free particle $\psi\in L^2(\mathbb{R}^3,\mathbb{C}^4)$ is given by $\psh{D_0\psi}{\psi}$, while the spectrum of $D_0$ is $(-\infty,-mc^2]\cup[mc^2,+\infty)$. According to Dirac's interpretation, all the negative energy states are already occupied by "virtual" electrons, the so-called Dirac sea. By the Pauli principle a real electron can only have positive energy.

%In the one-particle theory of Dirac, the energy of a free particle (with wave function $\psi\in L^2(\mathbb{R}^3,\mathbb{C}^4)$) is $\psh{D_0\psi}{\psi}$. The spectrum of $D_0$ is $(-\infty,-mc^2]\cup[mc^2,+\infty)$ and one cannot exclude negative energy state \text{a priori}. To explain why electrons with negative energies are not observed, Dirac postulated all the negative energy states are already occupied by "virtual" electrons, the so-called Dirac sea. By the Pauli principle a real electron can only have positive energy.

In this paper we study the Bogoliubov-Dirac-Fock (BDF) model which is a mean-field approximation of Quantum Electrodynamics (QED). This model, introduced by Chaix and Iracane in \cite{CI}, enables us to consider a system of relativistic electrons interacting with the vacuum in the presence of an electrostatic field.% (\emph{e.g.} that one created by some nucleus). 
This paper is a continuation of previous works by Hainzl, Gravejat, Lewin, S\'er\'e, Siedentop, Solovej \cite{HaiSied,ptf,Sc,mf,at,gs} and Sok  (unpublished work \cite{sok}). In this paper we will extend some results of \cite{gs} and of \cite{at}.

We use relativistic units $\hbar=c=4\pi\eps_0=1$ and set the bare particle mass equal to $1$. The fine structure constant is written $\alpha$. The free Dirac operator is written $D^0=-i\boldsymbol{\alpha}\cdot \nabla+\beta$, furthermore we write $\mathfrak{H}:=L^2(\mathbb{R}^3,\mathbb{C}^4)$ and define $P^0_-$ (resp. $P^0_+$) as the negative (resp. positive) spectral projector of $D_0$. %on % where $\beta,\boldsymbol{\alpha}_j\in\mathcal{M}_4(\mathbb{C})$ are the Dirac matrices defined by

We will not recall here how the BDF energy is derived from QED but refer the reader to \cite{CI} or \cite[Appendix]{ptf}. Let us just say that the starting point is the Hamiltonian of QED $\mathbb{H}_{\text{QED}}$, defined on the electronic Fock space $\mathcal{F}_{el}$. The mean-field approximation consists in restricting the Hamiltonian of QED $\mathbb{H}_{\text{QED}}$ to "Hartree-Fock" states, the so-called BDF states.
%Let us just say that the mean-field approximation consists in restricting the Hamiltonian of QED $\mathbb{H}_{\text{QED}}$ to Hartree-Fock states, the so-called BDF states. 
%This model is derived from full QED by making several approximations: the starting point is the full Hamiltonian $\mathbb{H}$ defined on the Fock space $\mathcal{F}_{el}\otimes\mathcal{F}_{pho}$ (the tensor product of that one of the electrons and that one of the photons) and the first approximation is to neglect the photons and work only with $\mathcal{F}_{el}$ (see \cite{CI} for more details). %We will just recall here what the BDF energy is and introduce the corresponding variational problems.

These BDF states are fully characterized by their one-body density matrix (1pdm) $P$, an orthogonal projector of $L^2(\mathbb{R}^3,\mathbb{C}^4)$. For instance, the projector $P^0_-$ is the 1pdm of the free vacuum $\Omega_0$ of the Fock space $\mathcal{F}_{el}$. Taking $P^0_-$ as a reference state, we consider the reduced 1pdm $Q:=P-P^0_-$. Not all projectors are admissible: a projector $P$ defines a BDF states if and only if the difference $P-P^0_-$ is Hilbert-Schmidt.

\begin{remark}
We recall that a Hilbert-Schmidt operator is a compact operator $Q$ whose integral kernel $Q(x,y)$ is square-integrable, or equivalently whose singular values form a sequence in $\ell^2$. If this sequence is in $\ell^1$, then the corresponding operator is trace-class. 
\end{remark}
%These BDF states are defined in the appendix of \cite{ptf}, it can be shown that a projector $P$ is the 1pdm of a BDF state $\Omega_P$ \emph{iff} $Q$ is Hilbert-Schmidt. 

%By algebraic computation, it can be shown that 
Let $\Omega_P$ be a BDF state with 1pdm $P$. The \emph{formal} difference of the energy $\langle \Omega_P| \mathbb{H}| \Omega_P\rangle$ of the state $\Omega_P$ and that of $\Omega_0$ gives a function of $Q$, the so-called BDF energy.

We assume the presence of an external density of charge $\nu$ (real-valued) of finite Coulomb norm:
\begin{equation}
D(\nu,\nu)=\ncc{\nu}^2:=4\pi\dint\frac{|\wh{\nu}(k)|^2}{|k|^2}dk=\diint\frac{\nu(x)\nu(y)^*}{|x-y|}dxdy.
\end{equation}
The last equality holds for suitable $\nu$ (for instance $\nu\in \mathcal{C}\cap L^{6/5}(\mathbb{R}^3)$).

Formally the BDF energy of a state with reduced 1pdm $Q$ is:
\begin{equation}\label{formel}
\left\{\begin{array}{l}
\ttr_{P^0_-}(D_0 Q)-\alpha D(\rho_Q,\nu)+\frac{\alpha}{2}\Big(D(\rho_Q,\rho_Q)-\text{Ex}[Q]\Big),\\
\ttr_{P^0_-}(D_0 Q):=\ttr\big\{P^0_- (D_0 Q)P^0_-+P^0_+(D_0 Q)P^0_+\big\},\\
\text{Ex}[Q]:=\diint \frac{|Q(x,y)|^2}{|x-y|}dxdy.
\end{array}\right.
\end{equation}
Here, $\alpha>0$ is the coupling constant, $Q(x,y)$ the integral kernel of the operator $Q$ and $\rho_Q$ is its density: $\rho_Q(x)=\ttr_{\mathbb{C}^4}(Q(x,x))$. We recognize the kinetic energy, the interaction energy with $\nu$, the direct term and the exchange term as in Hartree-Fock theory. 

This expression is not always well defined, in particular the formula for the density $\rho_Q$ makes sense \emph{a priori} only if $Q$ is (locally) trace-clas.

An ultraviolet cut-off $\La>0$ is needed: many choices are possible. In \cite{ptf,Sc,mf,at}, Hainzl \emph{et al.} have considered a "sharp" cut-off in which $L^2(\RR,\CC)$ is replaced by its subspace $\hl$ made of functions whose Fourier transforms vanish outside a ball $B(0,\La)$.

In \cite{mf}, Hainzl \emph{et al.} proposed another BDF energy based on an altered Dirac operator $\D$ and on its spectral projectors% different from $D_0$ is introduced in \cite{mf} with projectors
\begin{equation}
\mathcal{P}^0_{\pm}:=\chi_{\mathbb{R}^{*}_{\pm}}\big(\D\big)
\end{equation}
%\PP:=\chi_{(-\infty,0)}(\D)\text{\ and\ }\PPP:=\chi_{(0,+\infty)}(\D).
%\]
%(
In fact Hainzl \emph{et al.} studied the periodized Hamiltonian $\mathbb{H}_{\text{L}}$ in a finite box $[-\tfrac{\text{L}}{2},\tfrac{\text{L}}{2})$ (with periodic boundary conditions). Setting an ultraviolet cut-off, the problem becomes finite dimensional: for $\text{L}$ large enough they prove there exists a unique ground state which tends to $\PP$ as $\text{L}$ tends to $+\infty$. Thus the BDF energy with respect to this minimizer ("substracting $\langle \Omega_{\PP}|\mathbb{H}|\Omega_{\PP}\rangle$") gives a more relevant model.

The operator $\D$ has the same structure as the Dirac operator: $\D:=\boldsymbol{\alpha}\cdot\mathbf{g}_1(-i\nabla)+\beta g_0(-i\nabla)$ and it satisfies the following equation:
\begin{equation}
\D=D_0+\dfrac{\alpha}{2}\dfrac{\text{sgn}(\D)(x,y)}{|x-y|}.
\end{equation}
Here $g_0$ and $\mathbf{g}_1$ are smooth functions of $B(0,\La)$.

In this paper the energy functional $\mathcal{E}_{\text{BDF}}^\nu$ is defined on a subspace $\mathcal{K}$ of $\mathfrak{S}_2(\hl)$, made of convex combinations of reduced 1pdm's of form $P-\PP$. The set $\mathcal{K}$ is properly defined in the next section and $\mathcal{E}_{\text{BDF}}^\nu$ is defined as in \eqref{formel} except that we replace the $P^0_-$-trace by a $\PP$-trace:
%\[
%\mathcal{I}=\{ Q\in\mathfrak{S}_2(\hl),Q^*=Q,\ -\PP\le Q\le \PPP\}.
%\]
%The set $\mathcal{I}$ is the convex hull in $\mathfrak{S}_2(\hl)$ of the reduced 1pdm's and $\mathcal{K}$ is defined in the next section. 
%In the BDF energy we replace the kinetic energy $\ttr_{P^0_-}(D_0 Q)$ by $\tr(\D Q)$ defined by:
\begin{equation}\label{cutoff}
\begin{array}{l}
\ttr_{0}(\D Q):=\ttr\big\{\PP (\D Q)\PP+\PPP(\D Q)\PPP\big\},\\
\end{array}
\end{equation}

A global minimizer of $\mathcal{E}_{\text{BDF}}^\nu$ is interpreted as the polarized vacuum in the presence of $\nu$. 

The charge of a state $Q\in\mathcal{K}$ is given by $\ttr_{0}(Q)$. Thus the ground state of a system with $M$ electrons is given by a minimizer of $\mathcal{E}_{\text{BDF}}^\nu$ over the corresponding charge sector.

Furthermore, we define then the energy functional for $q\in\mathbb{R}$:
\[
\left\{\begin{array}{rl}
        E_{\text{BDF}}^\nu(q)&:=\inf\,\{\mathcal{E}_{\text{BDF}}^\nu(Q),\,Q\in\mathcal{Q}(q)\},\\
        \mathcal{Q}(q)&:=\{Q\in\mathcal{K},\,\ttr_{0}(Q)=q\}.
       \end{array}
\right.
\]

%An important question is that of the existence of a minimizer for $E_{\text{BDF}}^\nu(q)$. 
The question becomes: does there exist a minimizer for $E_{\text{BDF}}^\nu(q)$?

In \cite{at}, Hainzl \emph{et al.} proved that a sufficient condition for the existence is the validity of binding inequalities at level $q$:
\begin{equation}\label{HVZst}
 \forall\,q'\in\mathbb{R}\backslash\{0,q\},\ E_{\text{BDF}}^{\nu}(q)<E_{\text{BDF}}^{\nu}(q-q')+E_{\text{BDF}}^{0}(q').
\end{equation}
%It is possible to adapt the proof in our model (Theorem \ref{HVZ} stated here, proved in \cite{these}).

A much more difficult task is to check that these inequalities hold. 

In \cite{at}, the authors showed the following.

Let a density $\nu\in L^1(\RR,\mathbb{R}_+)\cap \mathcal{C}$, an integer $0\le M<\int\nu+1$ and a cut-off level $\La_0>0$ be given, then there exists minimizer for $E_{\text{BDF}}^\nu(M)$ provided $\alpha\le \eps_0(\nu,\La_0)$ for some number $\eps_0(\nu,\La_0)>0$.

In \cite{sok} we proved that $E_{\text{BDF}}^0(1)$ admits a minimizer provided that $\alpha,\La^{-1}$ and $L:=\alpha\llo$ are small enough. In other words, surprisingly an electron can bind alone in the Dirac sea without any external density, due to the vacuum polarisation.

In both cases the results hold in the non-relativistic regime $\alpha\ll 1$.

\medskip

Let $M\in\mathbb{Z}$: a minimizer for $E_{\text{BDF}}^\nu(M)$ satisfies a self-consistent equation of the form \cite{at}
\begin{equation}\label{eq_dcursif}%[
 Q+\PP=\chi_{(-\infty, \mu]}\Big(\D+\alpha((\rho_Q-\nu)*\tfrac{1}{|\cdot|}-\tfrac{Q(x,y)}{|x-y|})\Big)=:\chi_{(-\infty, \mu)}(D_Q).
\end{equation}%)
Here, $\mu$ is a Lagrange multiplier due to the charge constraint $M$, interpreted as a chemical potential. For $M>0$, it is positive, the projector $\chi_{(-\infty,0)}(D_Q)$ is interpreted as the 1pdm of the polarized vacuum while $\chi_{[0,\mu]}(D_Q)$ is the 1pdm of the "real" electrons. For $\alpha$ sufficiently small, the last projector is indeed of rank $M$. Furthermore in the limit $\alpha\to 0$, $\La_0>0$ fixed, its scaling by $\alpha^{-1}$ tends (up to extraction) to a minimizer of the Hartree-Fock energy $\mathcal{E}_{\text{HF}}^Z$ for $M$ electrons and $Z:=\int\nu$, restricted to $L^2(\RR,\mathbb{C}^2\oplus 0)$.

In \cite{sok}, a similar result is obtained with a minimizer for $\mathcal{E}_{\text{BDF}}^0(1)$ in the non-relativistic limit $\alpha\to 0,\ \alpha\llo:=L_0$ fixed, the limit is then the Choquard-Pekar model \cite{L}. %The amplitude of the vacuum polarization is of order $\mathcal{O}(L/(1+L))$: fixing $L$ is a way to keep track of it and this is done in cite{sok}.

In this paper we show that, assuming $L=\alpha\llo\le L_0$, there exists a minimizer for $E_{\text{BDF}}^\nu(M)$ as soon as $M<\int\nu+1$ and $\alpha\le \alpha_1(\nu,L)$. The nonrelativistic limit is an altered Hartree-Fock model: writing $Z=\int\nu$ and $a=(\tfrac{2}{3\pi}L)/(1+\tfrac{2}{3\pi}L)<1$ the energy is
%In $\cite{at}$ in the nonrelativistic limit we get the Hartree-Fock model while in cite{sok} we get the Choquard-Pekar model \cite{L}. In fact $L$ is roughly the amplitude of the vacuum polarization: we keep track of it in the second case. We aim here to see what happens in the regime of cite{sok} for $E_{\text{BDF}}^\nu(M)$. We obtain that in the limit, the model is an altered Hartree-Fock model where the polarization effects remain. If we write $Z=\int\nu$ and $a=\tfrac{2}{3\pi}L$ we have:
\[
\begin{array}{l}
\forall\,\G\in\mathfrak{S}_1(L^2(\RR,\mathbb{C}^4)),\,0\le\G\le 1,\, \ttr(\G)=M:\\
\ \ \ \ \ \ \ \ \ \ \ \ \ \ \ \mathcal{E}_{nr}^Z(\G):=\tfrac{1}{2}\ttr(-\Delta \G)-Z(1-a)\ttr\Big(\tfrac{1}{|\cdot|}\G\Big)+\tfrac{1}{2}\Big\{\ncc{\rho_\G}^2-\text{Ex}[\G]\Big\}-\frac{a}{2}\ncc{\rho_\G}^2.
\end{array}
\]
%The last term must be thought of as $-\tfrac{a}{2}\ttr(\rho_\G*\tfrac{1}{|\cdot|}\G)$. 
The vacuum polarizes due to the presence of $\nu$ and the electrons: the positive charge $\nu$ attracts a cloud of negative charge which makes it appear smaller (hence the term $Z(1-a)$) while the electrons repelled them resulting to an attractive well created by the distortion (hence the term $-\tfrac{a}{2}\ncc{\rho_\G}^2$ like in a polaron model). This result gives a wider range of existence of ground state in the space of parameters $(\alpha,\La)$ compared to that of \cite{at}, where the quantity $\alpha\log(\La_0)$ is neglected and considered as $\underset{\alpha\to 0}{o}(1)$.

To prove it, it is necessary to have a good understanding of a minimizer $Q_0$ and of its density $\rho_{Q_0}$. In \cite{gs} the authors proved that, in the simplified model without the exchange term, the density of a minimizer is integrable. This is a natural result: the distortion of the vacuum due to a finite number of charged particles with finite Coulomb energy should also be finite.

Mathematically speaking however this is a non-trivial fact because a minimizer for $E_{\text{BDF}}^\nu(M)$ is \emph{not} trace-class. As in \cite{gs} we prove that, assuming that $L$ is small enough and $M,\ncc{\nu}^2\apprle \llo$, then the density $\rho_Q$ of a minimizer $Q$ is in $ L^1\cap \mathcal{C}$. Moreover, the following \emph{charge renormalisation} formula holds:
\begin{equation}\label{charge_ren}
  \dint(\rho_Q-\nu)=:Z_3(M-Z)\simeq\dfrac{M-Z}{1+\tfrac{2}{3\pi}L},
\end{equation}
where $Z_3$ is interpreted as the \emph{renormalization constant} \cite{renorm}. This means that the total observed charge $\int(\rho_Q-\nu)$ is different from the real charge $M-Z$ of the system.

The quantity $L=\alpha\llo$ is related to $Z_3$. In the reduced BDF model where the exchange term is neglected, Gravejat \emph{et al.} showed in \cite{gs} that the density $\rho_Q$ of a minimizer of the reduced energy $E_{\text{rBDF}}^\nu(M)$ is radial as soon as $\nu$ is radial and that, in this case, away from the origin, the electrostatic potential of the system is
\[
\alpha (\rho_Q-\nu)*\dfrac{1}{|\cdot|}(x)\underset{x\to+\infty}{\sim}\dfrac{\alpha Z_3(M-Z)}{|x|}.%,\ Z_3\simeq\dfrac{1}{1+\tfrac{2}{3\pi} L}.
\]
In the full model we were unable to prove such behaviour at infinity but we think this is true. Taking $L$ small corresponds then to considering $Z_3$ close to $1$.

The main contribution of this paper is the integrability result stating that the density of a minimizer is in $L^1$ together with the charge renormalisation formula \eqref{charge_ren}. It cannot be easily obtained from \cite{gs}, the presence of the exchange term complicates the study. In our results, we were unable to remove the technical conditions $M,\ncc{\nu}^2\apprle \llo$. We emphasize here that we can prove the same results with another choice of cut-off considered in \cite{gs}, the one consisting in replacing $\D$ by $D_0(1-\tfrac{\Delta}{\La^2})$ in $L^2(\RR,\mathbb{C}^4)$.

The paper is organized as follows: in the next section we properly define the variational problem $\mathcal{E}_{\text{BDF}}^\nu$ and states the main results. 

In Section \ref{des}, we derive two fixed point schemes from the equation satisfied by a minimizer, using the Cauchy expansion. Moreover \emph{a priori} estimates are proved in Subsection \ref{secest}. 

In Section \ref{q10first} we prove important estimates on a term of the Cauchy expansion ($"Q_{1,0}"$) and prove Theorem \ref{alafurry}. 

Section \ref{fixedpointmethod} is devoted to prove estimates for the fixed point method and apply it to prove that the density of a minimizer is in $L^1$ (under some assumptions). 

We prove the formula of charge renormalization (Theorem \ref{density}) and the existence of minimizers close to the nonrelativistic limit (Theorem \ref{existence}) in Section \ref{mainsproof}. 

The nonrelativistic energy is studied in Appendix \ref{nonrel}. The very technical Appendix \ref{flaflafla} is devoted to prove Proposition \ref{profla}. We prove Lemma \ref{exch} which is used for Sections \ref{q10first} and \ref{fixedpointmethod} in Appendix \ref{appA}.

\begin{remark}[Fourier transform]
Throughout this paper, the Fourier transform $\mathscr{F}$ is defined as the extension of
\[
\forall\ f\in\,L^2(\RR)\cap L^1(\RR):\ \wh{f}(p):=\frac{1}{(2\pi)^{3/2}}\dint_{\mathbb{R}^3}f(x)e^{-ip\cdot x}dx.
\]
\end{remark}

\begin{remark}[Form of $\D$]
The operator $\D$ was first studied by Lieb and Siedentop in \cite{ls} in another context. We know $\mathbf{g}_1(-i\nabla)=\tfrac{-i\nabla}{|-i\nabla|}g_1(-i\nabla)$ and $g_0,g_1$ are radial functions satisfying
\begin{equation}\label{gstar}
\forall p\in B(0,\La),\ |p|\le g_1(p)\le g_0(p)|p|\text{\ and\ }1\le g_0(p)\le 1+\text{Cst}\times \alpha\llo.
\end{equation}
We define
\begin{equation}
m:=\inf\,\sigma\big( |\D|\big).
\end{equation}
For $\alpha\llo$ and $\alpha$ sufficiently small, $m$ is equal to $g_0(0)$.

Useful estimates on $g_0,\mathbf{g}_1$ are proved in \cite{sok}.
\end{remark}

\section{Description of the model and main results}

\paragraph{BDF Energy}
%\noindent\textbf{BDF Energy.}\ 
We assume there is an external density of charge $\nu$ (real-valued) of finite Coulomb norm ($\ncc{\nu}<+\infty$).
%\[
%D(\nu,\nu)=\ncc{\nu}^2:=\frac{4\pi}{(2\pi)^3}\dint\frac{|\wh{\nu}(k)|^2}{|k|^2}dk=\diint\frac{\nu(x)\nu(y)^*}{|x-y|}dxdy.
%\] 
%The last equality holds for sufficiently smooth $\nu$ (for instance $\nu\in \mathcal{C}\cap L^{6/5}(\mathbb{R}^3)$).

%Formally the BDF energy for a BDF state with reduced 1pdm $Q$ is:
%\[
%\ttr(D_0 Q)-\alpha D(\rho_Q,\nu)+\frac{\alpha}{2}\bigg(D(\rho_Q,\rho_Q)-\diint \frac{|Q(x,y)|^2}{|x-y|}dxdy\bigg).
%\]
%Above $\alpha>0$ is the coupling constant, $Q(x,y)$ the kernel of the operator $Q$ and $\rho_Q$ is its density: $\rho_Q(x)=\ttr_{\mathbb{C}^4}(Q(x,x))$. We recognize the kinetic energy, the interaction energy with $\nu$, the direct term and the exchange term as in Hartree-Fock theory. This expression does not always make sense even if $Q$ is Hilbert-Schmidt (that is if $\iint |Q(x,y)|^2dxdy<+\infty$), in particular it is not always possible to define $\rho_Q$. We will change a bit the expression and impose some conditions on $Q$ to get a well-defined energy.

Let us recall our choice of cut-off: following \cite{at}, we replace $D_0$ by $\D$ and work in $\hl$, defined by
\[
\hl:=\big\{\psi\in L^2(\RR,\CC),\ \text{supp}\,\wh{\psi}\subset B_{\RR}(0,\La)\big\},\ \La>0.
\]
%\[
%\D:=D_0\big(1-\frac{\Delta}{\La^2}\big)\text{\ with\ domain\ }D(\D)=H^{3/2}(\mathbb{R}^3,\mathbb{C}^4).
%\]
%This choice is done in order to get control in the Direct space of objects defined in the Fourier space. %In this paper
%We remark that $\chi_{(-\infty,0)}(\D)=P^0_-$. 
We write $\mathfrak{S}_p(\hl)$ the Schatten class of compact operators $A$ in $\hl$ such that $\ttr(|A|^p)<+\infty$ \cite{Sim}. The set of $\PP$-trace operators is \cite{at}:
\begin{equation}
\mathfrak{S}_1^{\PP}:=\{Q\in\mathfrak{S}_2(\hl), Q^{++},Q^{--}\in\mathfrak{S}_1(\hl)\}
\end{equation}
where $Q^{\eps_1 \eps_2}:=\mathcal{P}^0_{\eps_1} Q\mathcal{P}^0_{\eps_2}$. This set is a Banach space with
\begin{equation}
\lVert Q\rVert_{\mathfrak{S}_1^{\PP}}:=\ns{2}{Q^{+-}}+\ns{2}{Q^{-+}}+\ns{1}{Q^{--}}+\ns{1}{Q^{++}}.
\end{equation}

We recall that $\ttr_0\big(|\D|(Q^{++}-Q^{--})\big)$ is the kinetic energy functional.

%\begin{equation}
%\ttr_{P^0_-}(\D Q):=\ttr(|\D|^{1/2}(Q^{++}-Q^{--})|\D|^{1/2}).
%\end{equation}
We work in a subset of this space, namely
\begin{equation}
\mathcal{K}:=\{Q,\ -\PP\le Q\le \PPP\}\cap\mathfrak{S}_1^{P^0_-}\subset\{Q,\ Q^*=Q\}\cap\mathfrak{S}_1^{P^0_-}.
\end{equation}
It is the closed convex hull of the $P-\PP\in\mathfrak{S}_1^{\PP}$, where $P$ is an orthogonal projection.

The density $\rho_Q$ must be defined consistently with the usual formula when $Q$ is (locally) trace-class and it must also be of finite Coulomb energy.

Let $Q$ be in $\mathfrak{S}_1^{\PP}$, then $\rho_Q$ is defined by duality:
\begin{equation}
\forall\ V\in\mathcal{C}',\ QV\in\mathfrak{S}_1^{\PP}\ \mathrm{and}\ \ttr_{0}(QV)={}_{\mathcal{C}'}\psh{V}{\rho_Q}_{\mathcal{C}}.
\end{equation} 
The map $Q\in \mathfrak{S}_1^{\PP}\mapsto \rho_Q\in \mathcal{C}$ is continuous \cite[Proposition 2]{gs}. %Furthermore $\rho_Q(x)$ is well defined for $Q$ is locally trace-class.%$D(\rho_Q,\rho_Q)$ is the so-called \emph{direct term}.% or Lemma. (mettre le lemme en appendice)

%%%%%%% with norms 

\noindent The exchange term is well defined: thanks to Kato's inequality \cite{stab,mf,ptf}
\begin{equation}
\begin{array}{rl}
\dfrac{2}{\pi}\diint \frac{|Q(x,y)|^2}{|x-y|}dxdy&\le \ttr(|\nabla| Q^2)\le \ttr(|D_0|Q^2)=\ttr\{|D_0|^{1/2}Q^2|D_0|^{1/2}\}\\
    \mathrm{and\ for}\ Q\in\mathcal{K}:      &\le \ttr \{  |D_0|^{1/2}(Q^{++}-Q^{--})|D_0|^{1/2}\}\le \ttr_{P^0_-}(\D Q),
\end{array}
\end{equation}

%\begin{notation}
%For any operator $Q$ with kernel $Q(x,y)$, we define $R_Q(x,y):=\frac{Q(x,y)}{|x-y|}$.
%\end{notation}
The BDF energy is defined as follows:% for any external field $\nu\in \mathcal{C}:$
\begin{equation}
\mathcal{E}^{\nu}_{\text{BDF}}(Q):=\ttr_{P^0_-}(\D Q)-\alpha D(\nu,\rho_Q)+ \frac{\alpha}{2}\Big(D(\rho_Q,\rho_Q)-\diint\frac{|Q(x,y)|^2}{|x-y|}dxdy\Big),\ Q\in\mathcal{K}.
\end{equation}

%Minimisation under charge constraint $\ttr_{P^0_-}(Q)=q\in\mathbb{R}$ and 
As said in the introduction we define the energy functional $E_{\text{BD}}^\nu(q)$ by the infimum over $\mathcal{Q}(q)=\{Q\in\mathcal{K},\ \ttr_{P^0_-}(Q)=q\}$.
For $M\in\mathbb{N}^*$, let us say that the problem $E^{\nu}_{\text{BDF}}(M)$ has a minimizer: as pointed out in \cite{at,gs} such a minimizer $\g'=\g+N$ must be of the following form:
\begin{equation}\label{eqmin}
\left\{
\begin{array}{l}
\g+P^0_-=\chi_{(-\infty,0)}\big\{\D+\alpha ((\rho[\g']-\nu)*\tfrac{1}{|\cdot|}-R[\g']) \big\}=:\chi_{(-\infty,0)}(D_{\g'}),\\
N=\chi_{(0,\mu]}\big\{\D+\alpha\big((\rho_{\g'}-\nu)*\tfrac{1}{|\cdot|}- (R_{\g'})\big) \big\}=\sum_{j=1}^{M_0}\ket{\psi_j}\bra{\psi_j},\\
\text{so\ }D_{\g'} \psi_j=\mu_j \psi_j\text{\ and\ we\ write:}\,n:=\rho_N=\sum_j |\psi_j|^2.
\end{array}
\right.
\end{equation}%)
We choose $0\le\mu_1\le \mu_2\le \cdots\le \mu_{M_0}= \mu<m$. \textit{A priori} $M_0\neq M$ but in our regime they are equal (Lemma \ref{rank}). Indeed in the spirit of \cite{ptf} the equation of the dressed vacuum $\g$ enables us to say that $(\g',\rho_{\g'}-\nu)$ is the only fixed point of some function $F^{(1)}$ defined in (a ball of) the Banach space $\mathcal{X}_1=\mathbf{Q}_1\times \mathcal{C}$ where
\[
\lvert \lvert Q \rvert \rvert_{\mathbf{Q}_1}^2=\nqkino{Q}^2:=\diint (\ed{p}+\ed{q})|\wh{Q}(p,q)|^2dpdq.
\]

\begin{notation}
For a density $\rho\in\mathcal{C}$ we write: $ v_\rho=v[\rho]:=\rho*\frac{1}{|\cdot|}.$

For an operator $Q\in\mathfrak{S}_1^{P^0_-}$ with integral kernel $Q(x,y)$ we define the operator $R_Q=R[Q]$ by the formula:
 \[
  R_Q(x,y):=\frac{Q(x,y)}{|x-y|}.
 \]

We remark that $\text{Ex}[Q]=\ttr(R_Q^*Q)=:\nqq{Q}^2$.
 
Moreover we write
 \begin{equation}\label{mf_op}
  B_Q:=v[\rho_Q-\nu]-R_Q\text{\ and\ }D_Q:=\D+\alpha B_Q.
 \end{equation}

\end{notation}

\paragraph{The Cauchy expansion}
%\noindent \textbf{Cauchy's expansion:} 
Let $\g'=\g+N$ be a minimizer for $E^{\nu}_{\text{BDF}}(M)$, the decomposition being that of \eqref{eqmin}. 

\begin{notation}
Throughout this paper $n:=\rho_N$, moreover we write $\rho_{\g}'$ for $\rho_{\g'}$ and the double prime means $-\nu$ is added: 
\[
\rho_{\g}'':=\rho_\g+n-\nu,\ n''=n-\nu.
\]
We also write $B'_\g=B_{\g'}:=\rho_{\g}''*\tfrac{1}{|\cdot|}-R[\g']$.
\end{notation}

By functional calculus, we expand $\chi_{(-\infty,0)}(D_Q)-\PP$ in power of $\alpha$: this is the Cauchy expansion \cite{ptf}
%\noindent Writing the Cauchy expansion \cite{ptf} in \eqref{eqmin} we get:
\begin{equation}\label{def_f1_eq}
\left\{
\begin{array}{rl}
\g+N&=N-\dfrac{1}{2\pi}\dint_{-\infty}^{+\infty}d\eta\Big( \frac{1}{D_{\g'}+i\eta}-\frac{1}{\D+i\eta}\Big)=\ssum_{j=1}^{+\infty}\alpha^j Q_j(\g',\rho_{\g}''),\\
Q_j(\g',\rho_\g'')&:=-\dfrac{1}{2\pi}\dint_{-\infty}^{+\infty}d\eta\frac{1}{\D+i\eta}\Big( B_{\g'}\frac{1}{\D+i\eta} \Big)^j.
\end{array}
\right.
\end{equation}

We define $Q_{k,l}$ as the part of $Q_{k+l}(Q,\rho)$ which is a homogeneous polynomial of degree $k$ in $R_{Q}$ and of degree $l$ in $\rho$; $\rho_{k,l}(Q,\rho)$ denotes its density. For $\ell\ge 1$ and $(Q,\rho)\in \mathfrak{S}_2(H^{1/2})\in\mathcal{C}$, $\widetilde{Q}_\ell[Q,\rho]$ is the operator:
\[
\widetilde{Q}_\ell[Q,\rho]:=\ssum_{j=\ell}^{+\infty}\alpha^{j-\ell}Q_j[Q,\rho].
\]

As shown in \cite{ptf,gs} we have
\begin{equation}\label{def_f1_rho}
\rho_{0,1}[\rho]=-\mathscr{F}^{-1}(B_\La)* \rho
\end{equation}
where $\mathscr{F}^{-1}(B_\La)$ is a radial $L^1$ function.

In the following Lemma, we refer to the Banach spaces $\mathbf{Q}_w$ and $\mathfrak{C}_w$: they are defined below \eqref{q_w}. This Lemma is proved in Section \ref{q10first}.
\begin{lem}\label{TbfS}
$F_{1,0}:Q\mapsto Q_{1,0}(Q)$ is a bounded linear map of $\mathfrak{S}_p$ for $p=1$ and $p=2$ with respective norms $\mathcal{O}(\llo)$ and $\mathcal{O}(\sqrt{\llo})$. By interpolation $F_{1,0}$ is in $\text{L}(\mathfrak{S}_p)$ for $1<p=1+\eps<2$ with norm $\mathcal{O}((\llo)^{1-\tfrac{\eps}{2}})$.

Moreover it is also a bounded operator in $\text{L}(\mathbf{Q}_w)$ with norm $\mathcal{O}(1)$, and the function
\[
\rho F_{1,0}:Q\in\mathbf{Q}_w\mapsto \rho\big(F_{1,0}[Q]\big)\in\mathfrak{C}_w
\]
is bounded with norm $\mathcal{O}(\sqrt{\llo})$. Provided that $\alpha\llo$ is sufficiently small, the operator $(\text{Id}-\alpha F_{1,0})$ is invertible with inverse $\Tbf$ in all those Banach spaces with norm $\mathcal{O}(1)$. The function $\mathfrak{t}:Q\in\mathbf{Q}_w\mapsto \rho\big(\Tbf[Q]-Q\big)\in\mathfrak{C}_w$ is bounded and 
\[\ncg{\mathfrak{t}_Q}\apprle \sqrt{L\alpha}\nqbfg{Q}.
\]
\end{lem}

%\begin{remark}
%It is a bounded linear map in other Banach spaces and among them $\mathbf{Q}_1$. In particular if $Q\in \mathbf{Q}_1$, then $\rho[Q_{1,0}(Q)]\in\mathcal{C}$.
%\end{remark}
We write
\begin{equation}
\mathfrak{T}:=\Tbf-\mathrm{Id},\ \tau_Q:=\rho_{\Tbf(Q)},\tau_{j,k}:=\rho_{\Tbf(Q_{j,k})}\ \mathrm{and}\ \mathfrak{t}_Q:=\rho_{\mathfrak{T}(Q)}.
\end{equation}
If $Q\in \mathbf{Q}_{w=1}\cap \mathfrak{S}_1^{P^0_-}$ then $\tau_Q\in\mathcal{C}$ and if $(Q,\rho_Q)\in\mathbf{Q}_w\times\mathfrak{C}_w$ then $\tau_Q\in\mathfrak{C}_w$.

The self-consistent equation \eqref{eqmin} is rewritten as follows:
\[
(\text{Id}-\alpha F_{1,0})(\g')=N+\alpha Q_{0,1}(\rho_{\g}'')+\ssum_{j=2}^{+\infty} Q_j(\g',\rho_{\g}'').
\]
Taking the inverse $\Tbf$, we get:
% (which is possible as long as $\alpha \llo=L$ is sufficiently small), we obtain:
\begin{equation}\label{inverse}
\g'=\Tbf\Big\{ N+\alpha Q_{0,1}(\rho_{\g}'')+\ssum_{j=2}^{+\infty} Q_j(\g',\rho_{\g}'')\Big\}.
\end{equation}
%\begin{notation}
%We write $\tau_j$ for $\rho[\Tbf(Q_j)]$ and $\tau_{k,l}$ for $\rho[\Tbf(Q_{k,l})]$.
%\end{notation}
The important proposition holds:
\begin{pro}\label{profla}
For $\rho\in\mathcal{C}$ we have $\alpha\tau_{0,1}(\rho)=-\check{f}_\La* \rho$ where $\check{f}_\La$ is a radial $L^1$ function whose $L^1$-norm is $\mathcal{O}(\alpha\llo)$.
\end{pro}
Its technical proof is in Appendix \ref{flaflafla}. 

There holds a Theorem \emph{{\`a} la Furry} \cite{Fu,ptf}:
\begin{theo}\label{alafurry}
There exists $K>0$ such that for any $\rho_0,\rho_1$ (say in $\mathcal{C}$) and $\alpha\sqrt{\llo}\le K$ there holds:
\begin{equation}
\rho\big\{\Tbf(Q_{0,2}(\rho_0))\big\}=\rho\Big\{\Tbf\big(Q_{1,1}(\Tbf Q_{0,1}(\rho_1),\rho_0)\big)\Big\}=0.
\end{equation}
\end{theo}
\begin{remark}
$\Tbf(Q_{0,2}(\rho_0))$ and $\Tbf(Q_{1,1}(\Tbf(Q_{0,1}(\rho_1)),\rho_0))$ may not vanish but their density do due to the fact that the trace $\mathrm{Tr}_{\mathbb{C}^4}$ is taken.
The smallness of $\alpha\sqrt{\llo}$ is to ensure the $\Tbf$ operator is well defined on $\mathbf{Q}_1$.
\end{remark}

\paragraph{Main Theorems}
%\noindent \textbf{Computation of $\int_{\mathbb{R}} \rho_\g(x)dx$:} 
%In the following theorem, $\fla$ refers to a function defined in Section \ref{minfps} and studied in Appendix \ref{flaflafla}. Let us just emphasize here that $\fla(0)=\mathcal{O}(\alpha\llo)$ and that $\Fla:=\fla/(1+\fla)$.
\begin{theo}[Computation of $\int_{\mathbb{R}} \rho_\g(x)dx$]\label{density}
Let $M$ be in $\mathbb{N}$ and $\g'=\g+N$ be a minimizer of $E_{\text{BDF}}^{\nu}(M)$ and assume $M,\ncc{\nu}^2\apprle \llo$ and \eqref{para}, the decomposition of $\g'$ is that of \eqref{eqmin}. Then $\rho_\g\in L^1$ and
\begin{equation}%changement densit\'e
\dint \rho_\g(x)dx=-\frac{\alpha\fla(0)}{1+\alpha \fla(0)}(M-Z)
\end{equation}
\end{theo}

%\paragraph{Existence of minimizers}
%[
%\noindent\textbf{Existence of minimizers.}
\begin{theo}[Existence of minimizers]\label{existence}
There exists $K_0>0$ satisfying the following result: 

\noindent for any non-negative function $\nu\in\mathcal{C}\cap L^1$ with $Z=\int\nu$ and $0<L\le 1/(MK_0)$, there exists $\alpha_1=\alpha_1(\nu,L)>0$ such that if $\alpha\le \alpha_1$ then for any integer $0\le M<Z+1$ the problem $E^{\nu}_{\text{BDF}}(M)$ admits a minimizer. 

Let $\g'=\chi_{(0,\mu]}(D_{\g'})$ be a minimizer, decomposed as in \eqref{eqmin} and let $U_\alpha$ be defined as follows: 
\[
U_\alpha:\begin{array}{rll}L^2(\mathbb{R}^3,\mathbb{C}^4)&\to&L^2(\mathbb{R}^3,\mathbb{C}^4)\\ \phi(x)&\mapsto& \alpha^{-3/2}\phi(\tfrac{x}{\alpha})\end{array}.
\]%[
We write $\tfrac{\fla(0)}{1+\fla(0)}=a$, then as $\alpha$ tends to $0$, $U_{\alpha}^*\chi_{(0,\mu]}(D_{\g'})U_{\alpha}$ tends to a minimizer of
\[
\begin{array}{rl}
\mathcal{E}_{nr}^Z(\G)&:=\frac{1}{2}\ttr(-\Delta \G)-Z(1-a)\ttr\big(\frac{1}{|\cdot|} \G\big)+\frac{1}{2}(D(\rho_\G,\rho_\G)-\mathrm{Ex}[\G])-\frac{a}{2}D(\rho_\G,\rho_\G),\\
&\ \ \ \ \ \ \ \ \ \mathrm{for}\ 0\le \G\le 1,\ \ttr(\G)=M\mathrm{\ and\ }\frac{1+\beta}{2}\G=0.
\end{array}
\]
%)
\end{theo}
\begin{remark}

%The difference between the present result and Theorem 3 of \cite{at} is that here the cut-off $\La$ is not fixed, we just ask for smallness of the quantity $\alpha \llo$. In fact, the proof will be essentially \emph{the same}, this theorem is based on a more precise description of the density of an approximate minimizer. It appears that the non-relativistic limit is not the Hartree-Fock model: we study it in Appendix \ref{nonrel}.

Thanks to Section \ref{flaflafla} and \cite{gs} we have
\[
\dfrac{\fla(0)}{1+\fla(0)}=\frac{\tfrac{2}{3\pi}\alpha\llo}{1+\tfrac{2}{3\pi}\alpha\llo}+\mathcal{O}(\alpha+(\alpha\llo)^2).
\]
\end{remark}

\paragraph{Banach spaces}
%\noindent\textbf{Banach spaces.} 
We use several Banach spaces. For $p\in[1,+\infty]$, $s\ge 0$, $\nlp{p}{\cdot}$ (resp.$\nso{s}{\cdot}$) is the norm of the usual $L^p$ (resp. Sobolev) space. We write $\ns{p}{\cdot}$ for the norm of Schatten class operators $\mathfrak{S}_p$ \cite{Sim}. The norm of bounded linear operator in $\mathfrak{H}$ is written $\nb{\cdot}$. We recall $\nqq{\cdot}$ and $\ncc{\cdot}$ have already been defined in Sections 1 and 2 and $\nqbfg{\cdot},\ncg{\cdot}$ are defined in Remark \ref{qgcg}.

\begin{notation}\label{qgcg}
From now on, for any $w:\RR\to [1,+\infty)$ satisfying the condition
\begin{equation*}\label{Cnd}
\exists\,K_{(w)}>0\ |\ \forall\,p,q,p_1\in \RR,\ w(p-q)\le K_{(w)}(w(p-p_1)+w(p_1-q)),
\end{equation*}
we define two Hilbert spaces:
\begin{equation}\label{q_w}
\begin{array}{rl}
\mathbf{Q}_w&:=\big\{Q\in\mathfrak{S}_2,\  \diint (\sqrt{1+|p|^2}+\sqrt{1+|q|^2})w(p-q)|\wh{Q}(p,q)|^2dpdq<+\infty\big\},\\
\mathfrak{C}_w&:=\big\{\rho\in\mathcal{S}'(\RR),\ \dint \frac{w(k)}{|k|^2}|\wh{\rho}(k)|^2dk<+\infty\big\}.
\end{array}
\end{equation}
The letter $w$ always refers to a function of this kind. The case $w\equiv 1$ gives the space $\mathbf{Q}_1$ of operators $Q$ with $\ttr(|\D||Q|^2+Q^* |\D| Q)<+\infty$ and $\mathfrak{C}_1=\mathcal{C}$. Typically, we consider $w(p-q):=E(p-q)^a$ for $a>1$.% (see Lemma \ref{ineqgs}).
\end{notation}

By the fixed point method we may estimate together 
\begin{itemize}
\item $\nqkino{F_Q(Q,\rho)}$ and $\ncc{F_\rho(Q,\rho)}$,
\item In general $\nqbfg{F_Q(Q,\rho)}$ and $\ncg{F_\rho(Q,\rho)}$. We define $\mathcal{X}_g:=\mathbf{Q}_w\times \mathfrak{C}_w$
\end{itemize}
%\begin{remark}
%As in \cite{ptf} we will use $\nqbf{\cdot}$ for $\nqbfg{\cdot}$ with $g(p-q)=E(p-q)^2$ and $\nc{\cdot}$ for $\ncg{\cdot}$ with the same $g$, $\mathcal{X}_g$ is then denoted by $\mathcal{X}$.% If the sharp cut-off of \cite{ptf} is used we may take $E(p+q)$ (or $E(p)+E(q)$) and if not we take the other.
%\end{remark}

\paragraph{Notations}

\begin{notation}[On $D_0$ and $\D$]\label{spbf_def}
%We also write $P^0_+:=\chi_{(-\infty,0)}(D^0)=\text{Id}-P^0_-$ the projector on the positive spectral subspace of $D_0$ (and $\D$). 
The operator $\text{sign}(\D)$ is a Fourier multiplier that we write $\mathbf{s}_p=\frac{\wh{\D}(p)}{\sqrt{g_0(p)^2+g_1(p)^2}}$ . We also write 
\begin{equation}
E(p):=\sqrt{1+|p|^2}\text{\ and\ }\ed{p}:=\sqrt{g_0(p)^2+g_1(p)^2}.
\end{equation}
\end{notation}

\begin{remark}[Regime]\label{regime}
We will work in the regime  
\begin{equation}\label{para}
\alpha\le \alpha_0 \ll 1\text{\ and\ }L:=\alpha\llo\le L_0\ll 1.
\end{equation} 
We consider systems with $M$ electrons and an external charge density $\nu\ge 0$ with $\ncc{\nu},Z:=\nlp{1}{\nu}<+\infty$. We will often consider $M=\mathcal{O}(Z)$ and $\ncc{\nu}^2+M=\mathcal{O}(\llo)$.

Throughout this paper the letter $K$ denotes a constant independent of the parameters $\alpha,\La,M,\nu$. $K(M,\nu)$ is a constant depending on $M,\nu$ and so on. The inequality $a\apprle b$ means that $a\le K b$ for $a,b>0$. When $a>1$ is some integer, then as in \cite{ptf} we write
\begin{equation}\label{KM}
 K_a:=\frac{1}{2\pi}\dint_{-\infty}^{+\infty}\frac{d\eta}{E(\eta)^a}=\underset{a\to+\infty}{\mathcal{O}}(a^{-1/2}).
\end{equation}

%du coup contrôle !!!
\end{remark}

\begin{notation}[On $\widetilde{Q}_{k,\ell}$]\label{q_k_ell}
For $(\eps_1,\cdots,\eps_{J+1})\in \{+,-\}^{J+1}$ we define $Q_J^{\eps_1\,\cdots\,\eps_{J+1}}$ with the same formula as in \eqref{def_f1_eq} except that we replace the $J+1$ operators $(\D+i\eta)^{-1}$'s by $P^0_{\eps_j}/(\D+i\eta)$. We define $Q_{k,\ell}^{\eps_1\cdots\eps_{J+1}}$ analogously.

We write $Q_{k,\ell}^{\eps_1 a_1 \eps_2\cdots a_{J}\eps_{J+1} }$ with $a_j\in \{ v,R\}$ for the operator
\[
-\dfrac{1}{2\pi}\dint_{-\infty}^{+\infty}d\eta \dfrac{P^0_{\eps_1}}{\D+i\eta} A_1 \dfrac{P^0_{\eps_2}}{\D+i\eta}\cdots A_J \dfrac{P^0_{\eps_{J+1}}}{\D+i\eta},
\]
where $A_j=v=\rho_{\g}''*\tfrac{1}{|\cdot|}$ if $a_j=v$ or $A_j=-R(\g')$ if $a_j=R$.
\end{notation}

\begin{notation}[On $\fla$]
We introduce the function $\Fla:=\frac{\fla}{1+\fla}$, studied in Appendix \ref{flaflafla}. We prove in particular that $\check{\Fla}\in L^1$ and that $\nlp{1}{\Fla}\apprle L$.
\end{notation}

\section{Description of minimizers}\label{des}

%[
\subsection{Minimizers and fixed point schemes}\label{minfps}

Let $\g'=\g+N$ be a minimizer for $E_{\text{BDF}}^\nu(M)$. From Eq. \eqref{def_f1_eq} and \eqref{def_f1_rho}, we define a fixed-point scheme:

\begin{center}$F^{(1)}=F^{(1)}_Q\times F^{(1)}_\rho:\mathcal{X}_1\to \mathcal{X}_1,$\end{center}
\begin{subequations}
\begin{equation}
F_Q^{(1)}(Q',\rho'')=N+\ssum_{\ell=1}^{\infty}\alpha^\ell Q_\ell(Q',\rho''),
\end{equation}
\begin{equation}
\mathscr{F}(F^{(1)}_\rho(Q',\rho'');k)=\frac{1}{1+\alpha B_\La(k)}\wh{n}''(k)+\frac{1}{1+\alpha B_\La(k)}\Big(\alpha \wh{\rho}_{1,0}(Q';k)+\ssum_{\ell=2}^\infty\alpha^\ell \wh{\rho}_\ell(Q',\rho'';k)\Big)
\end{equation}
\end{subequations}
To prove $F^{(1)}$ is well-defined we use the following Lemma proved in Section \ref{fixedpointmethod}.
\begin{lem}\label{contlem}
Let $w$ be some function satisfying \eqref{Cnd}, with constant $K_{(w)}>0$. There exists $\text{C}_0>0$ such that for any $J\ge 2$, the linear operator:
\[
(Q,\rho)\in \textbf{Q}_w\times \mathfrak{C}_w\mapsto (Q_J(Q,\rho),\rho_J(Q,\rho))\in \textbf{Q}_w\times \mathfrak{C}_w
\]
is bounded with norm lesser than $2K_{(w)}^J \text{C}_0^J J^{1/2}$.
\end{lem}

We apply the Banach-Picard Theorem.
\begin{lem}\label{rank}
Let $\g'=\g+N$ be a minimizer for $E_{\text{BDF}}^\nu(M)$. In the regime of Remark \ref{regime} the following holds:
\begin{enumerate}
\item $F^{(1)}:B_{\mathcal{X}_1}(0,R_0)\to B_{\mathcal{X}_1}(0,R_0)$ is well-defined for some $R_0>0$ and this restriction is a Lipschitz function with constant lesser than $1$.
\item $(\g',\rho_{\g}'')$ is in the previous ball and so is the unique fixed point of $F^{(1)}$, moreover:
\[\lVert F^{(1)}(\g',\rho_{\g}'')-(N,n'')  \rVert_{\mathcal{X}_1}=o(1).\]
\item As a consequence $N=\chi_{(0,\mu]}(D_Q)$ has rank $M_0=M$.
\end{enumerate}%[%)
\end{lem}
\noindent\textbf{Proof of part 3.} If we assume the first two points, the last one is clear. Indeed on the one hand we have: $|\ttr_0(\g)|\apprle \ns{2}{\g}^2 =o(1)$, on the other hand, as  $\g$ is a difference of an orthogonal projector and $\PP$, it must be an integer \cite[Lemma 2]{ptf}. Thus $\ttr_0(\g)=0$ and 
\[
\ttr(N)=\tr(N)=\tr(\g')-\tr(\g)=M.
\]
\hfill{\footnotesize$\Box$}

To prove that $\rho_\g$ is integrable we need another fixed point scheme.

We see $\rho_{\g}''$ as the fixed point of a function $F^{(2)}$ defined in (a ball of) $\mathcal{C}$ and also in (a ball of) $\mathcal{C}\cap L^1$. We write:
\begin{equation}\label{F30}
\left\{
\begin{array}{rll}
h_{2}&=&\alpha^2 \tau_{1,1}\Big\{\Tbf[N]+\alpha^2\big\{\alpha\Tbf \widetilde{Q}_3(\g',\rho_{\g}'')+ \Tbf Q_{2,0}(\g',\rho_{\g}'')\big\},\rho_{\g}''\Big\}+\alpha^2\tau_{2,0}(\g')\\
F^{(2)}_{2}(\rho'')&=&\alpha^2\big(\tau_{1,1}\big\{\alpha^2\big[\Tbf Q_{1,1}(\g',\rho'')+\Tbf Q_{0,2}(\rho'')\big],\rho''\big\}\big)\\
h_{3}&=&\alpha^4\tau\big(\widetilde{Q}_4(\g',\rho_{\g}'')\big)+\alpha^3\big\{\tau_{3,0}(\rho_{\g}'')+\tau_{2,1}(\g',\rho_{\g}'')\big\}\\
F^{(2)}_{3}(\rho'')&=&\alpha^3\tau_{0,3}(\rho'')+\alpha^3\tau_{1,2}(\g',\rho'')
\end{array}
\right.
\end{equation}

\begin{equation}\label{F3}
\mathscr{F}\big\{F^{(2)}(\rho'')\big\}=\frac{1}{1+\fla(\cdot)}\wh{n}''+\frac{1}{1+\fla(\cdot)}\Big\{ \wh{h}_2+ \mathscr{F}\big\{F^{(2)}_{2}\big\}+\wh{h}_3+\mathscr{F}\big\{F^{(2)}_{3}\big\}\Big\}(\rho'')
\end{equation}
\begin{remark}
The definition of $F^{(2)}$ may appear complicated. It is built on the following self-consistent equation:
\[
\rho_{\g}'=\tau\Big\{N+\alpha Q_{0,1}(\rho''_\g)+\alpha^2 \big(\wit{Q}_{2}(\g',\rho''_\g)-Q_{1,1}(\g',\rho''_\g)\big)\Big\}+\alpha^2\tau\big[Q_{1,1}(F_Q^{(1)}(\g',\rho''_\g),\rho''_\g)\big].
\]
%The decomposition between $h_2$ and $F^{(2)}_2$, $h_3$ and $F^{(2)}_3$ are easily explained: $h_2$ and $h_3$ are densities of trace-class operators and their norms $\ns{1}{\cdot}$ is well estimated thanks to \cite{ptf}. $F^{(2)}_2$ and $F^{(2)}_3$ are densities of trace-class operators provided $\nlp{4}{v[\rho'']}$ is finite, which is the case if $\ncc{\rho''}+\nlp{\infty}{\wh{\rho''}}<+\infty$.
\end{remark}

\begin{lem}\label{ff2}
Let $\g'=\g+N$ be a minimizer for $E_{\text{BDF}}^\nu(M)$ and $F^{(2)}$ the function \eqref{F30}. In the regime of Remark \ref{regime}, there exists $R_0>0$ such that $F^{(2)}$ is well-defined in $B_{\mathcal{C}}(0,R_0)$ and in $B_{\mathcal{C}\cap L^1}(0,R_0)$. 

Furthermore these balls are $F^{(2)}$-invariant and $F^{(2)}$ is a contraction on them; $\rho''_\g$ is the only fixed point in both Banach spaces. In particular $\rho_\g\in L^1$.
%\begin{enumerate}
%\item First we take the minimizer of $E_{\text{BDF}}^{\nu}(M)$ and show as in cite{sok} that $(\g',\rho_{\g}'')$ is the fixed point of $F^{(1)}$ in (a ball of) $\mathbf{Q}_1\times \mathcal{C}$.
%\item
%Then we will show that $\rho_{\g}''$ is the fixed point of $F^{(2)}$ in (a ball of) $\mathcal{C}$ and in (a ball of) $\mathcal{C}\cap L^1$: so $\rho_\g\in L^{1}$ as $n-\nu\in L^1$.
%\end{enumerate}
\end{lem}
%\begin{remark}
%In the case of the sharp cut-off of \cite{ptf,at}, we can only apply the $F^{(2)}$-scheme in a ball of $\mathcal{C}\cap\mathscr{F}^{-1}(L^\infty)$ and get that $\wh{\rho}_\g\in\mathscr{C}^0_0$. The main difficulty appearing when one tries to apply $F^{(2)}$ in $\mathcal{C}\cap L^1$ is that it is not know whether $\mathscr{F}^{-1}(\fla)$ is in $L^1(\mathbb{R}^3)$ or not. It is believed \emph{it is} however this is not clear from its definition (\textit{cf} Appendix \ref{flaflafla}).
%\end{remark}

\begin{remark}
The \emph{linear} response of the vacuum to the presence of electrons $N$ and the external potential $\nu$ is:
\[
\left\{\begin{array}{rl}
\g&=\alpha\Tbf[Q_{0,1}((\delta_0-\check{F}_\La)*(n-\nu+\mathfrak{t}_N))]+\mathfrak{T}_N+\cdots\\
\rho_\g&=-\check{F}_\La*(n-\nu)+(\delta_0-\check{F}_\La)*\mathfrak{t}_N+\cdots
\end{array}\right.
\]
\end{remark}

%\begin{remark}
%Results of Lemma \ref{contlem} and \ref{rank} are already in \cite{sok} but we have chosen to rewrite the proof here.% because we use the same estimates to prove Lemma \ref{ff2} in a more difficult way.
%\end{remark}

\subsection{\emph{A priori} estimates}\label{secest}

\begin{lem}[Estimates on the energy]\label{apriori1}
Let $M\in\mathbb{N}$ and $Q$ a test function for $E^{\nu}_{\text{BDF}}(M)$. We assume: $\mathcal{E}^\nu_{\text{BDF}}(Q)\le E^{\nu}_{\text{BDF}}(M)+\eps$ where $0<\eps=o(\alpha \ncc{\nu}^2)$. 

Then we have $\ns{2}{Q}^2\apprle M+\alpha \ncc{\nu}^2$ and
\[
\begin{array}{l}
\ttr(|\nabla| Q^2)\apprle \alpha\ncc{\nu}^2+\alpha^{1/2}M+\sqrt{\alpha M}\ncc{\nu},\\
\alpha \ncc{\rho_Q-\nu}^2\apprle \alpha\ncc{\nu}^2+\alpha^{3/2}M+\sqrt{\alpha M}\alpha\ncc{\nu}.
\end{array}
\]
\end{lem}

As a corollary we get the following.
\begin{lem}[Estimates on the mean-field operator]\label{apriori2}
In the regime of Remark \ref{regime} and for $Q$ as in Lemma \ref{apriori1} we have in the sense of self-adjoint operator: %Let $D_Q$ be $D_Q:=\D+\alpha B=\D+\alpha ((\rho_Q-\nu)*\tfrac{1}{|\cdot|}-R_Q)$. 
% we have
\begin{equation}
(1-o(1))|\D|\le |\D+\alpha B_Q| \le (1+o(1))|\D|.
\end{equation}
Both $o(1)$ are $\mathcal{O}\big(\alpha\ncc{\nu}+\alpha^{5/4}M^{1/2}+(\alpha M)^{1/4}\alpha\ncc{\nu}^{1/2}\big)$.
\end{lem}

%\begin{dem}

%\noindent\textbf{\emph{A priori} estimates of a minimizer} 
\begin{lem}[\emph{A priori estimates of a minimizer}]\label{apriori3}
Let $\g'=\g+N$ be a minimizer for $E_{\text{BDF}}^{\nu}(M)$, decomposed as in \eqref{eqmin}. Then we have in the regime \eqref{para}
\[
\begin{array}{rll|rll}
\ttr(|\D|N)&\apprle& \llo,&\nqkino{\g}&\apprle&\apprle L,\\
\ncc{n''}&\apprle& \sqrt{\llo},&\ncc{\rho_\g}&\apprle& L\sqrt{\llo}.
\end{array}
\]
\end{lem}

\noindent\textbf{Proof of Lemma \ref{apriori1}:}
It is known that $E^{\nu}_{\text{BDF}}(M)\le M$ \cite{at}. There holds:
\[
\begin{array}{rl}
M+\eps+\frac{\alpha}{2}\ncc{\nu}^2&\ge \mathcal{E}^{\nu}_{\text{BDF}}(Q)+\frac{\alpha}{2}\ncc{\nu}^2\ge \big(1-\alpha\frac{\pi}{4}\big)\ttr_0(\D Q)+\frac{\alpha}{2}\ncc{\rho_Q-\nu}^2\\
 &\ge  \big(1-\alpha\frac{\pi}{4}\big)\ns{2}{Q}^2+\frac{\alpha}{2}\ncc{\rho_Q-\nu}^2.
\end{array}
\]
Furthermore:
\begin{equation}
\begin{array}{rl}
\ttr_0(\D Q)-M&=\ttr(|\D|^{1/2}(Q^{++}-Q^{--})|\D|^{1/2})-\ttr_0(Q)\\
                           &\ge \ttr(|\D|^{1/2}Q^2 |\D|^{1/2})-\ttr(Q^2)\\
                           &\ge \dfrac{1}{(2\pi)^3}\diint (\ed{p}-1)|\wh{Q}(p,q)|^2dpdq,
\end{array}
\end{equation}
and $\ed{p}-1\ge \tfrac{1}{2}\tfrac{p^2}{E(p)}$. Then thanks to Kato's inequality \eqref{kathardy}: 

\noindent$\ttr(Q R_Q)\le \tfrac{\pi}{2}\ttr(|\nabla| Q^2)$ which leads to:
\[
\frac{1}{2}\ttr\Big( \frac{-\Delta}{|D_0|} Q^2\Big)+\frac{\alpha}{2} \ncc{\rho_Q-\nu}^2\le \eps+\alpha\big(\frac{\ncc{\nu}^2}{2}+ \frac{\pi}{4}\ttr(|\nabla| Q^2)\big).
\]
Splitting at level $r_0=\tfrac{\alpha \pi}{\sqrt{1-(\alpha \pi)^2}}$ (to get $\alpha \tfrac{|p|\pi}{4}\le \tfrac{1}{4}\tfrac{|p|^2}{E(p)}$ for $|p|\ge r_0$) we obtain:
\begin{equation}
\ttr\Big( \frac{-\Delta}{|D_0|} Q^2\Big)\apprle \alpha(\ncc{\nu}^2+M),
\end{equation}
thus by the Cauchy-Schwartz inequality: $\ttr(|\nabla| Q^2)\apprle \alpha\ncc{\nu}^2+\sqrt{\alpha} M+\sqrt{\alpha M}\ncc{\nu}$.\hfill{\footnotesize$\Box$}

\medskip

\noindent\textbf{Proof of Lemma \ref{apriori2}:}

For all $f\in\ \hl$ we have:
\begin{equation}
\psh{\,|\D|^2 f}{f}(1-\alpha \nb{|\D|^{-1}B})^2\le \psh{|\D+\alpha B|^2f}{f}\le \psh{\,|\D|^2 f}{f}(1+\alpha \nb{|\D|^{-1}B})^2.
\end{equation}
However thanks to Ineq.\eqref{vphi} and the second point of Lemma \ref{exch}:

\[\nb{R_Q |\nabla|^{-1/2}}\apprle\sqrt{\ttr(Q R_Q)}\text{\ and\ }\nb{(\rho_Q-\nu)*\tfrac{1}{|\cdot|}|\nabla|^{-1/2}}\apprle \nlp{6}{(\rho_Q-\nu)*\tfrac{1}{|\cdot|}}\apprle \ncc{\rho_Q-\nu}.\]
%by the Kato-Seiler-Simon and Sobolev inequalities \cite{Sim,LL} which are recalled in Appendix \ref{appA}.%!!!!!!(mettre en appendice)

As the square root is monotone, there holds%with $Q$ a minimizer there holds:
\begin{equation}
(1-\alpha \nb{|\D|^{-1}B_Q})|\D|\le |\D+\alpha B_Q|\le (1+\alpha \nb{|\D|^{-1}B_Q})|\D|,
\end{equation}
and in the regime of Remark \ref{regime}, this gives $(1-o(1))|\D|\le |\D+\alpha B_Q| \le (1+o(1))|\D|$.
% in the sense of self-adjoint unbounded operator of domain $H^3$ (\textit{cf} Kato-Rellich Theorem in \cite{ReedSim} Vol II). By density of $H^3$ in $H^{3/2}$ these inequalities hold for any $f\in H^{3/2}=D(\D)$. 
This $o(1)$ is of order $\mathcal{O}(\alpha(\ncc{\rho_Q-\nu}+\ns{2}{\,|\nabla|^{1/2} Q}))$, that is of order 

$\mathcal{O}\big(\alpha\ncc{\nu}+\alpha^{5/4}M^{1/2}+(\alpha M)^{1/4}\alpha\ncc{\nu}^{1/2}\big)$.\hfill{\footnotesize$\Box$}

\medskip

\noindent\textbf{Proof of Lemma \ref{apriori3}:} For $E_{\text{BDF}}^{\nu}(M)$ with $M,\ncc{\nu}^2\apprle \llo$, we have thanks to Lemma \ref{apriori1}:
\[
\alpha(\ncc{\rho''_\g}+\sqrt{\ttr(|\nabla| \g')})\apprle \sqrt{\alpha}(\alpha^{1/2}\ncc{\nu}+\alpha^{3/4} M^{1/2}+(\alpha M)^{1/4}\alpha^{1/2}\ncc{\nu}^{1/2})=:\alpha^{1/2}\ell.
\]
We have $\ell=O(\sqrt{L})$. Using Eq.~\eqref{inverse} and assuming Lemma \ref{contlem} and Proposition \ref{profla} above we get that:
\[
\ncc{\rho_\g}\le \ncc{\check{\Fla}*n''}+\ncc{(\delta_0-\check{\Fla})*(\mathfrak{t}_{N}+\underset{j\ge 2}{\sum}\alpha^j \tau_j)}\apprle L\ncc{n''}+\sqrt{L\alpha}\nqkino{N}+O(L\alpha).
\]
As $\ncc{n''}\le \ncc{\rho''_\g}+\ncc{\rho_\g}$ we get 
\[
%\ncc{n''}\apprle \ncc{\nu}+(\alpha M)^{1/4}(\sqrt{M}+\sqrt{\ncc{\nu}})+O(\alpha \e\ll^2)\apprle \ncc{\nu}+(\alpha M)^{1/4}(\sqrt{M}+\sqrt{\ncc{\nu}})+O(\alpha \ell^2).
\ncc{n''}\apprle \ncc{\nu}+(\alpha M)^{1/4}(M^{1/4}+\sqrt{\ncc{\nu}})+\sqrt{L\alpha M}+O(\alpha \ell^2)\apprle \sqrt{\llo}.%+L^{1/4}\sqrt{\llo}+O(L\alpha)\apprle \sqrt{\llo}.
\]
Thanks to the equations $\D \psi_j=\mu_j\psi_j-B\psi_j$, there holds:
\[
\ttr(|\D|N)\apprle M(1+O(\sqrt{\alpha} \ell))\apprle \llo.
\]
Finally we have
\begin{equation}\label{apriorimin}
\begin{array}{rl}
\nqkino{\g}&\apprle \sqrt{L\alpha} \ncc{n''}+\alpha\sqrt{\ttr(|\nabla| Q^2)}+O(L\alpha)\apprle L+O(L\alpha)\apprle L\\
\ncc{\rho_\g}&\apprle L\ncc{n''}+\sqrt{L\alpha M}+O(L\alpha)\apprle L\sqrt{\llo}.
\end{array}
\end{equation}
\hfill{\footnotesize$\Box$}

\section{The operator $F_{1,0}$}\label{q10first}

\begin{remark}
\begin{itemize}
\item [\textbullet] If $Q$ is a nonnegative operator then so is $R_Q$ when it is well defined. Moreover if $Q$ is self-adjoint then so is $R_Q$.

\item [\textbullet] The $R_\cdot$ operator commutes with Fourier multiplier of the form $g(p-q)$, indeed we have
\[
\wh{R_Q}(p,q)=\frac{1}{2\pi^2}\dint \frac{\wh{Q}(p-l,q-l)}{|l|^2}.
\]
In particular there holds:
\begin{equation}\label{commut}
[\partial_j, R_Q]=R([\partial_j,Q]).
\end{equation}
\end{itemize}
\end{remark}

\begin{lem}\label{exch}
Let $Q$ be in $\mathcal{S}(\RR\times\RR)$ (Schwartz class). 
%The following arrows $\to$ mean a continuous linear operator.
\begin{enumerate}
%\item There holds for all $-\frac{1}{2}<a<\frac{3}{2}$
%\[
%\begin{array}{rlcl}
%R_\cdot &:\overset{\boldsymbol{\cdot}}{H}^a_dH^s_m&\to&H^{a-1}_dH^s_m.\\
%\end{array}
%\]

%\item There holds for any $\tfrac{1}{2}<a<1$
%\[
%\begin{array}{rlcl}
%|D_0|^{-1} R_\cdot &:H^{\tfrac{1}{2}}_dH^s_m&\to&\mathfrak{S}_2,\\
%|\D|^{-1/2}R_\cdot&:H^{\tfrac{1}{2}}_dH^s_m&\to &H^{0}_dH^s_m\ (\mathrm{with\ norm\ }\mathcal{O}(\sqrt{\log(\Lambda)})),\\
%|\D|^{-a}R_\cdot&:H^{\tfrac{1}{2}}_dH^s_m&\to &H^{0}_dH^s_m\ (\mathrm{with\ norm\ }\mathcal{O}(2a-1)^{-1/2}).
%\end{array}
%\]

\item We have:
\[
\begin{array}{l}
%|D_0|^{-1} R_\cdot :H^{\tfrac{1}{2}}_dH^s_m\to\mathfrak{S}_2,\\
\ns{2}{\,|\nabla|^{-1/2}R_Q}\apprle \sqrt{\ttr(R_Q^* Q)}.\\
\end{array}
\]
In particular for any $w\ge 1$ there holds:%There holds for any $\tfrac{1}{2}<a<1$
\[
\diint \frac{w(p-q)}{|p|}|\wh{R}_Q(p,q)|^2dpdq\apprle\diint |p+q|w(p-q)|\wh{Q}(p,q)|^2dpdq.
%\begin{array}{rlcl}
%|\nabla|^{-1/2}R_\cdot&:H^{\tfrac{1}{2}}_dH^s_m&\to &H^{0}_dH^s_m\ (\text{with\ norm\ }\mathcal{O}(1)).
%|\D|^{-a}R_\cdot&:H^{\tfrac{1}{2}}_dH^s_m&\to &H^{0}_dH^s_m\ (\mathrm{with\ norm\ }\mathcal{O}(2a-1)^{-1/2}).
%\end{array}
\]

\item There exists $K>0$ such that for all $0<\epsilon \le 1$
\[
%\ns{4}{|D_0|^{-\tfrac{1+\eps}{2}}R_Q}&\le K(-\log(\epsilon))\ns{4}{Q}\\
\begin{array}{rl}
%\ns{2}{|\nabla|^{-1/2} R_Q }&\le \frac{K}{\eps^{1/2}}\ns{2}{Q},\\
\ns{1}{\,|D_0|^{-\tfrac{1+\eps}{2}}R_Q |D_0|^{-\tfrac{1+\eps}{2}}}&\le \frac{K}{\eps}\ns{1}{Q},\\
\ns{2}{\,|D_0|^{-(1+\eps)}R_Q }&\le \frac{K}{\sqrt{\eps}}\ns{2}{Q}.
\end{array}
\]
For $Q\in\mathfrak{S}_2(\hl)$, we can replace $|D_0|^{-(1+\eps)/2}$ by $|\D|^{-1/2}$, provided that $\eps^{-1}$ is replaced by $\llo$. 
%and 
%\[\begin{array}{rl}
%\ns{2}{|\nabla|^{-1/2}R_Q}&\apprle \sqrt{\ttr(R_Q^* Q)},\\
%\end{array}
%\]
\end{enumerate}
By density, these inequalities hold for $Q$ in the Banach spaces corresponding to the norms in the r.h.s.
\end{lem}
We prove this Lemma in Appendix A.
%\begin{remark}
%We emphasize here that one of our main (technical) ingredients is Lemma \ref{exch} (Appendix \ref{appA}): it gives essential results on the $R_\cdot$ operator:
%\[
%R_{\cdot}:Q(x,y)\mapsto \dfrac{Q(x,y)}{|x-y|}
%\]
%in several Banach spaces.
%\end{remark}
\subsection{Proof of Lemma \ref{TbfS}}
\paragraph{In the Schatten norms} We recall $F_{1,0}$ is defined as
\begin{equation}
F_{1,0}:Q\mapsto Q_{1,0}(Q):=-\frac{1}{2\pi}\dint_{-\infty}^{+\infty}d\eta \frac{1}{\D+i\eta}R_Q\frac{1}{\D+i\eta}.
\end{equation}
The integral kernel of its Fourier transform is \cite{ptf}:
\begin{equation}\label{q10}
\wh{Q}_{1,0}(p,q)=\frac{1}{2}\frac{1}{\ed{p}+\ed{q}}\left(\wh{R}(p,q)-\sbf{p}\wh{R}(p,q)\sbf{q}\right).
\end{equation}
It corresponds to a difference of two operators which are in $\mathfrak{S}_p$ if $Q$ is in $\mathfrak{S}_p$ for both cases $p=1$ and $p=2$ (see below). By interpolation, for $p\in[1,2]$, if $Q\in\mathfrak{S}_p$ then so is $F_{1,0}(Q)$. Let us show the $\mathfrak{S}_1$-norm is $\mathcal{O}(\log(\Lambda))$ while the $\mathfrak{S}_2$-norm is $\mathcal{O}(\sqrt{\log(\La)})$. Indeed
\[
\frac{1}{f(p)+f(q)}=\dint_{s=0}^{+\infty}e^{-sf(p)-sf(q)}ds,
\]
therefore if $Q$ is nonnegative, then so is
\[
\dint_{s=0}^{+\infty}\tfrac{\D}{|\D|}\mathscr{F}^{-1}(e^{-s\ed{\cdot}}) R_Q \mathscr{F}^{-1}(e^{-s\ed{\cdot}})\tfrac{\D}{|\D|}ds.
\]
Writing $Q=\tfrac{Q+Q^*}{2}+\tfrac{Q-Q^*}{2}$ and splitting each self-adjoint operator into nonnegative and nonpositive part, we may assume that $Q\ge 0$. Then from Eq. \eqref{q10}, we get:
\[
\ns{1}{F_{1,0}(Q)}\le K\llo \ns{1}{Q}.
\]

As $(\ed{p}+\ed{q})^{-1}\le \ed{p}^{-1/2}\ed{q}^{-1/2}$, it follows that
\[
\begin{array}{rl}
\ns{2}{\,|\D|^{-\tfrac{1}{2}}R(\mathscr{F}^{-1}(|\wh{Q}(p,q)|)) |\D|^{-\tfrac{1}{2}}}&\le K\sqrt{\llo}\ns{2}{\mathscr{F}^{-1}(|\wh{Q}(p,q)|)}\\ &=K\sqrt{\llo}\ns{2}{\wh{Q}}=K\sqrt{\llo}\ns{2}{Q}.
\end{array}
\]

By interpolation ($1<p=1-\eps+2\eps<2$), there exists $K_{(1,0)}^{\mathfrak{S}}>0$
\begin{equation}\label{interpolation}
\ns{p}{Q_{1,0}(Q)}\le K_{(1,0)}^{\mathfrak{S}}(\log(\Lambda))^{1-\tfrac{\eps}{2}}\ns{p}{Q},
\end{equation}
%we write:
%\begin{equation}
%\text{C}_{(1,0),\mathfrak{S}}^\La:=K_{(1,0)}^{\mathfrak{S}}\llo.
%\end{equation}

\begin{remark}
The operators $Q_{1,0}(Q_0)$ (and $Q_{0,1}(\rho_0)$) can be rewritten as
\begin{subequations}
\begin{equation}
\mathcal{J}_t(x-y):=\mathscr{F}^{-1}(\exp(-t\ed{p}))(x-y)
\end{equation}
\begin{equation}
\left\{
\begin{array}{rl}
Q_{1,0}(Q_0)&=\frac{1}{2} \dint_{t=0}^{+\infty}(\mathcal{J}_t R_{Q_0} \mathcal{J}_t-\mathcal{J}_t\tfrac{\D}{|\D|}R_{Q_0}\tfrac{\D}{|\D|}\mathcal{J}_t)dt\\
Q_{0,1}(\rho_0)&=-\frac{1}{2} \dint_{t=0}^{+\infty}\big(\mathcal{J}_t \, v_{\rho_0}\, \mathcal{J}_t-\mathcal{J}_t\tfrac{\D}{|\D|}\,v_{\rho_0}\,\tfrac{\D}{|\D|}\mathcal{J}_t\big)dt
\end{array}
\right.
\end{equation}
\end{subequations}
\end{remark}

\paragraph{$\rho[Q_{1,0}(\cdot)]$}
We show here inequalities needed to estimate $\Tbf(Q_\ell(Q,\rho))$ and $\tau_\ell (Q,\rho)$ in norms $\nqbfg{\cdot},\ncg{\cdot}$. There exists a constant $C_R$ (defined in \cite{ptf}) such that for any function $w\ge 0$
\begin{equation}\label{schema_q10}
\diint (\ed{p}+\ed{q})w(p-q)|\wh{Q}_{1,0}(Q,p,q)|^2dpdq\le C_R^2\diint w(p-q)E(p+q)|\wh{Q}(p,q)|^2dpdq.
\end{equation}
By Cauchy-Schwartz inequality (\textit{cf} \cite{ptf} and inequality \eqref{trick1}):
\begin{equation}
|\wh{\rho}_{1,0}(Q,k)|^2\apprle |k|^2\underset{B(0,\La)}{\dint} \frac{|\wh{R}(u+\tfrac{k}{2},u-\tfrac{k}{2})|^2}{1+\ed{u,k/2}}du\underset{B(0,\La)}{\dint} \frac{du}{1+\ed{u,k/2}}\frac{1}{1+|u|^2+|k|^2/4},
\end{equation}
where $\ed{u,k/2}:=\max(\ed{u+k/2},\ed{u-k/2})$. Thus we have:
\begin{equation}\label{schema_r10}
|\wh{\rho}_{1,0}(Q,k)|^2\le \text{C}_{(1,0)}\dint E(2u)|\wh{Q}(u+\tfrac{k}{2},u-\tfrac{k}{2})|^2du,
\end{equation}
where $0<\text{C}_{(1,0)}=\text{C}_{(1,0)}(\La)$ satisfies $\text{C}_{(1,0)}\apprle \log(\La)$.

\paragraph{Well-definedness of $\Tbf$ and $\tau$}

Thanks to \eqref{interpolation} we can prove Lemma \ref{TbfS}: for $\alpha \llo$ sufficiently small the function $\Tbf$ is a linear bounded operator in $\mathrm{L}(\mathfrak{S}_p)$ for $1\le p=1+\eps \le 2$ with norm lesser than
\[
\text{C}_{\Tbf,\mathfrak{S}}^{(p)}:=\ssum_{\ell=0}^{+\infty} \big(\alpha K_{(1,0)}^{\mathfrak{S}}(\llo)^{1-\tfrac{\eps}{2}}\big)^\ell=\frac{1}{1-\alpha (\llo)^{1-\tfrac{\eps}{2}}K_{(1,0)}^{\mathfrak{S}}}
\]
which is finite as soon as $\alpha\llo$ is sufficiently small. We write $\text{C}_{\Tbf,\mathfrak{S}}:=\text{C}_{\Tbf,\mathfrak{S}}^{(1)}$.

As $\Tbf=(\mathrm{Id}-\alpha F_{1,0})^{-1}=\sum_{\ell=0}^{+\infty}\alpha^\ell F_{1,0}^{\circ (\ell)}$, let us show that $\alpha F_{1,0}$ is a bounded operator in $\mathrm{L}(\mathbf{Q}_w)$ with norm lesser than $1$. Thanks to inequality \eqref{schema_q10}, $\alpha F_{1,0}$ is bounded with norm lesser than $\alpha C_R$. Thus $\Tbf$ is a bounded linear operator with norm lesser than
\begin{equation}
\mathrm{C}_{\Tbf,\mathbf{Q}_w}:=\frac{1}{1-\alpha C_R}.
\end{equation}

Then thanks to Ineq. \eqref{schema_q10} and \eqref{schema_r10}, for $\ell\ge 1$ we have:
\[
|\wh{\rho}(F_{1,0}^{\circ (\ell)}(Q);k)|^2\le \alpha^{2\ell}\text{C}_{(1,0)}^\ell |k|^2\dint E(2u)|\wh{Q}(u+\tfrac{k}{2},u-\tfrac{k}{2})|^2du
\]
Therefore:
\begin{equation}
\dint\frac{g(k)}{|k|^2}|\wh{\rho}(F_{1,0}^{\circ (\ell)};k)|^2\le \alpha^{2\ell}\text{C}_{(1,0)}^\ell \diint g(p-q)E(p+q)|\wh{Q}(p,q)|^2dpdq
\end{equation}
and $\mathfrak{t}$ is a bounded linear operator in $\mathrm{L}(\mathfrak{C}_w)$ with norm lesser than 
\begin{equation}
\mathrm{C}_{\mathfrak{t},\mathfrak{C}}:=\ssum_{\ell=1}^{+\infty} (\alpha \sqrt{\text{C}_{(1,0)}})^\ell=\mathcal{O}(\alpha \sqrt{\llo})
\end{equation}
for $\alpha \sqrt{\llo}$ sufficiently small.
%\end{dem}

\begin{notation}
Let us define for $1\le p=1+\eps\le 2$:
\begin{equation}\label{Y}
Y_{\alpha,\La}(p)=Y(p)\apprle \text{C}_{\Tbf,\mathfrak{S}}^{(p)},
\end{equation}
which is an upper bound of the $\mathrm{L}(\mathfrak{S}_p)$-norm of $Q\mapsto |D_0|^{-7/12}R(\Tbf[Q])|D_0|^{-7/12}$: \emph{cf} Lemma \ref{exch} in Appendix \ref{sexch}.
\end{notation}

We have thus proved:
\begin{equation}
\left\{
\begin{array}{rl}
\nqbfg{\Tbf(Q)}&\le \mathrm{C}_{\Tbf,\mathbf{Q}_w}\nqbfg{Q}=\frac{\nqbfg{Q}}{1-\alpha C_R},\\
\ncg{\tau_Q}&\le \mathrm{C}_{\mathfrak{t},\mathfrak{C}}\nqbfg{Q}.
\end{array}
\right.
\end{equation}

\subsection{Proof of Theorem \ref{alafurry}}
First we recursively define the function $A_{J}^{(\ell_j)_{j=1}^J}$ as follows:

\begin{equation}\label{recur}
\left\{\begin{array}{ll}A_1^{\ell_1}\wh{Q}(p,q)&:=\wh{Q}(p-\ell_1,q-\ell_1)-\sbf{p}\wh{Q}(p-\ell_1,q-\ell_1)\sbf{q},\\
A_{J}^{(\ell_1,\mathbf{L})}\wh{Q}(p,q)&:=A_1^{\ell_1}\big(A_{J-1}^{\mathbf{L}}Q\big)(p,q)\text{\ with\ }J\in\mathbb{N}^*,\ell_j\in\mathbb{R}^3.
\end{array}\right.
\end{equation}
These functions appear in the Fourier transform of $Q_{1,0}^{\circ J}[Q]$ (see Appendix \ref{flaflafla}).

\begin{dem}
It is based on the following fact:
\begin{lem}\label{tracev}
The trace $\mathrm{Tr}_{\CC}$ of the product of an \textit{odd} number of Dirac matrices (that is $\boldsymbol{\alpha}_1,\boldsymbol{\alpha}_2,\boldsymbol{\alpha}_3,\beta$) vanishes.
\end{lem}
Writing $\langle a_1,\ldots,a_M \rangle$ the algebra spanned by the $a_j$'s, we define:
\begin{equation}
\left\{\begin{array}{ll}
\mathcal{A}_D&:=\langle \boldsymbol{\alpha}_1,\boldsymbol{\alpha}_2,\boldsymbol{\alpha}_3,\beta \rangle,\\
\mathcal{A}_D^+&:=\langle \mathrm{Id},(1-\delta_{jk})\boldsymbol{\alpha}_j\boldsymbol{\alpha}_k, \beta \boldsymbol{\alpha}_j\rangle\\
\mathcal{A}_D^-&:=\boldsymbol{\alpha}_1\mathcal{A}_D^++\boldsymbol{\alpha}_2\mathcal{A}_D^++\boldsymbol{\alpha}_3\mathcal{A}_D^++\beta\mathcal{A}_D^+
\end{array}\right.
\end{equation}
It is clear that $\mathcal{A}_D=\mathcal{A}_D^++\mathcal{A}_D^-$ and Lemma \ref{tracev} just says that 
\[\forall M\in \mathcal{A}_D^-: \mathrm{Tr}_{\CC}(M)=0.\]

Remark \ref{importantremark} and Appendix \ref{flaflafla} implies that for almost all $(p,q)\in\mathbb{R}^3\times \mathbb{R}^3$:
\begin{itemize}
\item[\textbullet] $\wh{F_{1,0}^{\circ J}}(Q_{0,1}(\rho);p,q)\in \mathcal{A}_D^+$,
\item[\textbullet] if $\wh{Q}(p,q)\in \mathcal{A}_D^\eps$ then so is $\wh{F_{1,0}^{\circ J}}(Q;p,q)$.
\end{itemize}

Now let us study $Q_{0,2}(\rho)$:
\[
Q_{0,2}=-\frac{1}{2\pi}\dint_{\eta=-\infty}^{+\infty}\frac{d\eta}{\D+i\eta}v_\rho\frac{1}{\D+i\eta}v_\rho\frac{1}{\D+i\eta}.
\]
where $Q_{0,2}^{\eps_1\,\eps_2\,\eps_3}$ is defined in Notation \ref{q_k_ell} (as $Q_{1,2}^{\eps_1, R,\eps_2, v,\eps_3}$ and so on). By the residuum formula in the case $\eps_1=\eps_2=\eps_3$ the term vanishes. We deal with $Q_{0,2}^{+--}$ and $Q_{0,2}^{-++}$ together, like $Q_{0,2}^{+-+}$ and $Q_{0,2}^{-+-}$, $Q_{0,2}^{--+}$ and $Q_{0,2}^{++-}$.
We compute the first couple with $A=Q_{0,2}^{+--}$ and $B=Q_{0,2}^{-++}$:
\[
\begin{array}{rl}
A&=\dfrac{1}{2\pi}\dint_{-\infty}^{+\infty}d\eta\dint_{p_1}dp_1 \frac{P^0_+(p)}{\ed{p}+i\eta}\wh{v}(p-p_1)\frac{P^0_-(p_1)}{-\ed{p_1}+i\eta}\wh{v}(p_1-q)\frac{P^0_-(q)}{-\ed{q}+i\eta}\\
&=\dint_{p_1}\dfrac{dp_1}{8}\dfrac{1}{\ed{p}+\ed{p_1}}\dfrac{1}{\ed{p}+\ed{q}}(1+\sbf{p})\wh{v}(p-p_1)(1-\sbf{p_1})\wh{v}(p_1-q)(1-\sbf{q}),\\
B&=\dfrac{1}{2\pi}\dint_{-\infty}^{+\infty}d\eta \dint_{p_1}dp_1\frac{P^0_-(p)}{-\ed{p}+i\eta}\wh{v}(p-p_1)\frac{P^0_+(p_1)}{\ed{p_1}+i\eta}\wh{v}(p_1-q)\frac{P^0_+(q)}{\ed{q}+i\eta}\\
&=-\dint_{p_1}\dfrac{dp_1}{8}\dfrac{1}{\ed{p}+\ed{p_1}}\dfrac{1}{\ed{p}+\ed{q}}(1-\sbf{p})\wh{v}(p-p_1)(1+\sbf{p_1})\wh{v}(p_1-q)(1+\sbf{q}),
\end{array}
\]
However
\begin{equation}\label{calcul}
\begin{array}{l}
\tfrac{1}{2}\big((1+\sbf{p})\wh{v}(p-p_1)(1-\sbf{p_1})\wh{v}(p_1-q)(1-\sbf{q})-(1-\sbf{p})\wh{v}(p-p_1)(1+\sbf{p_1})\wh{v}(p_1-q)(1+\sbf{q})\big)\\
=\sbf{p}\wh{v}(p-p_1)\sbf{p_1}\wh{v}(p_1-q)\sbf{q}+\sbf{p}\wh{v}(p-p_1)\wh{v}(p_1-q)-\wh{v}(p-p_1)\wh{v}(p_1-q)\sbf{q}-\wh{v}(p-p_1)\sbf{p_1}\wh{v}(p_1-q).
\end{array}
\end{equation}
%\begin{remark}
%We have considered $\wh{v}(p-p_1)$ and $\wh{v}(p_1-q)$ as matrices, Thinking of $Q_{1,2}(Q,\rho)$
%\end{remark}
In \eqref{calcul} there only remains matrices in $\mathcal{A}_D^-$. Symmetrically, the other two couples give:
\begin{equation}\label{calcul2}
\begin{array}{l}
\bullet\ \tfrac{1}{2}\big((1+\sbf{p})\wh{v}(p-p_1)(1-\sbf{p_1})\wh{v}(p_1-q)(1+\sbf{q})-(1-\sbf{p})\wh{v}(p-p_1)(1+\sbf{p_1})\wh{v}(p_1-q)(1-\sbf{q})\big)\\
=-\sbf{p}\wh{v}(p-p_1)\sbf{p_1}\wh{v}(p_1-q)\sbf{q}+\sbf{p}\wh{v}(p-p_1)\wh{v}(p_1-q)+\wh{v}(p-p_1)\wh{v}(p_1-q)\sbf{q}-\wh{v}(p-p_1)\sbf{p_1}\wh{v}(p_1-q),\\
\bullet\ \tfrac{1}{2}\big((1-\sbf{p})\wh{v}(p-p_1)(1-\sbf{p_1})\wh{v}(p_1-q)(1+\sbf{q})-(1+\sbf{p})\wh{v}(p-p_1)(1+\sbf{p_1})\wh{v}(p_1-q)(1-\sbf{q})\big)\\
=\sbf{p}\wh{v}(p-p_1)\sbf{p_1}\wh{v}(p_1-q)\sbf{q}-\sbf{p}\wh{v}(p-p_1)\wh{v}(p_1-q)+\wh{v}(p-p_1)\wh{v}(p_1-q)\sbf{q}-\wh{v}(p-p_1)\sbf{p_1}\wh{v}(p_1-q).
\end{array}
\end{equation}

Therefore for almost all $(p,q)$: $\wh{Q}_{0,2}(\rho;p,q)\in \mathcal{A}_D^-:$ its trace $\mathrm{Tr}_{\CC}$ vanishes. Furthermore for all $J\ge 1$:%%!!!!!
\begin{equation}\label{Jgeun}
\begin{array}{rl}
\wh{\rho}(F_{1,0}^{\circ J}(Q_{0,2}(\rho));k)&=\mathrm{Cst}\underset{u,\ell_1}{\diint}\cdots\underset{\ell_J}{\dint}\dfrac{dud\boldsymbol{\ell}}{\underset{1\le j\le J}{\prod}|\ell_j|^2}\mathrm{Tr}_{\CC}\dfrac{A_J^{(\ell_j)_{j=1}^{J}}\wh{Q}_{0,2}(\rho)(u+\tfrac{k}{2},u-\tfrac{k}{2})}{\underset{0\le j\le J}{\prod}(\ed{u+k/2-L_j}+\ed{u-k/2-L_j})}
\end{array}
\end{equation}
where for almost all $(p,q,\ell_j)$: $\mathrm{Tr}_{\CC}\Big\{A_{J}^{(\ell_j)_{j=1}^{J}}\wh{Q}_{0,2}(\rho;p,q)\Big\}=0$ because these matrices are in $\mathcal{A}_D^-$. Thus $\wh{\rho}(F_{1,0}^{\circ J}(Q_{0,2}(\rho));k)=0$ for almost all $k\in\mathbb{R}^3$ and so $\wh{\tau}_{0,2}(\rho;k)=0$ for almost all $k\in\mathbb{R}^3$. In other words $\wh{\tau}_{0,2}(\rho)=0$.

There remains to prove that $\tau_{1,1}\Big(\alpha\Tbf(Q_{0,1}(\rho_0)),\rho_1\Big)=0$: it suffices to show that for all $J,J'\ge 0$: $\rho\Big\{F_{1,0}^{\circ J}\big[Q_{1,1}\big(\alpha F_{1,0}^{\circ J'}[Q_{0,1}(\rho_0)],\rho_1\big)\big]\Big\}$ vanishes. As before we treat together
\begin{itemize}
\item[\textbullet] $Q_{1,1}^{+R-v-}(F_{1,0}^{\circ J'}(Q_{0,1}(\rho_0)),\rho_1)$ and $Q_{1,1}^{-R+v+}(F_{1,0}^{\circ J'}(Q_{0,1}(\rho_0)),\rho_1)$,
\item[\textbullet] then $Q_{1,1}^{+v-R-}(F_{1,0}^{\circ J'}(Q_{0,1}(\rho_0)),\rho_1)$ and $Q_{1,1}^{-v+R+}(F_{1,0}^{\circ J'}(Q_{0,1}(\rho_0)),\rho_1)$, and so on.
\end{itemize}
As $\wh{F_{1,0}^{\circ J'}}(Q_{0,1}(\rho_0);p,q)\in  \mathcal{A}_D^+$ for almost all $p,q$, then
 $\wh{Q}_{1,1}^{+R-v-}(F_{1,0}^{\circ J'}(Q_{0,1}(\rho_0);p,q),\rho_1)+\wh{Q}_{1,1}^{-R+v+}(F_{1,0}^{\circ J'}(Q_{0,1}(\rho_0)),\rho_1;p,q)\in \mathcal{A}_D^-$ for almost all $p,q$ thanks to \eqref{calcul} and \eqref{calcul2}. So its trace $\mathrm{Tr}_{\CC}$ vanishes. The same result holds for the other cases: $Q_{1,1}^{+v-R-}+Q_{1,1}^{-v+R+}$, $Q_{1,1}^{+-+}+Q_{1,1}^{-+-}$ and $Q_{1,1}^{--+}+Q_{1,1}^{++-}$. Finally as in \eqref{Jgeun} we have:
 \[
 \wh{\rho}(F_{1,0}^{\circ J}(Q_{1,1}(F_{1,0}^{\circ J'}(\rho_0),\rho_1));k)=0\ \mathrm{for\ almost\ all\ }k.
 \]
\end{dem}

\section{The fixed point method}\label{fixedpointmethod}
We prove here Lemmas \ref{contlem}, \ref{rank} and \ref{ff2} and start with some inequalities.

\subsection{Tools}

%\textbf{Some inequalities}
%\textbullet\ 
%A non-sharp analog of the trick of \cite{ptf}: $(E(p)^2+\eta^2)(E(q)^2+\eta^2)\ge \tfrac{1}{4}E(p+q)^2E(\eta)^2$ (\textit{cf} $(58)$ \cite{ptf}) is:
%\begin{equation}
%\begin{array}{rl}
%(\ed{p}^2+\eta^2)(\ed{q}^2+\eta^2)&\ge \ed{p}^2\ed{q}^2+(\ed{p}^2+\ed{q}^2)\eta^2\\
 %                   &\ge \frac{1}{2}(E(p+q)^2+E(p+q)^2\eta^2)\ge \frac{1}{2}E(p+q)^2E(\eta)^2,
%\end{array}
%\end{equation}
%leading to
%\begin{equation}\label{trick2}
%\frac{1}{\sqrt{(\ed{p}^2+\eta^2)(\ed{q}^2+\eta^2)}}\le \frac{\sqrt{2}}{E(p+q)E(\eta)},
%\end{equation}
%It is used to get a term of the form $\tfrac{|\wh{R}(p,q)|}{E(p+q)^{1/2}}$.

%\textbullet\ 
%There holds for any $a,b\ge 0$:
%\begin{equation}\label{root}
%\tfrac{a+b}{\sqrt{2}}\le \sqrt{a^2+b^2}\le a+b.
%\end{equation}

%We first give elementary estimates based on $D_0$ and $E(\cdot)$.

%We have:
%\begin{equation}\label{trick_la}
%|\wh{\D}(p)|=\ed(p)\Big(1+\frac{|p|^2}{\La^2}\Big)\ge e^{-1}E(p)^{1+\tfrac{1}{\llo}}.
%\end{equation}
%We write $\eps_\La:=\tfrac{1}{\llo}$. It is equivalent to $E(p)^{\eps_\La}\le 1+\tfrac{|p|^2}{\La^2}$ and this last inequality is clear if we consider the two cases $|p|<\La$ and $|p|\ge \La$.

%\begin{remark}\label{bahbah}
%We will consider constants $C_{(k,\ell)}$ that will depend on the value of $g(p-q)=E(p-q)^{2s}$. For all $0<s\le 1$ they are the same but because of \eqref{bah} the constant $C_{(k,\ell)}^{(s)}$ with $s>1$ equals $2^{(s-1)}$ times $C_{(k,\ell)}^{(1)}$
%\end{remark}

%\textbullet\ We recall 
%\begin{equation}
%
%\end{equation}

\textbullet\ We recall the following Sobolev inequalities in $\mathbb{R}^3$: for suitable $f$ --say $H^1$-- we have
\begin{equation}\label{sobolev}
\begin{array}{lll}
\nlp{6}{f}\apprle \nlp{2}{\nabla f},&\nlp{4}{f}\apprle \nlp{2}{|\nabla|^{3/4}f},&\nlp{3}{f}\apprle \nlp{2}{|\nabla|^{1/2}f}.
\end{array}
\end{equation}

We use them  to prove the following inequalities: for $\rho\in\mathcal{C}$, $v_\rho:=\rho*\tfrac{1}{|\cdot|}$ and $\phi\in H^{1/2}$:
\begin{equation}\label{vphi}
\nlp{2}{v_\rho \phi}\apprle \nlp{6}{v_\rho}\nlp{3}{\phi}\apprle \ncc{\rho}\nlp{2}{|\nabla|^{1/2}\phi}.
\end{equation}
\begin{equation}\label{L4}
\nlp{4}{\rho*\tfrac{1}{|\cdot|}}\apprle \nlp{2}{|\nabla|^{3/4}\rho*\tfrac{1}{|\cdot|}}\apprle \sqrt{\dint \frac{|\wh{\rho}(k)|^2}{|k|^{5/2}}dk}\apprle \Big(\underset{\eps>0}{\mathrm{\inf}}\Big\{2\pi \eps^{1/2}\nlp{\infty}{\wh{\rho}}^2+\eps^{-1/2}\ncc{\rho}^2\Big\}\Big)^{1/2}.
\end{equation}

With $v_\rho:=\rho*\tfrac{1}{|\cdot|}$ Eq. \eqref{L4} is used in:
\begin{equation}\label{S4}
\ns{4}{\tfrac{1}{\D+i\eta}v_\rho},\ns{4}{\tfrac{1}{|\D+i\eta|^{1/2}}v_\rho \tfrac{1}{|\D+i\eta|^{1/2}}}\le \frac{K_2^{1/4}}{E(\eta)^{1/4}}\nlp{4}{\rho*\tfrac{1}{|\cdot|}}
\end{equation}

\noindent\textbullet\ We recall Kato's and Hardy's inequalities for $\phi\in L^2(\mathbb{R}^3)$:
\begin{equation}\label{kathardy}
\left\{
\begin{array}{rl}
\dint_{\mathbb{R}^3}\frac{|\varphi(x)|^2}{|x|}dx&\le \frac{\pi}{2}\psh{|\nabla| \varphi}{\varphi},\\
\dint_{\mathbb{R}^3}\frac{|\varphi(x)|^2}{|x|^2}dx&\le 4\psh{(-\Delta) \varphi}{\varphi},
\end{array}\right.
\end{equation}
and the Kato-Seiler-Simon's inequality (KSS) for compact operators in $\mathcal{B}(L^2(\mathbb{R}^3))$:
\begin{equation}
\forall\,2\le p\le +\infty:\ \ns{p}{f(-i\nabla)g(x)}\le (2\pi)^{-3/p}\nlp{p}{f}\nlp{p}{g}.
\end{equation}

\textbullet\ We recall that for any $p,q\in B(0,\La)$ we have (see \cite{sok}.)
\begin{equation}\label{trick_p}
\big|\wh{\PP}(p)-\wh{\PP}(q)\big|=\big|\wh{\PPP}(p)-\wh{\PPP}(q)\big|\apprle \dfrac{|p-q|}{\text{max}(\ed{p},\ed{q})}.
\end{equation}

By Ineq. \eqref{trick_p} we get the following.
\begin{lem}\label{vpm}
Let $\rho\in\mathcal{C}$, then there exists $K>0$ such that for any $a>1/2$ and $\eps\in\{+,-\}$ we have:
\[
\ns{2}{\mathcal{P}^0_\eps v_\rho \mathcal{P}^0_{-\eps} |D_0|^{-a}}\le \dfrac{K}{\sqrt{2a-1}} \ncc{\rho}.
\]
\end{lem}
\begin{dem}
It is obvious once we have seen that the norm of the integral kernel of its Fourier transform is lesser than:
\[
K\dfrac{|\wh{\rho}(p-q)|}{|p-q|}\dfrac{1}{E(q)^a\max(E(q),E(p))}.
\]
\end{dem}

\subsection{Estimate on $Q_{0,1}$}
We estimate $\nqbfg{Q_{0,1}}$ as in \cite{ptf}. We have
\begin{equation}
\begin{array}{rl}
\underset{B(0,\La)}{\dint}\dfrac{du}{E(u+\eps k/2)^2}\dfrac{\ed{u+k/2}+\ed{u-k/2}}{(\ed{u+k/2}+\ed{u-k/2})^2}&\le 4\pi\dint_0^{\La}\frac{du}{\sqrt{1+r^2}}\\
       &\le 4\pi (1+\llo)\apprle \llo,
\end{array}
\end{equation}
leading to:
\begin{equation}
\diint w(p-q)(\ed{p}+\ed{q})|\wh{Q}_{0,1}(\rho;p,q)|^2dpdq\apprle (1+\llo)\ncg{\rho}^2,
\end{equation}
where we have used \eqref{trick_p}.

\subsection{Proof of Lemma \ref{contlem}}

%\noindent\textbf{}
%\noindent\textbf{Method:} 
We recall that for $J\ge 1$:
\[
Q_{J}(Q,\rho):=\frac{1}{2\pi}\dint_{-\infty}^{+\infty}\frac{d\eta}{\D+i\eta}\underset{1\le j\le J}{\prod}\left((v_\rho-R_Q)\frac{1}{\D+i\eta}\right)
\]

We write
\[
\abs(Q):=\mathscr{F}^{-1}(|\wh{Q}|)\text{\ and\ }\abs(\rho):=\mathscr{F}^{-1}(|\wh{\rho}|).
\]
It is clear that $|\wh{Q}_{k,\ell}(p,q)|$ is lesser than the integral kernel of the Fourier transform of
\[
\abs(Q_{k,\ell}):=\frac{1}{2\pi}\dint_{-\infty}^{+\infty}\dfrac{d\eta}{\sqrt{|\D|^2+\eta^2}}\Big( \abs(\rho)*\tfrac{1}{|\cdot|}+R[\abs(Q)]\Big)^J.
\]
We write $\abs(v_\rho)=v_{\abs(\rho)}$ and $\abs(R_Q):=R_{\abs(Q)}$ and $d_\eta:=\sqrt{|\D|^2+\eta^2}$. We have:
\[
\begin{array}{|lrlll}
\nlp{6}{\abs(v_\rho)}&\apprle& \nlp{2}{\nabla \abs{v_{\rho}}}&\apprle& \ncc{\abs(\rho)}=\ncc{\rho},\\
\nlp{4}{\abs(v_\rho)}&\apprle& \nlp{2}{\,|\nabla|^{3/2}\abs(v_\rho)}&\apprle& \nlp{\infty}{\wh{\abs(\rho)}}+\ncc{\abs(\rho)}=\nlp{\infty}{\wh{\rho}}+\ncc{\rho},\\
\ns{2}{\tfrac{1}{|\cdot|^{1/2}} \abs(R_Q)}&\apprle& \nqq{\abs(R_Q)}&\apprle &\nqkino{\abs(Q)}=\nqkino{Q}.
\end{array}
\]
By the KSS inequality, there exist $\text{C}_6,\text{C}_4>0$ such that:
\begin{equation}\label{kss_use}
\begin{array}{rl}
\ns{6}{d_\eta^{-1/2} v_\rho d_\eta^{-1/2}}&\le \text{C}_6E(\eta)^{-1/2}\ncc{\rho},\\
\ns{4}{d_\eta^{-5/12} v_\rho d_\eta^{-7/12}}&\le \text{C}_4E(\eta)^{-1/4}\nlp{4}{v_\rho}.
\end{array}
\end{equation}

As $w$ satisfies \eqref{Cnd}, we have:
\[
w(p-q)\wh{\abs}(Q_J(Q,\rho);p,q)\le JK_{(w)}^J\wh{\abs}\Big(Q_J\big[\mathscr{F}^{-1}(w(p'-q')\wh{Q}(p',q')), \mathscr{F}^{-1}(\rho)\big];p,q\Big).
\]
It suffices to check that for $p_0=p,p_{J+1}=q$ and $p_1,\cdots,p_J\in \mathbb{R}^3$ we have:
\[
w(p-q)\le \ssum_{j=1}^{J+1}K_{(w)}^{j} w(p_{j-1}-p_j)\le JK_{(w)}^J\prod_{j=1}^{J+1}w(p_{j-1}-p_j).
\]
In the definition of $\nqbfg{\cdot}$, there remains to multiply by $\ed{p}^{1/2}+\ed{q}^{1/2}$. We use the first or the last $d_\eta^{-1}$ to get:
\[
\dfrac{\ed{r}^{1/2}}{\sqrt{\ed{r}^2+\eta^2}}\le \dfrac{1}{(\ed{r}^2+\eta^2)^{1/4}}\text{\ with\ }r\in\{ p,q\}.
\]
For the terms $Q_{J}(Q,\rho)$ with $J\ge 3$ we get that:
\[
\nqbfg{\abs{Q_J(Q,\rho)}}\le \dfrac{JK_{(w)}^J }{2\pi}\Big( \ns{2}{\tfrac{1}{|\cdot|^{1/2}} R[\abs(Q)]}+\text{C}_6\ncc{\rho}\Big)^J\dint_{-\infty}^{+\infty}\dfrac{d\eta}{\ed{\eta}^{(J+1)/2}}.
\]
For $J=2$, we treat $Q_{0,2}(\rho)$ in another way because the product of two operators in $\mathfrak{S}_6$ is not necessarily Hilbert-Schmidt. By the Cauchy expansion we have \cite{ptf}
\[
Q_J^{+\cdots +}=Q_J^{-\cdots -}=0.
\]
So it suffices to treat $Q_{0,2}^{\eps_1,\eps_2,\eps_3}$ with $(\eps_1,\eps_2,\eps_3)\neq (+++),(---)$. In particular there is a change of sign $+-$ or $-+$. By Hölder inequality and Lemma \ref{vpm} we have for $\eps\in\{+,-\}$:
\[\ns{2}{d_\eta^{-1/2} v_\rho^{\eps,-\eps} d_{\eta}^{-1/4}}\apprle \ncc{\rho}\Big\{\dint\frac{dq}{E(q)^{7/2}}\Big\}^{1/2}\apprle \ncc{\rho}.\]

Hence using the above inequality and \eqref{kss_use} we get:
\[
\nqbfg{Q_{0,2}(\rho)}\apprle \ncc{\rho}^2\dint_{-\infty}^{+\infty}\dfrac{d\eta}{E(\eta)^{1+4^{-1}}}.
\]
%ancien
By \eqref{KM}, there exists $K>0$ such that
\[
\nqbfg{Q_J(Q,\rho)}\le J^{1/2}\big(K\times K_{(w)} (\nqbfg{Q}+\ncg{\rho})\big)^J.
\]

To deal with $\rho_J$, we use the same method as in \cite{ptf} and estimate $\ncc{\rho_J}$ by duality. We take a Schwartz function $\zeta\in\mathcal{S}(\mathbf{R^3})$ and prove that for any $k,\ell\ge 0$ with $k+\ell\ge 2$ we have:
\[
\big|\ttr(Q_{k,\ell} \zeta)\big|\le K(Q,\rho,k,\ell)\sqrt{\dint \dfrac{|p|^2|\wh{\zeta}(p)|^2}{g(p)^2}dp}=K(Q,\rho,k,\ell)\ncgp{\zeta}.
\]

We emphasize that by Furry's Theorem \cite{Fu,ptf} we have $\rho_{0,2J}=0$ for any $J\in\mathbb{N}^*$.

First we must prove that $Q_{k,\ell}\zeta$ is trace-class. We use the same method as in \cite{ptf}:
\[
\ns{1}{Q_{k,\ell}\zeta}\le \ns{2}{Q_{k,\ell}|\D|^2}\ns{2}{\tfrac{1}{|\D|^2}\zeta}\apprle E(\La)^2\ns{2}{Q_{k,\ell}}\nlp{2}{\zeta}.
\]
%\[
%\begin{array}{rl}
%\ns{1}{Q_{0,2}^{--+}\zeta}&\le \ns{6}{\,|\D|^{-1/2}v_\rho}\ns{2}{\,|D_0|^{-\tfrac{1}{2}-\tfrac{\eps_\La}{4}} v^{-+}}\ns{3}{\,|D_0|^{-1-\tfrac{\eps_\La}{2}}\zeta}\dint_{\mathbb{R}}\dfrac{d\eta}{2\pi}\dfrac{1}{E(\eta)^{1+3\eps_\La /2}}\\ 
%&\le \text{Cst}(\La)\ncc{\rho}^2\nlp{3}{\zeta}<+\infty.
%\end{array}
%\]
%We prove $Q_{k,\ell}\zeta\in\mathfrak{S}_1$ in the same way.

It is clear that $|\wh{Q_{k,\ell}\zeta}(p,p)|\le|\wh{\abs(Q_{k,\ell}) \zeta}|.$

Writing $d_\eta(p):=\sqrt{\ed{p}^2+\eta^2}$, $p_0=p$ and $\mathbf{m}=(m_1,\cdots,m_J)\in\{v_\rho,R_Q\}^J$ we have:
\[
2\pi|\wh{\abs(Q_{k,\ell}^{\mathbf{m}}) \zeta}(p,p)|\le\underset{\mathbb{R}}{\dint}\text{d}\eta\underset{(B(0,\La))^J}{\dint}\dfrac{\text{d}\mathbf{p}}{ d_\eta(p)}\prod_{j=1}^J |\wh{m_j}(p_j,p_{j-1})|d_\eta(p_j)^{-1} |\wh{\zeta}(p_J-p)|
\]
We replace $|\wh{\zeta}(p_J-p)|$ by:
\begin{equation}\label{trick_zeta}
|\wh{\zeta}(p_J-p)|\times \dfrac{w(p_J-p)}{w(p_J-p)}\le JK_{(w)}^J\dfrac{|\wh{\zeta}(p_J-p)|}{w(p_J-p)}\prod_{j=1}^J w(p_{j}-p_{j-1})=:JK_{(w)}^J|\wh{\zeta'}(p_J-p)|\prod_{j=1}^J w(p_{j}-p_{j-1}).
\end{equation}
We write $R':=R\big[\mathscr{F}^{-1}(w(p-q)|\wh{Q}(p,q)|)\big]$ and $V':=v\big[ \mathscr{F}^{-1}(w(p) |\wh{\rho}(p)|)\big]$. 

\noindent For $(k,\ell)$ different from $(0,3),(1,1),(0,2J)$ we have:
\[
\begin{array}{rl}
|\ttr(Q_{k,\ell} \zeta)|&\le \dfrac{(k+\ell)K_{(w)}^{k+\ell}}{2\pi}\binom{k+\ell}{k}\underset{\mathbb{R}}{\dint}d\eta\ns{6}{d_\eta^{-1/2} \zeta' d_\eta^{-1/2}} \ns{2}{d_\eta^{-1/2} R' d_\eta^{-1/2}}^k\ns{6}{d_\eta^{-1/2} V' d_\eta^{-1/2}}^\ell\\
  &\apprle \dfrac{(k+\ell)K_{(w)}^{k+\ell}}{2\pi}\binom{k+\ell}{k}K^{k+\ell} \underset{\mathbb{R}}{\dint}\dfrac{d\eta}{E(\eta)^{(1+j+\ell)/2}}\nqbfg{Q}^k\ncg{\rho}^\ell.
\end{array}
\]
To deal with $\rho_{1,1},\rho_{0,3}$ we use the same method as the one used for $\nqbfg{Q_{0,2}}$. We treat the case of $\rho[Q_{1,1}^{+R-v-}]$ as an example and the other cases are similar and left to the reader.
%dealt with in the thesis of the author (to appear in 2014). We have:
\[
|\ttr_{\mathbb{C}^4}(\wh{Q}_{1,1}^{+R-v-}(p_0,p_2)\wh{\zeta}(p_2-p_0))|\le \underset{\mathbb{R}}{\dint}\underset{(B(0,\La))^3}{\dint}\dfrac{\text{d}\eta \text{d}p_1\text{d}p_2|\wh{R}_Q(p_0,p_1)||\wh{v}(p_1-p_2)|}{d_\eta(p_0)d_\eta(p_1)d_\eta(p_2)}|\wh{\zeta^{-+}}(p_2-p_0)|.
\]
Using Lemma \ref{vpm} and \eqref{trick_zeta} we get that:
\[
|\ttr(Q_{1,1}^{+R-v-}\zeta)|\apprle \nqbfg{Q}\ncg{\rho} K_{5/4}\ncgp{\zeta}.
\]
\hfill{\footnotesize$\Box$}

\subsection{Estimates for $F^{(2)}$}

We consider $\g'=\g+N$ a minimizer of $E_{\text{BDF}}^{\nu}(M)$ and define the function $F^{(2)}$ \eqref{F3}. Two Banach spaces will be considered: first $\mathcal{C}$ and then $\mathcal{C}\cap L^1$. We recall that for $\eta\in\mathbb{R}$ we write $d_\eta=\sqrt{|\D|^2+\eta^2}$.

\subsubsection{Estimates on the $\mathcal{C}$-norm}
%We can do the same for $E^{\nu}(M)$ in the $\mathcal{C}$-norm, this is left to the reader. We refer to the Introduction for the notation $\nqbf{\cdot},\nc{\cdot}$.

\noindent Thanks to previous estimates (Lemmas \ref{apriori1}, \ref{apriori2}, \emph{a priori} estimates \eqref{apriorimin} and estimates in the $\ncg{\cdot}$-norm), in the regime $M,\ncc{\nu}\apprle \llo$ there hold the following \emph{non-sharp} estimates:
\begin{equation}
\left\{
\begin{array}{rl}
\ncc{h_2}&\apprle \alpha^2\Big\{ \ncc{\rho''_\g}\big[\nqkino{N}+\alpha^2 (\nqkino{\g'}+\ncc{\rho''_\g})^2\big]+\nqkino{\g'}^2\Big\}\\
% \mathrm{C}_{\mathfrak{C},h_2}\big\{(\nqbf{N}+\alpha^2\nx{(\g',\rho_{\g}')}^2)\nc{\rho_{\g}'}+\alpha^2 \nqbf{\g'}^2\big\}\\
       %&\le \tfrac{\alpha^2}{\sqrt{c}}  \ov{\mathrm{C}}_{\mathfrak{C},h_2}\\
       &\apprle \alpha^2\times \llo=L\alpha \\
 \ncc{h_3}&\apprle \alpha^3 (\nqkino{\g'}+\ncc{\rho_{\g}''})^3\apprle (L\alpha)^{3/2}.%\tfrac{\alpha^3}{c^{3/2}}\ov{\mathrm{C}}_{\mathfrak{C},h_3}.
\end{array}
\right.
\end{equation}
Then $F^{(2)}_2(\rho'')$ and $F^{(2)}_3(\rho'')$ are at most cubic in $\rho''$:
\begin{equation}
\left\{
\begin{array}{rl}
\ncc{F^{(2)}_2(\rho')}&\apprle \alpha^4 (\nqkino{\g'}+\ncc{\rho''})\ncc{\rho''}^2\\
\ncc{F^{(2)}_3(\rho')}&\apprle \alpha^3(\ncc{\rho''}+\nqkino{\g'})\ncc{\rho''}^2\\
% \mathrm{C}_{\mathfrak{C},h_3}(\tfrac{S_6C_6}{4\pi})^2\big\{\tfrac{S_6C_6}{4\pi}\nc{\rho'}^3+C_R\sqrt{2}\nqbf{\g'}\nc{\rho'}^2\big\}\\
\norm{\text{d}F^{(2)}_2(\rho')}_{\mathrm{L}(\mathcal{C})}&\apprle \alpha^4 (\nqkino{\g'}\ncc{\rho''}+\ncc{\rho''}^2)\\
%\alpha^4\mathrm{C}_{\mathfrak{C},h_2}\big\{ C_R\sqrt{2}\nqbf{\g'}+2\tfrac{S_6C_6}{4\pi}\nc{\rho'}\big\}\\
\norm{\text{d}F^{(2)}_3(\rho')}_{\mathrm{L}(\mathcal{C})}&\apprle \alpha^3(\nqkino{\g'}\ncc{\rho''}+\ncc{\rho''}^2).
%\alpha^3 \mathrm{C}_{\mathfrak{C},h_3}(\tfrac{S_6C_6}{4\pi})^2\big\{3\tfrac{S_6C_6}{4\pi}\nc{\rho'}^2+2C_R\sqrt{2}\nqbf{\g'}\nc{\rho'}\big\}
\end{array}
\right.
\end{equation}
%hhhhhhhh

\subsubsection{Estimates on the $L^1$-norm}\label{estimL1}
Our aim in this part is to prove Lemma \ref{estf(2)} below which states  that $F^{(2)}$ is a well-defined $\mathscr{C}^1$ function of $\mathcal{C}\cap L^1$ (differentiable with a continuous differential).

\noindent\textbullet\ We first prove that $h_2,h_3\in L^1$ (we recall they are defined in \eqref{F30}). In fact they are densities of trace-class operators: to see this we use the methods of the proof of Lemma \ref{contlem}.

\begin{enumerate}
\item $N=\sum_j \ket{\psi_j}\bra{\psi_j}\in \mathfrak{S}_1$ so $\Tbf[N]\in\mathfrak{S}_1$ and
\begin{equation}
\nlp{1}{\tau_N}\le \ns{1}{\Tbf[N]}\le \text{C}_{\Tbf,\mathfrak{S}}\ns{1}{N}.
\end{equation}

\item $Q_{2,0}(\g')\in\mathfrak{S}_1:$
We have:
\begin{equation}
\ns{1}{Q_{2,0}(\g')}\apprle \nqq{\g'}^2K_2.
\end{equation}

\item $Q_{0,\ell}(\rho_{\g}'')$ with $\ell\ge 4$. As $Q_{0,\ell}^{+\cdots +}=Q_{0,\ell}^{-\cdots -}=0$ there is at least one change of sign $+-$ or $-+$. Then with the help of Lemma \ref{vpm} and Estimates \eqref{kss_use} we have
\[
\ns{1}{Q_{0,\ell}(\rho_\g'')}\apprle \ncc{\rho''_\g}^\ell K_{\tfrac{\ell+1}{2}+\tfrac{1}{4}},
\]
the product of $\ell-1$ operators in $\mathfrak{S}_6$ and one in $\mathfrak{S}_2$ is trace-class.
%\[
%\begin{array}{rl}
%\ns{2}{\tfrac{1}{|\D|^{1/2}} P^0_+ v(\rho_{\g}'')P^0_- \tfrac{1}{|\D|^{1/2}}}^2&\le 2\frac{(4\pi)^2}{(2\pi)^3}\diint dudk\frac{|\rho_{\g}''(k)|^2}{|k|^2}\frac{dudk}{2\ed{u+\tfrac{k}{2}}^2E(u)^2}\\
% &\le \frac{4\pi}{(2\pi)^3}\dint\frac{du}{E(u)^2\ed{u}^2} \ncc{\rho_{\g}''}^2\le \ncc{\rho_{\g}''}^2,
%\end{array} 
%\]
%where we used:
%\begin{equation}\left\{
%\begin{array}{cl}
%\{\ed{u+k/2}\ed{u-k/2}\}^{-1}&\le \ed{u+\tfrac{k}{2}}^{-2}+\ed{u-\tfrac{k}{2}}^{-2}\\
%|P^0_-(p)P^0_+(q)|&\le \frac{|p-q|}{\sqrt{2}E((p+q)/2)}.
%\end{array}\right.
%\end{equation}
%Then $d^{-1/2}v(\rho_{\g}'')d^{-1/2}$ is treated as in \cite{ptf} and:
%\begin{equation}
%\left\{
%\begin{array}{rl}
%\ns{1}{Q_{0,\ell}(\rho_{\g}'')}& \le 2\ell \left(\frac{S_6 C_6}{\sqrt{4\pi}}\right)^{\ell-1}\ncc{\rho_{\g}''}^{\ell}K_{1+(\ell-1)/2},\ \ell=4,5\\
%\ns{1}{Q_{0,\ell}}(\rho_{\g}'')&\le \left(\frac{S_6 C_6}{\sqrt{4\pi}}\right)^{\ell}\ncc{\rho_{\g}''}^\ell K_{1+\ell/2},\ \ell\ge 6
%\end{array}
%\right.
%\end{equation} 
%where we recall that there is a constant $4\pi$ in the definition of the $\ncc{\cdot}$-norm.

\item Similarly $Q_{k,\ell}(\g',\rho_{\g}'')\in\mathfrak{S}_1$ with $k\ge 2$ or $k\ge 1$ and $\ell\ge 3:$
\begin{equation}
\ns{1}{Q_{k,\ell}(\g',\rho_{\g}'')}\apprle \binom{k+\ell}{k}(K\nqkino{\g'})^k(K\ncc{\rho_{\g}''})^\ell K_{1+(k+\ell)/2}.
\end{equation}

\item Thanks to Furry's Theorem and Theorem \ref{alafurry}:
\begin{equation}
\tau\big\{Q_{0,2}(\rho_{\g}'')\big\}=\tau_{1,1}\big\{ \Tbf[Q_{0,1}(\rho_{\g}'')],\rho_{\g}''\big\}=0.
\end{equation}

\item By the same methods as before we have $Q_{0,3}(\rho_{\g}''),Q_{1,2}(\g',\rho_{\g}'')\in \mathfrak{S}_{6/5}$ with:
\[
\ns{6/5}{Q_{0,3}(\rho_\g'')}\apprle \ncc{\rho''_\g}^3 K_{2+1/4}\text{\ and\ }\ns{6/5}{Q_{1,2}(\g',\rho_{\g}'')}\apprle \nqkino{\g'}\ncc{\rho''_\g}^2K_{1+3/2}.
\]
%We use Lemma \ref{exch}, \eqref{Y} and the equality $\text{id}=\tfrac{|D_0|^{5/8}}{|D_0|^{5/8}}$. We have $d^{-3/8}vd^{-5/8}\underset{1/6}{\to}(-1/2)\nlp{6}{\rho''_\g}$:
%\[
%\begin{array}{rl}
%d^{-1}Rd^{-1}vd^{-1}&\to d^{-3/8}(d^{-5/8} R d^{-5/8})(d^{-3/8}v d^{-5/8})d^{-3/8}\\
 %     &\underset{5/6+1/6}{\longrightarrow} (-3/2)\ns{6/5}{\Tbf[Q_{k,\ell}]}\ncc{\rho_{\g}''}.
%\end{array}
%\]
Furthermore the following inequalities hold (we recall that $Y$ is defined in \eqref{Y}):
\[
\ns{6}{d_\eta^{-3/8} v_\rho d_\eta^{-5/8}\,}\apprle E(\eta)^{-1/2} \ncc{\rho}\text{\ and\ }\ns{6/5}{d_\eta^{-5/8} R\big(\Tbf[Q]\big)d_\eta^{-5/8}}\apprle Y(\tfrac{6}{5})\ns{6/5}{Q}.
\]
Thus
\begin{equation}
\left\{
\begin{array}{rl}
\ns{1}{\Tbf_{1,1}\big\{ \Tbf Q_{0,3}(\rho_{\g}''),\rho_{\g}''\big\}}&\apprle 2\text{C}_{\Tbf,\mathfrak{S}}K_{5/4}\ncc{\rho''_\g}\big(Y(\tfrac{6}{5})\ncc{\rho''_\g}^3K_{2+1/4}\big),\\
%\le 2\Big(\left(\frac{S_6 C_6}{\sqrt{4\pi}}\right)^2 Y(\tfrac{6}{5})\text{C}_{\Tbf,\mathfrak{S}}K_{5/2}\ncc{\rho_{\g}''}^3\Big)\\
  %&\ \ \ \ \ \ \times\{\tfrac{S_{6} C_6}{\sqrt{4\pi}}\ncc{\rho_{\g}''}\}K_{3/2}\\
\ns{1}{\Tbf_{1,1}\big\{ \Tbf Q_{1,2}(\g',\rho_{\g}''),\rho_{\g}''\big\}}&\le 2\text{C}_{\Tbf,\mathfrak{S}}K_{5/4}\ncc{\rho''_\g}\big(3Y(\tfrac{6}{5})\nqkino{\g'}\ncc{\rho''_\g}^2 K_{1+3/2}\big)\\
%2\Big(\left(\frac{S_6 C_6}{\sqrt{4\pi}}\right)^2K_{5/2}Y(\tfrac{6}{5})\text{C}_{\Tbf,\mathfrak{S}}\ncc{\rho_{\g}''}^2C_R\sqrt{2}\nqkino{\g'}\Big)\\
 %&\ \ \ \ \ \  \times\{\tfrac{S_{6} C_6}{\sqrt{4\pi}}\ncc{\rho_{\g}''}\}K_{3/2}\\
 \ns{1}{\Tbf_{1,1}\big\{ \Tbf N,\rho_{\g}''\big\}}&\apprle 2\text{C}_{\Tbf,\mathfrak{S}}K_{5/4}\ncc{\rho''_\g}Y(\tfrac{6}{5})M.
 %2\Big( Y(\tfrac{6}{5}) \text{C}_{\Tbf,\mathfrak{S}}M\Big)\times\{\tfrac{S_{6} C_6}{\sqrt{4\pi}}\ncc{\rho_{\g}''}\}K_{3/2}.
\end{array}
\right.
\end{equation}

\item We apply $\Tbf$, $h_2$ (resp. $h_3$) is the density of $Q(h_2)$ (resp. $Q(h_3)$) with
\begin{equation*}\left\{
 \begin{array}{rl}
  Q(h_2)&=\alpha^2\Big\{ \Tbf Q{1,1}\big[\Tbf N+\alpha^2\Tbf[Q_{2,0}(\g')+\widetilde{Q}_3(\g',\rho''_\g)];\rho''_\g\big]+\Tbf Q_{2,0}(\g')\Big\}\\
  Q(h_3)&=\alpha^3\Big\{\Tbf Q_{3,0}(\g')+\Tbf Q_{2,1}(\g',\rho''_\g)+\alpha \widetilde{Q}_4(\g',\rho_\g'')\Big\}
 \end{array}\right.
\end{equation*}

The previous estimates lead to a sequence of numbers $(b_\ell)_{\ell\ge 2}$ with the following asymptotic behaviour:
\begin{equation}
b_\ell=\mathcal{O}_{\ell\to+\infty}(\ell^{1/2})
\end{equation}
and a constant $\text{C}_0>0$ such that:
\begin{equation}
\begin{array}{l}
\bigg|\bigg| \alpha^2Q_{2,0}(\g')+\alpha^3[Q_{3,0}+Q_{2,1}](\g',\rho_{\g}'')+\alpha^4\widetilde{Q}_{4}(\g',\rho_{\g}'')\bigg|\bigg|_{\mathfrak{S}_1}\\
+\alpha^3\bigg|\bigg|  Q_{0,3}(\rho_{\g}'')+Q_{1,2}(\g',\rho_{\g}'')\bigg|\bigg|_{\mathfrak{S}_{6/5}}\le \ssum_{\ell=2}^{+\infty}b_\ell(\alpha\text{C}_0)^\ell(\ncc{\rho_{\g}''}+\nqkino{\g'})^\ell=:\text{A}_{h,\mathfrak{S}}.
\end{array}
\end{equation}
We have:
\begin{equation}
 \ns{1}{Q(h_2)}\apprle \alpha^2\text{C}_{\Tbf,\mathfrak{S}} \big(2K_{5/4}Y(\tfrac{6}{5})(M+\text{A}_{h,\mathfrak{S}})+\nqbfg{\g'}^2\big)
\end{equation}
and write $B_{h_2,\mathfrak{S}}$ this upper bound.
%\begin{equation}
%\left\{
%\begin{array}{rl}
%\mathrm{A}_{h_2,\mathfrak{S}_1}&=\text{Cst}(\ns{6/5}{N}+\mathrm{A}_{h,\mathfrak{S}})  \\
%\ns{1}{Q(h_2)}&\le \alpha^2 \text{C}_{\Tbf,\mathfrak{S}}\big\{\tfrac{S_{6,15/4} C_6}{\sqrt{4\pi}}\ncc{\rho_{\g}''}\big\}\big\{Y(\tfrac{6}{5})\mathrm{A}_{h_2,\mathfrak{S}_1}\big\}K_{13/12}\\
%    &\ \ \ \ \ \ \ \ \ \ \ \ \ \ +\alpha^2 \text{C}_{\Tbf,\mathfrak{S}}(C_R\sqrt{2}\nqkino{\g'})^2K_{2}=:\mathrm{B}_{h_2,\mathfrak{S}_1}.
%\end{array}
%\right.
%\end{equation}
Similarly:
\begin{equation}
\ns{1}{Q(h_3)}\le  \text{C}_{\Tbf,\mathfrak{S}}\ssum_{\ell=3}^{+\infty}b_\ell(\alpha\text{C}_0)^\ell(\ncc{\rho_{\g}''}+\nqkino{\g'})^\ell=:\mathrm{B}_{h_3,\mathfrak{S}_1}.
\end{equation}
\end{enumerate}

\begin{remark}
The introduced numbers $\mathrm{A}_{h,\mathfrak{S}},\mathrm{B}_{h_2,\mathfrak{S}_1},\mathrm{B}_{h_3,\mathfrak{S}}$ are \emph{not} constants: they all depend on $\alpha$ and the minimizer $\g'$. As \emph{a priori} estimates hold(Lemma \ref{apriori1}), these upper bounds are small provided that we are in the regime of Remark \ref{regime}. Indeed we have
\[
 \big(1-\frac{\alpha\pi}{4}\big)\nqkino{\g'}^2+\dfrac{\alpha}{2}\ncc{\rho_\g''}^2\le \dfrac{\alpha}{2}\ncc{\nu}^2+M,
\]
so $\alpha (\nqkino{\g'}+\ncc{\rho_\g''})\apprle \alpha \ncc{\nu}+\sqrt{\alpha M}=\mathcal{O}((L\alpha)^{1/4})$. In particular those upper bounds are $o(1)$.
%\[
%\begin{array}{l|l}
%\mathrm{A}_{h,\mathfrak{S}}=\mathcal{O}(\alpha^2c^{-1})& \mathrm{B}_{h_2,\mathfrak{S}_1}=\mathcal{O}(\alpha^2c^{-1/2})\\
% \mathrm{A}_{h_2,\mathfrak{S}_1}=\mathcal{O}(1)& \mathrm{B}_{h_3,\mathfrak{S}}=\mathcal{O}(\alpha^3 c^{-3/2}).
%\end{array}
%\]
%For $E^\nu(M)$ those quantities are also small in the regime $\alpha(\ncc{\nu}+ M)=o(1)$.
\end{remark}

\noindent\textbullet\ Let us estimate the $L^1$-norm of $F^{(2)}_2(\rho'')$ and $F^{(2)}_3(\rho'')$ with $\rho''\in\mathcal{C}\cap L^1$.
To this end we use \eqref{S4} and \eqref{L4} at level $\eps=1$ for instance: there exists $K_{L^4}^{(v)}>0$ such that: 
\begin{equation}
\nlp{4}{v_{\rho''}}\le K_{L^4}^{(v)}\{\nlp{1}{\rho''}+\ncc{\rho''}\}.
\end{equation}
We use the second inequality of \eqref{kss_use} and Lemma \ref{vpm} with $a=7/12$. Using the method of the proof of Lemma \ref{contlem}, we obtain the following.
\begin{lem}\label{estf(2)}
Let $\rho''$ be in $\mathcal{C}\cap L^1$ and $\g'$ a minimizer for $E_{\text{BDF}}^\nu(M)$ with density $\rho'_\g$. We have:
\begin{equation}
\begin{array}{cll}
\ns{1}{\Tbf Q_{0,3}(\rho'')}&\apprle &  6K_{13/12} \text{C}_{\Tbf,\mathfrak{S}}\{\nlp{1}{\rho''}+\ncc{\rho''}\}^2\ncc{\rho''}\\
\ns{1}{\Tbf Q_{1,2}(\g',\rho'')}&\apprle &\binom{3}{1} K_2\text{C}_{\Tbf,\mathfrak{S}}\nqkino{\g'}\{\nlp{1}{\rho''}+\ncc{\rho''}\}^2\\
\ns{4/3}{Q_{0,2}(\rho'')}&\apprle &4K_{7/3}\ncc{\rho''}\{\nlp{1}{\rho''}+\ncc{\rho''}\}\\
\ns{4/3}{ Q_{1,1}(\g',\rho'')}&\apprle &2K_{7/4}\nqkino{\g'}\{\nlp{1}{\rho''}+\ncc{\rho''}\}\\
\ns{1}{\Tbf Q_{1,1}\big\{\Tbf Q_{0,2}(\rho''),\rho''  \big\}}&\apprle & 2 K_{13/12} Y(\tfrac{4}{3}) \text{C}_{\Tbf,\mathfrak{S}}\ns{4/3}{Q_{0,2}(\rho'')}\{\nlp{1}{\rho''}+\ncc{\rho''}\}\\
\ns{1}{\Tbf Q_{1,1}\big\{\Tbf Q_{1,1}(\g',\rho''),\rho''\big\}}&\apprle & 2K_{13/12}Y(\tfrac{4}{3})\text{C}_{\Tbf,\mathfrak{S}}\ns{4/3}{ Q_{1,1}(\g',\rho'')}K_{L^4}^{(v)}\{\nlp{1}{\rho''}+\ncc{\rho''}\}
\end{array}
\end{equation}
Similarly we can estimate $\norm{\text{d}F^{(2)}_j}_{\mathrm{L}(\mathcal{C}\cap L^1)}$. As $\nqkino{\g'}\apprle \sqrt{\llo}$ we have:
\begin{equation}
\left\{
\begin{array}{lll}
\norm{F^{(2)}_2(\rho'')}_{\mathcal{C}\cap L^1}&\apprle & \alpha^4\norm{\rho''}_{\mathcal{C}\cap L^1}^2\big\{\sqrt{\llo}+\norm{\rho''}_{\mathcal{C}\cap L^1}\big\}\\
\norm{F^{(2)}_3(\rho'')}_{\mathcal{C}\cap L^1}&\apprle &\alpha^3\norm{\rho''}_{\mathcal{C}\cap L^1}^2\big\{\sqrt{\llo}+\norm{\rho''}_{\mathcal{C}\cap L^1}\big\},\\
\norm{dF^{(2)}_2(\rho'')}_{\mathrm{L}(\mathcal{C}\cap L^1)}&\apprle &\alpha^4\norm{\rho''}_{\mathcal{C}\cap L^1}^2\big\{\sqrt{\llo}+\norm{\rho''}_{\mathcal{C}\cap L^1}\},\\
\norm{dF^{(2)}_2(\rho'')}_{\mathrm{L}(\mathcal{C}\cap L^1)}&\apprle &\alpha^3\norm{\rho''}_{\mathcal{C}\cap L^1}^2\big\{\sqrt{\llo}+\norm{\rho''}_{\mathcal{C}\cap L^1}\big\}.
\end{array}
\right.
\end{equation}
\end{lem}

%iiiiiiiii

\subsection{Application of the Banach fixed point theorem}\label{fifix}
\subsubsection{$F^{(1)}$}

With \emph{exactly} the same method of \cite{ptf} let us apply the Banach fixed point theorem to $F^{(1)}$ with the help of estimates of the previous subsections. We recall the different steps.

\noindent We define (where $K_{(w)}>0$ is defined in \eqref{Cnd} and $\text{C}_0>0$ is the constant of Lemma \ref{contlem})
\begin{equation}
\begin{array}{ll}
\mathcal{X}_w:=\mathbf{Q}_w\times\mathfrak{C}_w,&\ \mathrm{with}\ \nxg{(Q,\rho)}:=K_{(w)}\text{C}_0(\nqbf{Q}+\ncg{\rho}).
\end{array}
\end{equation}
Thanks to the previous estimates we can say that the function $F^{(1)}$ is well defined in a ball $B_{\mathcal{X}_g}(0,\ov{R})$ with $\ov{R}=O(\sqrt{\llo})$, say $\ov{R}=K_0\sqrt{\llo}$. Indeed:
\begin{equation}\label{trouverR}
\nxg{F^{(1)}(Q',\rho'')}\le \nxg{(N,n'')}+\alpha \kappab{1}(\La)\nxg{(Q',\rho'')} +\ssum_{\ell=2}^{+\infty}\alpha^\ell \kappab{\ell}\nxg{(Q',\rho'')}^\ell,
\end{equation}
where 
\begin{equation}
\left\{
\begin{array}{cl}
\kappab{1}(\La)&=\mathcal{O}_{\La\to+\infty}(\sqrt{\llo})\\
\kappab{\ell}&=\mathcal{O}_{\ell\to+\infty}(\ell^{1/2}).
\end{array}
\right.
\end{equation}
In particular the radius of convergence of the power series $f(x)=\sum_{\ell=2}^{+\infty}\kappab{\ell} x^\ell$ is  $1$ and:
\begin{equation}
\norm{\text{d}F^{(1)}(Q',\rho'')}_{\mathrm{L}(\mathcal{X}_g)}\le \alpha \kappab{1}(\La)+\alpha f'(\alpha \nxg{(Q',\rho'')}).
\end{equation}
For$\nxg{(N,n'')}\neq (0,0)$ it is clear that $F^{(1)}(0,0)=(N,\mathscr{F}^{-1}(-\tfrac{1}{1+\alpha B_\La(\cdot)}\wh{n}''))\neq 0$. So
\begin{equation}
\underset{(Q',\rho'')\in B_{\mathcal{X}_g}(0,\ov{R})}{\sup}\norm{\text{d}F^{(1)}(Q',\rho'')}_{\mathrm{L}(\mathcal{X}_g)}\le \alpha \kappab{1}(\La)+\alpha f'(\alpha \ov{R})=:\nu(\ov{R}).
\end{equation}
For $(Q',\rho'')\in B_{\mathcal{X}_g}(0,\ov{R})$ we have
\[
\begin{array}{rll}
\nxg{F^{(1)}(Q',\rho'')}&\le & \nxg{F^{(1)}(Q',\rho'')-F^{(1)}(0,0)}+\nxg{F^{(1)}(0,0)}\\
 &\le& \nu(\ov{R})\nxg{(Q',\rho'')}+\nxg{F^{(1)}(0,0)}.
 \end{array}
\]
Thus $B_{\mathcal{X}_g}(0,\ov{R})$ is invariant under $F^{(1)}$ provided that:
\begin{equation}\label{condition}
\nxg{F^{(1)}(0,0)}\le (1-\nu(\ov{R}))\ov{R}.
\end{equation}
As $F^{(1)}(0,0)\neq 0$ this gives $\nu(\ov{R})<1$.

Let us say that $\nxg{(N,n'')}=\eps_0 \ov{R}=\eps_0 K_0\sqrt{\llo},\,\eps_0<1$. We have:
\begin{equation}\label{f1zero}
\nxg{F^{(1)}(0,0)}\le \eps_0 \ov{R},
\end{equation}
it suffices to take $\alpha>0$ such that $\sqrt{L\alpha} K_0\ll 1$ and then take $\ov{R}$ accordingly. The constant $K_0$ depends on the constants in the conditions $M,\ncc{\nu}\apprle \sqrt{\llo}$: we get $\ov{R}=K_0\sqrt{\llo}$ and for sufficiently small $\alpha$ the Theorem can be applied on that ball.
%In the case where it is only known that $\nlp{2}{\nabla\psi}\apprle \alpha^{1/4},\nqkino{}$

\subsubsection{$F^{(2)}$}

We work with $(\mathcal{C},\ncc{\cdot})$ and $(\mathcal{C}\cap L^1,\max(\ncc{\cdot},\nlp{1}{\cdot}))$. In Appendix \ref{flaflafla} it is proved that $\nlp{1}{\check{f}_\La}\le K\alpha B_\La(0)$ where we can choose $K=2$ for $\alpha\llo$ sufficiently small.
Thus:
\[
\mathscr{F}^{-1}(\Fla)=\mathscr{F}^{-1}\left\{\frac{\fla}{1+\fla}\right\}=\ssum_{\ell=1}^{+\infty}(-1)^{\ell+1}\check{f}_\La^{*\ell}\in L^1
\]
and its $L^1$-norm is lesser than $\tfrac{2\alpha B_\La(0)}{1-2\alpha B_\La(0)}\le 4\alpha B_\La(0)$ as soon as $\alpha B_\La(0)\le 4^{-1}$. Moreover we can write
\[
\frac{1}{1+\fla}=1-\frac{\fla}{1+\fla};
\]
therefore if $\rho\in L^1$ then $\mathscr{F}^{-1}\{\tfrac{1}{1+\fla}\wh{\rho}\}^{-1}\in L^1$ and its $L^1$-norm is lesser than 

\[(1+4\alpha B_\La(0))\nlp{1}{\rho}\le 2\nlp{1}{\rho}.\]

In particular:
\[
\nlp{1}{\mathscr{F}^{-1}(\tfrac{1}{1+\fla}\wh{n}'')}\le 2(M+Z).
\]
So we have:
\begin{equation}
\left\{
\begin{array}{rl}
\norm{F^{(2)}(\rho'')}_{\mathcal{C}\cap L^1}&\le 2(M+Z)+\norm{h_2+h_3}_{\mathcal{C}\cap L^1}+K\alpha^3(\sqrt{\llo}+\norm{\rho''}_{\mathcal{C}\cap L^1})\norm{\rho''}_{\mathcal{C}\cap L^1}^2\\%+K\alpha^3(1+\norm{F^{(2)}(\rho')}^3_{\mathcal{C}\cap L^1})\\
\norm{\text{d}F^{(2)}(\rho'')}_{\mathrm{L}(\mathcal{C}\cap L^1)}&\le K\alpha^3\norm{\rho''}_{\mathcal{C}\cap L^1}(2\sqrt{\llo}+3\norm{\rho''}_{\mathcal{C}\cap L^1}).%+K\alpha^4(1+\norm{\rho''}_{\mathcal{C}\cap L^1})
\end{array}
\right.
\end{equation}
where the constants $K$ can be chosen indepently of $\alpha \le \alpha_0$ and $\alpha \llo\le L_0$ for $\alpha_0,L_0$ sufficiently small. The term $\sqrt{\llo}$ is due to $\nqkino{\g'}\apprle \sqrt{\llo}$ (see Lemma \ref{apriori1} and the regime of Remark \ref{regime}).
We get similar estimates for $F^{(2)}$ defined in $\mathcal{C}$. So it suffices to take $\ov{R}>2$ sufficiently large so that $B_{\mathcal{C}\cap L^1}(0,\ov{R})$ is invariant under $F^{(2)}$. This function is a contraction and we can apply the fixed point theorem. To end the proof we remark:
\begin{itemize}
\item There is only one fixed point of $F^{(2)}$ in $B_{\mathcal{C}}(0,\ov{R})$ by the Banach-Picard Theorem and $\rho_\g+n-\nu$ is a fixed point. Indeed by Section \ref{secest}, $(\g+N,\rho_\g+n-\nu)$ has norm $\mathbf{Q}_1\times \mathcal{C}$ bounded by $K\sqrt{\llo}$ in the regime of Remark \ref{regime} and is a fixed point of $F^{(1)}$. So it is a fixed point of $F^{(2)}$.

\item There is only one fixed point of $F^{(2)}$ in $B_{\mathcal{C}\cap L^1}(0,\ov{R})$ by the same theorem. In particular it is also a fixed point of $F^{(2)}$ in $B_{\mathcal{C}}(0,\ov{R})$ as $B_{\mathcal{C}\cap L^1}(0,\ov{R})\subset B_{\mathcal{C}}(0,\ov{R})$. By unicity $\rho_\g\in L^1$.
\end{itemize}

\section{Proofs of Theorems \ref{density} and \ref{existence}}\label{mainsproof}

\subsection{Proof of Theorem \ref{density}}

\begin{dem}
The fact that $\rho_\g\in L^1$ is a result of Section \ref{fifix}. We recall that if $Q\in\mathfrak{S}_1$, then $\int \rho_Q=\ttr(Q)=\ttr_{P^0_-}(Q)$. Writing
\begin{equation}
\begin{array}{rl| rl}
A&:=\alpha \Tbf[Q_{0,1}(\rho_{\g}'')] &C&:=\alpha^3 \Tbf \Big\{ Q_{1,1}\big[\Tbf[Q_{0,1}(\rho_{\g}'')],\rho_{\g}''\big] \Big\}\\
B&:=\alpha^2 \Tbf(Q_{0,2}(\rho_{\g}''))&S&:=\g-(A+B+C)
\end{array}
\end{equation}
it has been shown in Section \ref{fixedpointmethod} that $S\in\mathfrak{S}_1$. Theorem \ref{alafurry} says $\rho_B=\rho_C=0$. 

Let us show that $B^{++},B^{--},C^{++},C^{--}$ are trace-class. First for any $Q$ in $\mathfrak{S}_2$, we have 
\[
\PP Q_{1,0}(Q)\PP =\PPP Q_{1,0}(Q)\PPP =0.
\]
It follows that $B^{\pm\pm}=\alpha^2 Q_{0,2}(\rho_{\g}'')^{\pm \pm}$ and $C^{\pm \pm}=\alpha^3 Q_{1,1}\big(\Tbf Q_{0,1}(\rho_{\g}''), \rho_{\g}'' \big)^{\pm \pm}$. And as
\[
Q_{0,2}^{+++}=Q_{0,2}^{---}=Q_{1,1}^{+++}=Q_{1,1}^{---}=0
\]
there only remain $Q_{0,2}^{+-+},Q_{0,2}^{-+-},Q_{1,1}^{+-+},Q_{1,1}^{-+-}$. Using Lemma \ref{vpm} with $a=\tfrac{3}{4}$ and Cauchy-Schwartz inequality we have
%\[
%\ns{2}{\tfrac{1}{|\D|^{1/2}} P^0_\eps v_\rho P^0_{-\eps} \tfrac{1}{|\D|^{1/2}} }\apprle \ncc{\rho}.
%\]
%With exactly the same method, we can show that
\begin{equation}
\ns{2}{\tfrac{1}{|D_0|^{3/8}} \mathcal{P}^0_{\pm} v_\rho \mathcal{P}^0_{\mp} \tfrac{1}{|D_0|^{3/8} }}\apprle \ncc{\rho}\apprle \ncc{\rho}
\end{equation}
%as soon as $2a>\tfrac{1}{2}$: we choose $2a=\tfrac{3}{4}$. 
We recall that $\ns{2}{\tfrac{1}{|\nabla|^{1/2}} R_Q}\apprle \nqq{Q}$:
%\[
%\ns{2}{\tfrac{1}{|\D|^{1/2}} R(\Tbf [Q]) \tfrac{1}{|\D|^{1/2}} }\apprle \nqkino{\Tbf [Q]}\apprle \nqkino{Q},
%\]
these two estimates enables us to prove the following:
\[
\begin{array}{rll}
 \ns{1}{Q_{0,2}^{\pm\, \mp\,\pm}(\rho_\g'')}&\apprle& K_{3/2}\ncc{\rho''_\g}^2,\\
 \ns{1}{Q_{1,1}^{\pm\, \mp\,\pm}(\g',\rho_\g'')}&\apprle & K_{7/4}\nqq{\g'}\ncc{\rho_\g''}.
\end{array}
\]

\noindent As shown in Sections \ref{fixedpointmethod} and \ref{flaflafla} we have $Q_{0,1}^{++}=Q_{0,1}^{--}=0$ and $\rho_A=-\check{f}_\La*(\rho_{\g}')\in L^1$.
\[
\begin{array}{rl}
\dint \rho_\g &= \dint (\rho_{\g^{++}}+\rho_{\g^{--}})+ \dint \{\rho_{A^{+-}}+\rho_{A^{-+}}+\rho_{B^{+-}}+\rho_{B^{-+}}+\rho_{C^{+-}}+\rho_{C^{-+}} \}\\
         &= \ttr_{P^0_-}(\g)-\alpha \fla(0)\dint \big\{ \rho_\g + n-\nu\big\}-\dint \{ \rho_{B^{++}}+\rho_{B^{--}}+\rho_{C^{--}}+\rho_{C^{++}}\}\\
         &=0-\alpha \fla(0) \Big\{\dint \rho_\g +M-Z\Big\}-\ttr_{P^0_-}(B)-\ttr_{P^0_-}(C).
\end{array}
\]
To end the proof we have to show that $\ttr(B^{++}+B^{--})=\ttr(C^{++}+C^{--})=0$: this is straightforward when written in Fourier space (see \cite{ptf} for formulae).
\end{dem}

\subsection{Proof of Theorem \ref{existence}}\label{minimini}

We follow the method of \cite{at}. We apply a Lemma of Borwein and Preiss \cite[Theorem 4]{at} and consider an approximate minimizer $\g'_0=\g_0+N_0$ of  $E^{\nu}(M)$. 

Indeed, we can extend $\mathcal{E}^\nu_{BDF}$ to $\mathfrak{K}=\cap\{Q\in \mathfrak{S}_2\,: Q^*=Q,\ 0\le Q+P^0_-\le 1 \}$ by setting $\mathcal{E}^\nu_{BDF}(Q):=+\infty$ whenever $Q\notin \mathcal{K}$. This extension is lower semi-continuous and bounded from below in the $\mathfrak{S}_2$-topology and the set
\[
\mathcal{M}:=\{Q\in \mathfrak{K}, (Q+P^0_-)^2=Q+P^0_-, \ttr_0(Q)=M\}
\]
is closed in the same topology. Its convex closure in $\mathfrak{S}_2$ is 
\[\mathfrak{K}(M):=\{ Q\in \mathfrak{K},\ \ttr_0(Q)=M\}.\]

Applying the lemma, for each $\eps>0$ there exists a projector $P$ and $A\in \mathfrak{K}(M)$ such that $\g_0':=P-P^0_-$ minimizes the functional $\mathcal{E}^\nu_{\text{BDF}}+\eps\ttr((A-\cdot)^2)$ on $\mathcal{M}$ and
\[
\mathcal{E}^\nu_{\text{BDF}}(\g'_0)\le E_{\text{BDF}}^\nu(M)+\eps^2,\ \ns{2}{\g'_0-A}\le \sqrt{\eps}.
\]
As in \cite{at}, $\g'_0$ satisfies the self-consistent equation
\begin{equation}
\begin{array}{rl}%[
\g'_0+P^0_-&=\chi_{(-\infty,\mu_0]}( D_{\g'_0}+2\eps(\text{sgn}(D_0)-A))\\%)[
                      &=\chi_{(-\infty,\mu_0]}(\Dt +\alpha B_{\g'_0}-2\eps A)%)
\end{array}
\end{equation}
where $\mu_0\in\mathbb{R}$ and $\Dt:=\D +D_0\tfrac{2\eps}{|D_0|}$. We choose $\eps=\la^{-1}$ small \emph{e.g.} $\eps=\G(\tfrac{\La}{\alpha})^{-1}$. Using the proof of Lemma \ref{apriori1} we show that the following \emph{a priori} estimate holds for $\g'_0:$ 
\[
\ttr(|\nabla| (\g'_0)^2)+\alpha \ncc{\rho''_{\g_0}}^2\apprle \alpha \ncc{\nu}^2+\sqrt{\alpha} M+\sqrt{\alpha M}\ncc{\nu}.
\]
Using the Cauchy expansion, we can write
\[
\g_0=\ssum_{j=0}^{+\infty}\alpha^j O_{j}(\rho''_{\g_0},\g'_0)+\frac{2}{\la}W_\la(A,\alpha B(\g'_0)),
\]
where the $O_j$'s are defined as the $Q_j$'s with $\Dt$ replacing $\D$ (see \eqref{def_f1_eq}). By the same method as in Section \ref{fixedpointmethod} we have:
\[
\ns{2}{\,|\D|^{1/2} W_\la}+\ncc{\rho[W_\la]}\apprle \ns{2}{A}\big(1+\alpha [\ncc{\rho''_{\g_0}}+\ns{2}{\,|\nabla|^{1/2} \g'_0}]\big).
\]
Indeed it suffices to replace one $R[\g'_0]$ in the $O_j$'s by $A$ and remark that $A\in\mathfrak{S}_2$. Replacing $\D$ by $\Dt$ is harmless; as before, by defining some function $\widetilde{F}^{(1)}$ we can show that 

\noindent $\ttr_0(\g_0)=0$ (but with an alternative $B_\La$ \textit{cf} Section \ref{flaflafla}). 

 In particular we can write
\begin{equation*}
\rho_{\g_0}:=-\mathscr{F}^{-1}(\wit{F}_\La)*n_0''+(\delta_0-\mathscr{F}^{-1}(\wit{F}_\La))*\tau_{\text{rem}}\in\mathcal{C}
\end{equation*}
where $\ncc{\tau_{\text{rem}}}\apprle \ncc{\mathfrak{t}[N_0]}+\alpha^2\ncc{\wit{\tau}_2}+\ns{2}{A}/\la$ and $\wit{F}_\La$ is defined in Section \ref{flaflafla}. We write $\mathfrak{f}_\La:=\mathscr{F}^{-1}(\wit{F}_\La)$ for short. As in Section \ref{fixedpointmethod} we get:
\begin{equation}\label{chichi}
\begin{array}{|l}
\ns{2}{\g_0}\apprle \alpha(\ncc{\rho''_\g}+ \nqkino{\g'_0})\\
\ncc{\rho_{\g_0}+\mathfrak{f}_\La*n_0''-(\delta_0-\mathfrak{f}_\La)*\mathfrak{t}[N_0]}\apprle  \alpha^2(\nqkino{\g'_0}+\ncc{\rho_{\g_0}''})^2.\\
\nlp{1}{-\mathfrak{f}_\La*n_0''+(\delta_0-\mathfrak{f}_\La)*\mathfrak{t}[N_0]}\apprle L(Z+M).
\end{array}
\end{equation}

Let $(\psi_j)_{1\le j\le M}$ be an orthonormal family of eigenvectors of $\wit{D}+\alpha B_{\g'_0}+2/\eps(1-P^0_--A)$ spanning $\text{Ran}(N_0)$ (with eigenvalues $(\mu_j)$).

We then scale $\g'_0$ by $\alpha^{-1}$ (this procedure is emphasized by an underline)  as in \cite{at} we get:
\begin{equation}\label{eqalpha}
\Big[\big(\frac{g_0(-i\alpha\nabla)\beta}{\alpha^2} -\dfrac{ig_1(-i\alpha \nabla)}{\alpha^2}\boldsymbol{\alpha}\cdot \nabla\big)+\rho[\un{\g'_0}]*\frac{1}{|\cdot|}-R[\un{\g'_0}]+\frac{2}{\alpha^2\la}(\tfrac{1}{2}-\un{\PP}-\un{A})\Big]\un{\psi_j}=\frac{\mu_j}{\alpha^2}\un{\psi_j}.
\end{equation}
\begin{remark}
We have $U_{\alpha} \un{\psi}(x)=\alpha^{\tfrac{3}{2}}\un{\psi}(\alpha x)=\psi(x)$ and for an operator $S$ we define: 
\[\un{S}:=U_{\alpha}^* S U_{\alpha}.\]
\end{remark}
This mean-field operator $H_{\alpha^{-1}}$ is decomposed as follows: $H_{\alpha^{-1}}=H^{(1)}_{\alpha^{-1}}+h_{\text{rem}}$ where
\[
H^{(1)}_{\alpha^{-1}}:=\frac{\un{\D}}{\alpha^2}+(\delta_0-\un{\mathfrak{f}_\La})*\un{n}_0''-R[\un{N_0}],\ \un{n}_0''(x)=\alpha^{-3} n''(x/\alpha), \un{\wit{F}_\La}(k)=\wit{F}_\La(\alpha k).
\]
As in the Lemma 13 and 14 of \cite{at} we can show that there exists $\eps>0$ such that 

\noindent$\limsup_{\alpha \to 0}(\alpha^{-2}(\mu_j-1))<-\eps<0$ for all $1\le j\le M$ and that $(\un{\psi_j})_j$ is bounded in $H^1(\mathbb{R}^3,\mathbb{C}^4)^M$ (as $\alpha$ tends to $0$). Lemma 13 is based on a min-max description of eigenvalues in the gap of the mean-field operator $H_{\alpha^{-1}}$. We refer to this paper for the proofs. The only difference lies in the presence of $-\un{\mathfrak{f}_\La}*(\un{n_0}'')*\tfrac{1}{|\cdot|}$ and $(\delta_0-\un{\mathfrak{f}_\La})*\un{\mathfrak{t}_{N_0}}$: we deal with these terms in the following lemma, proved below.
\begin{lem}\label{der_des_der}
Let $\chi$ be a Schwartz function and for $R>0$: $\chi_R(x):=R^{-3/2}\chi(x/R)$. Then there holds:
\[
\begin{array}{l}
\Big| \psh{\un{\mathfrak{f}_\La}*(\un{n_0}'')*\tfrac{1}{|\cdot|}\chi_R-Z\wit{F}_\La(0) \frac{\chi_R}{|\cdot|}}{\chi_R}\Big|\apprle \frac{ZL}{R^2}\nlp{2}{\nabla \chi}^2\\
\ \ \ \ \ \ \ \ \ +\frac{1}{R}\nlp{2}{\nabla \chi}\nlp{2}{\chi}\Big(L\dint_{|y|>\tfrac{1}{\alpha}}\nu(y)dy+Z\dint_{|y|>\tfrac{1}{\alpha}}|\mathfrak{f}_\La(y)|dy\Big),
\end{array}
\]
and $\dint_{|y|>\tfrac{1}{\alpha}}|\mathfrak{f}_\La(y)|dy\apprle L\alpha^{1/2}$. Moreover for $r_0>0$ we have
\[
\begin{array}{l}
R\Big|\psh{(\delta_0-\un{\mathfrak{f}_\La})*\un{\mathfrak{t}_{N_0}}*\tfrac{1}{|\cdot|} \chi_R}{\chi_R}\Big|\apprle \frac{\alpha \nlp{1}{\mathfrak{t}_{N_0}}}{R}\dint\frac{|\mathscr{F}(|\chi|^2;k)|}{|k|}dk+\!\!\!\!\underset{|y|\ge r_0}{\dint}|\un{\mathfrak{t}_{N_0}}(y)|dy\dint\frac{|\chi(x)|^2}{|x|}dx\\
\ \ \ \ \ \ \ \ \ \ \ \ \ \ \ \ \ \ \ \ \ \ \ \ \ \ \ +\nlp{1}{\mathfrak{t}_{N_0}}\underset{|x|\le \tfrac{r_0}{R}}{\dint}\frac{|\chi(x)|^2}{|x|}dx.
%\Big|\psh{(\delta_0-\un{\mathfrak{f}_\La})*\un{\mathfrak{t}_{N_0}}*\tfrac{1}{|\cdot|} \chi_R}{\chi_R}\Big|\apprle \frac{L^2M}{R^2}\nlp{2}{\nabla \chi}^2\\
%\ \ \ \ \ \ \ \ \ +\frac{1}{R}\nlp{2}{\nabla \chi}\nlp{2}{\chi}\Big(L\dint_{|y|>\tfrac{1}{\alpha}}|\mathfrak{t}_{N_0}|(y)dy+LM\dint_{|y|>\tfrac{1}{\alpha}}|\mathfrak{f}_\La(y)|dy\Big).
\end{array}
\]
\end{lem}
\begin{remark}
This is because of the last term that the bound on $L$ depends on $M$. If we could prove that $\int_{|x|\ge r_0} |\mathfrak{t}_{N_0}(y)|dy$ tends to $0$ as $r_0\to+\infty$ \emph{uniformly in $\eps$} (the parameter of Borwein and Preiss's Lemma), then we could take $L\le L_0$ instead of $L\le 1/(K_0 M)$ in Theorem \ref{existence}.
\end{remark}

To prove $(\un{\psi_j})_j$ is $H^1$-bounded we show that:
\begin{equation}\label{but}
\dfrac{M}{\alpha^4}+\dfrac{\ttr(-\Delta \un{N_0})}{\alpha^2}\le \dfrac{\ttr(\un{\D}^2\un{N_0})}{\alpha^{4}}\le \dfrac{M}{\alpha^4}+K(M,\nu)\Big\{\ttr(-\Delta \un{N_0})+\dfrac{\ns{2}{\nabla \un{N_0}}}{\alpha^2}\Big\}.
\end{equation}
The lower bound is clear and the upper bound follows from Eq. \eqref{eqalpha}, Lemma \ref{apriori1} and Proposition \ref{www} (for estimations of $g_\star(\alpha p)^2$, $\star\in\{0,1\}$). We get:
%???????????????????????????????????\ref{estimates}
\[
\begin{array}{rl}
\Big\lvert\Big\lvert\rho[\un{\g_0}]+\rho[\un{\mathfrak{f}_\La}*\un{n''_0}-(\delta_0-\un{\mathfrak{f}_\La})*\un{\mathfrak{t}_{N_0}}] \un{\psi_j}\Big\rvert\Big\rvert_{L^2}&\apprle \alpha^{3/2}(\ncc{\rho''_{\g_0}}+\nqkino{\g'_0})^2\nlp{2}{\,|\nabla|^{1/2}\un{\psi_j}}\\
 &\apprle K(M,\nu)\nlp{2}{\,|\nabla|^{1/2}\un{\psi_j}}.
\end{array}
\]
Moreover:
\begin{equation}\label{presquefini}
\begin{array}{|l}
\nlp{2}{R[\un{\g_0}] \un{\psi_j}}\apprle \ns{2}{\g_0}\nlp{2}{\nabla \un{\psi_j}}\apprle \alpha^{3/4}K(M,\nu)\nlp{2}{\nabla \un{\psi_j}}\\
\nlp{2}{v\big[\un{\mathfrak{f}_\La}*\un{n''_0}-(\delta_0-\un{\mathfrak{f}_\La})*\un{\mathfrak{t}_{N_0}\big]}\un{\psi_j}}^2\le 4\nlp{2}{\nabla \un{\psi_j}}^2\nlp{1}{\rho[\un{\mathfrak{f}_\La}*\un{n''_0}-(\delta_0-\un{\mathfrak{f}_\La})*\un{\mathfrak{t}_{N_0}]}}^2\\
\ \ \ \ \ \ \ \ \ \ \ \ \ \ \ \ \ \ \apprle L^2(Z+M)^2\nlp{2}{\nabla \un{\psi_j}}^2\\
\big|\psh{v\big[\un{\mathfrak{f}_\La}*\un{n''_0}-(\delta_0-\un{\mathfrak{f}_\La})*\un{\mathfrak{t}_{N_0}\big]}\un{\psi_j}}{\un{\psi_j}}\big|\le \big|D(\rho\big[\un{\mathfrak{f}_\La}*\un{n''_0}-(\delta_0-\un{\mathfrak{f}_\La})*\un{\mathfrak{t}_{N_0}}\big],|\un{\psi_j}|^2)\big|\\
\ \ \ \ \ \ \ \ \ \ \ \ \ \ \ \ \ \apprle L(Z+M)\psh{|\nabla|\un{\psi_j}}{\un{\psi_j}}.
\end{array}
\end{equation}

Summing over $1\le j\le M$ the inequalities \eqref{presquefini} we get \eqref{but} because
\[
\ssum_{j=1}^M\nlp{2}{\nabla \un{\psi_j}},\ttr(|\nabla| \ov{N}_0)\le \sqrt{M}\sqrt{\ttr(-\Delta \un{N_0})}.
\]

We conclude as in \cite{at} (the proof uses \cite{pll}) provided that there hold binding inequalities for the non-relativistic limit: this is the result of Proposition \ref{binding} in Appendix \ref{nonrel}.

In particular there holds
\begin{equation}
\underset{\alpha\to 0}{\lim}\alpha^{-2}\big(E^\nu_{BDF}(M)-M+\frac{\alpha}{2}D(\check{\Fla}*\nu,\nu)\big)=E_{nr}(M),
\end{equation}
where $E_{nr}$ is the non-relativistic energy \emph{cf} Appendix \ref{nonrel}.

\noindent\textbf{Proof of Lemma \ref{der_des_der}}
With $f(x)=|\chi_R|^2*\mathscr{F}^{-1}(\un{\wit{F}_\La})$, we first estimate 

\noindent$|\iint f(x)\nu(y)(1/|x-\alpha y|-1/|x|)dxdy|$: it is lesser than
\[
\diint |f(x)|\nu(y)|\alpha y|\frac{dxdy}{|x||x-\alpha y|}.
\]
Splitting at level $\alpha^{-1}$ for $y$, we use Hardy's and Kato's inequalities:
\[
\left\{
\begin{array}{rl}
\dint_{|y|\le \tfrac{1}{\alpha}}\nu(y)dy\dint \frac{|f(x)|dx}{|x||x-\alpha y|}&\le (4Z\nlp{1}{\wit{F}_\La})\frac{\nlp{2}{\nabla\chi}^2}{R^2}\\
\dint_{|y|>\tfrac{1}{\alpha}} \nu(y)dy \dint | \alpha y|\frac{dx}{|x| |x-\alpha y|}|f(x)|&\le \frac{2\pi}{2}\dint_{|y|>\tfrac{1}{\alpha}} \nu(y)dy\nlp{1}{\wit{F}_\La}\frac{\nlp{2}{\nabla \chi}\nlp{2}{\chi}}{R}.
\end{array}\right.
\]
We estimate $Z\big|\iint |\chi_R(x)|^2\mathfrak{f}_\La(y)(1/|x-\alpha y|-1/|x|)dxdy\big|$ analogously, with the help of Lemma \ref{witfla}.
To treat the terms with $\un{\mathfrak{t}_{N_0}}$ we use the fact that:
\[
\nlp{1}{\un{\mathfrak{t}_{N_0}}}=\nlp{1}{\mathfrak{t}_{N_0}}\apprle LM\text{\ and\ }\dint\un{\mathfrak{t}_{N_0}}=\dint\mathfrak{t}_{N_0}=0.
\]
The first term in the upper bound corresponds to the error term that we get when we replace $\mathscr{F}^{-1}(\wit{F}_\La)*\tfrac{1}{|\cdot|}$ by $\wit{F}_\La(0)$. To see this, we write $a:=\wh{\un{\mathfrak{t}_{N_0}}}$ and $b:=\wh{|\chi|^2}$: we have
\[
\begin{array}{rll}
	\dint \frac{a^*(k)b(Rk)}{|k|^2}(\wit{F}_\La(\alpha k)-\wit{F}_\La(0))&=&\dint \frac{a^*\big(\frac{k}{R}\big)b(k)}{|k|^2}(\wit{F}_\La\big(\frac{\alpha k}{R}\big)-\wit{F}_\La(0)),\\
	\Big| \dint \frac{a^*(k)b(Rk)}{|k|^2}(\wit{F}_\La(\alpha k)-\wit{F}_\La(0))\Big|&\apprle& \frac{\alpha\nlp{1}{\mathfrak{t}_{N_0}}}{R}\dint\frac{|b(k)|}{|k|}dk.
\end{array}
\]
Let $\varrho$ be in $L^1$. Thanks to Newton's Theorem (for radial functions) we have
\[
\begin{array}{rll}
	R\times D\big(|\chi_R|^2,\varrho \big)&=&\dint \varrho(y)\Big(\frac{R}{|y|}\dint_{|x|\le \tfrac{|y|}{R}}|\chi(x)|^2dx+\dint_{|x|\ge \tfrac{|y|}{R}} \frac{|\chi(x)|^2}{|x|}dx\Big)\\
	 &=&\underset{|y|> r_0}{\dint}\varrho(y)\dint_{|x|\le \tfrac{|y|}{R}}|\chi(x)|^2\big(\frac{R}{|y|}-\frac{1}{|x|} \big)dx+\underset{|y|\le r_0}{\dint}\varrho(y)\Big(\dint_{|x|\le \tfrac{|y|}{R}}|\chi(x)|^2\big(\frac{R}{|y|}-\frac{1}{|x|} \big)dx \Big)\\
	 &\apprle& \nlp{2}{\,|\nabla|^{1/2}\chi}\underset{|y|> r_0}{\dint}|\varrho(y)|dy+\nlp{1}{\varrho}\dint_{|x|\le \tfrac{r_0}{R}}\frac{|\chi(x)|^2}{|x|}.
\end{array}
\]
%and use the same method.
\hfill{\footnotesize$\Box$}

\begin{appendix}

\section{Estimates and inequalities}\label{appA}
\begin{notation}\label{cs}
In Section \ref{appA} and \ref{flaflafla}, $\mathbf{e}$ refers to any unitary vector in $\mathbb{R}^3$ and for $p\in \mathbb{R}^3$, we write $\om_p:=\tfrac{p}{|p|}$.

We recall that $\sbf{p}=\mathscr{F}(\text{sign}(\D);p)$. There exists $C_s>0$ such that:
\begin{equation}\label{trick1}
\begin{array}{|l}
\mathrm{Id}-\sbf{p}\sbf{q}=\sbf{p}(\sbf{p}-\sbf{q})=(\sbf{p}-\sbf{q})\sbf{q}\\
|\mathrm{Id}-\sbf{p}\sbf{q}|\le |\sbf{p}-\sbf{q}|=\Big|\wh{\PP}(p)-\wh{\PP}(q)\Big|\le C_s\frac{|p-q|}{\max(\ed{p},\ed{q})}.
\end{array}
\end{equation}
\end{notation}

%\subsection{On the Dirac operator}

\subsection{Proof of Lemma \ref{exch}}\label{sexch}

We have \cite{LL} $\tfrac{1}{|\nabla|}(x-y)=\text{Cst}/|x-y|^2$. By Cauchy-Schwartz inequality there holds:
\[
\begin{array}{rl}
\mathrm{Tr}(R_Q^*|\nabla|^{-1}R_Q)&=\diiintdens \mathrm{Tr}_{\mathbb{C}^4}\frac{{}^t\ov{Q}(x,y)}{|x-y|}\frac{\text{Cst}}{|y-z|^2}\frac{Q(z,x)}{|z-x|}dxdydz\\
        &\apprle \diiintdens |Q(x,y)|^2\frac{1}{|y-z|^2}\frac{dxdydz}{|z-x|^2}\\
        &\apprle \diint \frac{|Q(x,y)|^2}{|x-y|}=\ttr(R_Q^*Q).\\
\end{array}
\]
We write $m(|p+q|)$ the multiplication in Fourier space by $|p+q|$: the operators $R_\cdot$ and $\tfrac{1}{|\nabla|^{1/2}}$ commute with the multiplication in Fourier space by $w(p-q)$ (written $m(w)$). By Kato's inequality we have
\[
\ns{2}{m(w)\cdot \tfrac{1}{|\nabla|^{1/2}} R[Q]}=\ns{2}{\tfrac{1}{|\nabla|^{1/2}} R[m(w)\cdot Q]}\apprle \ns{2}{m(|p+q|) m(w)\cdot Q}.
\]
Similarly for $a>0$ the operator $|D_0|^{-a}$ is a convolution operator associated to a \emph{positive} function $\phi_a$. Indeed there holds \cite{LL}: $\tfrac{1}{\om^2-\Delta}(x-y)=\dfrac{e^{-\om|x-y|}}{4\pi|x-y|},\om\ge 0$ and for any $0<\eps<1$ (see \cite[footnote p. 87]{stabilitymatter}):
\[
\dfrac{1}{|D_0|^{2\eps}}=\dfrac{\sin(\eps\pi)}{\pi}\dint_0^{+\infty} t^{p-1}\dfrac{1-\Delta}{t+1-\Delta}dt.
\]
Thus for $a=1+\eps>1$ we have by Cauchy-Schwarz inequality:
\[
\begin{array}{rl}
\ttr(R_Q^* \tfrac{1}{|D_0|^{2a}} R_Q)&\le \diint |Q(x,y)|^2\tfrac{1}{|\cdot|^2}*\phi_{2a}(x-y)dxdy,\\
     &\le \diint |Q(x,y)|^2\nlp{\infty}{\tfrac{1}{|\cdot|^2}*\phi_{2a}}\\
     &\apprle \diint |Q(x,y)|^2dxdy \dint\dfrac{dp}{|p| E(p)^{2a}}\apprle \dfrac{\ns{2}{Q}^2}{2(a-1)}.
\end{array}
\]

Let us consider a finite rank operator $Q(x,y)$. As $Q=\tfrac{Q+Q^*}{2}+\frac{Q-Q^*}{2}$ one may suppose it is self-adjoint, writing $Q=Q_+-Q_-$ one may suppose it is nonnegative: then so is $R_Q$ and $|D_0|^{-a/2}R_Q|D_0|^{-a/2}$. We have
\[
\begin{array}{rl}
\dint\frac{\mathrm{Tr}_{\mathbb{C}^4}(\wh{R}(p,p))}{E(p)^{2a}}dp&=\dfrac{1}{2\pi^2}\dint\!\!\!\!\dint\frac{d\ell}{|\ell|^2}\mathrm{Tr}(\wh{Q}(p-\ell,p-\ell))\frac{dp}{E(p)^{2a}}\\
               &=\dfrac{1}{2\pi^2}\dint dp \mathrm{Tr}(\wh{Q}(p,p))\dint\frac{d\ell}{|\ell|^2}\frac{1}{E(p+\ell)^{2a}}\\
               &\apprle \dfrac{\ns{1}{\wh{Q}}}{2a-2}.
\end{array}
\]
In Fourier space we have: $\mathscr{F}\big(|\D|^{-1/2}\big): f(p)\mapsto \chi_{|p|<\La}\frac{f(p)}{\ed{p}^{1/2}}$. Thus writing $\Pi_A$ the projection onto $\{f\in L^2,\ \text{supp}\,\wh{f}\subset\, B(0,A)\}$ we get
\[
\ns{2}{\,|\D|^{1/2} R_Q}\le \ns{2}{\,|\D|^{1/2} \Pi_{2\La}R_Q \Pi_{3\La}}.
\]
As $|\D|^{-1/2} \Pi_{2\La}\le e|D_0|^{-\tfrac{1}{2}-\tfrac{1}{2\llo}}$ for $\La\ge e$ we finally have:
\[
\ttr\big(\Pi_{3\La} R_{Q*} \frac{\Pi_{2\La}}{|\D|}R_Q \Pi_{3\La} \big)\le \ttr\big( R_{Q*} \frac{e}{|\D|^{\tfrac{1}{2}+\tfrac{1}{2\llo}}}R_Q\big)\apprle \llo \nqq{Q}^2.
\]

\section{The non relativistic limit}\label{nonrel}
We fix the value $\Fla(0)=a$. For any trace-class operator $0\le \G\le 1$ with density $\rho_\G$ the non-relativistic energy is
\begin{equation}
\begin{array}{rl}
\mathcal{E}_{nr}^Z(\G)&:=\frac{1}{2}\ttr(-\Delta \G)-Z(1-a)\ttr\big(\frac{1}{|\cdot|} \G\big)\\
&\ \ \ \ \ \ \ \ \ +\frac{1}{2}(D(\rho_\G,\rho_\G)-\text{Ex}[\G])-\frac{a}{2}D(\rho_\G,\rho_\G).
\end{array}
\end{equation}
If we drop the last term, this is exactly the Hartree-Fock energy $\mathcal{E}_{\text{HF}}$ with a nucleus of charge $Z_0:=Z(1-a)$ and if we drop $\ttr(\tfrac{1}{|\cdot|} Q)$ we get the Pekar-Tomasevitch energy $\mathcal{E}_{nr}^{0}=\mathcal{E}_{\text{PT}}[\tfrac{a}{2},U=\tfrac{1}{2}]$ (\textit{cf} \cite{bip}).%: this last functional is what we obtain by putting the system at infinity.

%We write $E_{\text{PT}}:=E_{\text{PT}}[1,U=a^{-1}]$, putting any test function at infinity we have $E_{nr}(M)\le \tfrac{a^2}{2}E_{\text{PT}}(M)$ and $E_{\text{PT}}(M)=ME_{\text{PT}}(1)$ for sufficiently small $a$, we refer to Corollary 1 in \cite{bip}: there exists $U_0$ such that if $U\overset{\text{here}}{=}\tfrac{1}{a}\ge U_0$ then there is no binding for $E_{\text{PT}}(M)$, $M\ge 2$.
%souci avec \cite{bip}
\begin{remark}
We can easily show stability of matter of the second kind for $a\le a_0$ by splitting the energy in two: a Hartree-Fock one and a Pekar-Tomasevitch  one,
\[
\begin{array}{l}
\mathcal{E}_{nr}^Z(\G)=\frac{x^2}{2}\ttr(-\Delta \G)+\frac{y^2}{2}(D(\rho_\G,\rho_\G)-\text{Ex}[\G])-Z(1-a)\ttr\big(\frac{1}{|\cdot|} \G\big)\\
\ \ \ +\frac{1-x^2}{2}\ttr(-\Delta \G)+\frac{1-y^2}{2}(D(\rho_\G,\rho_\G)-\text{Ex}[\G])-\frac{a}{2}D(\rho_\G,\rho_{\G})\text{\ with\ }0<x,y<1.
\end{array}
\]
Optimizing in $x$ and $y$ we get a lower bound $\mathcal{O}(K(a) M)$ for $M\ge 2Z_0+1$.
\end{remark}
We define
\[
\mathcal{G}(x)=\{\G\in\mathfrak{S}_1: \G^*=\G,\ 0\le \G\le 1,\sqrt{-\Delta} \G\in\mathfrak{S}_2\text{\ and\ }\ttr(\G)=x\}\text{\ with\ }x\in\mathbb{R}_+^*.
\]
$E_{nr}(M)$ corresponds to the infimum over $\mathcal{G}(M)$. We want to prove:
\begin{pro}\label{binding}
For any $M<Z+1$, the variational problem $E_{nr}^Z(M)$ admits a minimizer.
\end{pro}
\noindent By Lieb's method in \cite{L}, it is easy to see that there is a minimizer for $E_{nr}^Z(1)$.% if $\nu$ is radial, else we prove it by concentration-compactness \cite{pll}. 
 To prove binding for $2\le M\le Z(1-a)$ we can follow Lieb's and Simon's method \cite{lisim,pll}. We will however prove it with the method of concentration-compactness. We prove the problem $E_{nr}^Z(M)$ admits a minimizer by induction over $M$ by using:
\begin{pro}
For each $\ell>0$ the following assertions are equivalent
\begin{itemize}
\item[\textbullet] $\forall 0<k<\ell:\ E_{nr}^Z(\ell)<E_{nr}^Z(\ell-k)+E_{nr}^0(k)$.
\item[\textbullet] Each minimizing sequence for $E_{nr}^Z(\ell)$ is precompact in $H^1(\mathbb{R}^3\times \mathbb{R}^3)$.
\end{itemize}
In the case $\ell\in\mathbb{N}^*$, it suffices to prove binding inequalities for $K\in(0,\ell)\cap\mathbb{N}.$
\end{pro}
 This proposition is standard and we will not give the proof here but refer to \cite{geom,lewlen,pll}. In \cite{bip} Frank \emph{et al.} prove that  $E_{nr}^0(M_0)=M_0E_{nr}^0(1)$ for $M_0\in\mathbb{N}^*$ provided that $a$ is sufficiently small. Thus we just have to show
\[
E_{nr}^Z(M)<E_{nr}^Z(M-1)+E_{nr}^0(1).
\]
To this end, we exhibit a test function $Q$ whose energy is lesser than $E_{nr}^Z(M-1)+E_{nr}^0(1)$.

Lieb's variational principle still holds (\textit{cf} \cite[Proposition 3]{at}). In fact for any orthonormal family $(\phi_1,\phi_2)$ , with $P_\phi:=\ket{\phi}\bra{\phi}$ and $0<t<1$, we have
\begin{equation}\label{phi1phi2}
\begin{array}{l}
\mathcal{E}_{nr}^Z(\G+t(P_{\phi_1}-P_{\phi_2}))-\mathcal{E}_{nr}^Z(\G)=\tfrac{t}{2}(\nlp{2}{\nabla\phi_1}^2-\nlp{2}{\nabla \phi_2}^2+2(1-a)D(\rho_\G,|\phi_1|^2-|\phi_2|^2))\\
-t\mathfrak{R}[\ttr(\G R[P_{\phi_1}-P_{\phi_2}])] -t^2\big\{D(|\phi_1|^2,|\phi_2|^2)-D(\phi_1^*\phi_2,\phi_1^*\phi_2)+\tfrac{a}{2}\ncc{|\phi_1|^2-|\phi_2|^2}^2\big\}.
\end{array}
\end{equation}
This shows that $E_{nr}^Z(m)$ is also the infimum of $\mathcal{E}^Z_{nr}$ over
\[
\{\G\in \mathcal{G}(m):\ \G=P+(m-[m])\ket{\phi}\bra{\phi}, P^2=P=P^*, \phi\in\text{Ker}(P) \}.
\]
Taking $\phi_2=0$ in \eqref{phi1phi2} shows that $E_{nr}^Z(\cdot)$ is concave in $[M_0,M_0+1]$ with $M_0\in\mathbb{N}$. It is also clear that $E^Z_{nr}$ is decreasing since large binding inequalities hold. 

We consider a minimizer of $E^Z_{nr}(M-1)$ of the form $\G=\underset{1\le j\le M-1}{\sum}\ket{\psi_j}\bra{\psi_j}$, each $\psi_j$ satisfying
\[
\frac{-\Delta}{2}\psi_j-\frac{Z_0}{|\cdot|}\psi_j+(1-a)\rho[\G]*\frac{1}{|\cdot|}\psi_j-R[\G]\psi_j+\eps_j\psi_j=0,\text{\ with\ }\eps_j>0.
\]
In particular we can easily show the $\psi_j$'s are in $H^2(\RR)$ and fast decaying.

We also consider a minimizer for $E_{nr}^0(1)$: this is a minimizer $\phi_{\text{CP}}$ of $E_{\text{PT}}(1)$
scaled by $a$: $\phi_0(x)=a^{3/2}\phi_{\text{CP}}(a x)$, we chose it to be radial \cite{L}. Following \cite{lewlen}, we take a Schwartz function $0\le \chi \le 1$ that satisfies $\chi(x)=1$ for $|x|\le 1$ and $\chi(x)=0$ for $|x|\ge 2$ and $\chi_R(x)=\chi(x/R)$ with $R>0$ to be chosen. 

We define the trial state as follows: for some $\mathbf{e}\in \mathbb{S}^2$ we write
\[
\G':=\chi_R \G \chi_R+\tau_{-5R\mathbf{e}}\ket{\chi_R \phi_0}\bra{\chi_R \phi_0}\tau_{5R\mathbf{e}}
\]
where $\tau_{x_0}\psi(x):=\psi(x-x_0)$. We have $0\le \G'\le 1$ and $\ttr(\G')\le M$, so 

\noindent $\mathcal{E}_{nr}(\G')\ge E_{nr}(M)$. As the wave functions $(\psi_j)$'s and $\phi_0$ are fast decaying, the following holds:
\[
\begin{array}{l}
\mathcal{E}_{nr}^Z(\G')=\mathcal{E}_{nr}^Z(\G)+\mathcal{E}_{nr}^{0}(\phi_0)+\dint\big(\rho[\G]*\frac{1}{|\cdot|}(x)-\frac{Z_0}{|x|}\big)|\tau_{5R\mathbf{e}}\phi_0(x)|^2dx\\
\ \ \ \ \ \ \ \ \ \ \ \ \ \ \ \ \ \ -aD(\rho[\G],|\tau_{5R\mathbf{e}}\phi_0|^2)+o(R^{-1}).
\end{array}
\]
As $R$ tends to infinity we get:
\[
\mathcal{E}_{nr}^Z(\G')\le E_{nr}^Z(M-1)+E_{nr}^0(1)+\frac{(M-1)(1-a)-Z_0}{5R}+o(R^{-1})<E_{nr}(M-1)+E_{nr}^0(1).
\]
%This proof works \textit{mutatis mutandis} to prove that $E_{nr}(m)<E_{nr}(m-k)+E_{nr}^0(k)$ in the case $E_{nr}(m-k)$ admits a minimizer f

\section{Proof of Proposition \ref{profla}}\label{flaflafla}
\begin{notation}
We write:
\[
\begin{array}{l}
E(u,k/2):=\max\big( E(u+k/2),E(u-k/2)\big)\ge \sqrt{1+|u|^2+\tfrac{|k|^2}{4}},\\
\ed{u,k/2}:=\max\big( \ed{u+k/2},\ed{u-k/2}\big)\ge E(u,k/2).
\end{array}
\]
\end{notation}
Our aim is to prove Proposition \ref{lalala} below.
\begin{pro}\label{lalala}
Let $\rho_0\in\mathcal{C}$. Then we have:
\[
\alpha\rho(\Tbf[Q_{0,1}(\rho_0)])=-\check{f}_\La*\rho_0
\]
where $\check{f}_\La\in L^1$ is a radial function. Moreover 
\[
\begin{array}{ll}
\fla=\ssum_{J=0}^{+\infty}\alpha^Jf_{\La,J},&f_{\La,0}=\alpha B_\La\quad\text{and}\quad g_\La:=\ssum_{J=1}^{+\infty}\alpha^Jf_{\La,J},
\end{array}
\] 
with 
\[
\nlp{1}{\check{f}_\La}\apprle L\text{\ and\ }\nlp{1}{\check{g}_\La}\apprle L\alpha. 
\]
\noindent In particular $\check{F}_\La:=\mathscr{F}^{-1}\big( \frac{\fla}{1+\fla}\big)\in L^1$.
\end{pro}
We also study an alternative function $\Fla$, needed for the proof of Theorem \ref{existence}, at the end of this section.

We need the following proposition.
\begin{pro}\label{www}
The function $\wh{\D}:\overline{B(0,\La)}\to \mathbb{R}^3$ is infinitely differentiable. In particular so is $\ed{\cdot}$ and there exists $L_0\ge 0$ such that if $L:=\alpha\llo\le L_0$ then for any $J\ge 1$ there exists $C_J>0$ such that:
\[
\nlp{\infty}{\text{d}^J g_0}\le \alpha C_J \text{\ and\ }\nlp{\infty}{\text{d}^J \mathbf{g}_1}\le \chi_{J=1}+L C_J.
%\Big\lvert \Big\lvert \text{d}^J \Big\rvert\Big\rvert
\]
\end{pro}
\begin{dem}In the spirit of \cite{sok}, we can prove it by induction over $J$: in \cite{mf} Hainzl \emph{et al.} proved that $\wh{\D}$ is infinitely differentiable. Thus the function
\[
|\wh{\D}(p)|=\sqrt{g_0(p)^2+\mathbf{g}_1(p)\cdot\mathbf{g}_1(p)},
\]
is infinitely differentiable and does not vanish on $\overline{B(0,\La)}$. Thanks to the self-consistent equation one has:
\[
\text{d}^J\wh{\D}(p)=\text{d}^J \wh{D_0}(p)+\dfrac{\alpha}{4\pi^2}\dfrac{1}{|\cdot|^2}*\text{d}^J\Big(\frac{\D}{|\D|}\Big)(p).
\]
\end{dem}

\noindent \textbf{Proof of Proposition \ref{lalala}:}
Throughout this proof we write $k:=r\mathbf{e}$.

\noindent $\mathfrak{1}.$ Let us first prove the following:
\[
\wh{\tau}_{1,0}(\rho)=-\fla(\cdot)\wh{\rho}(\cdot),
\]
We recall that for any $Q\in\mathfrak{S}_2(\hl)$ we have \eqref{q10}:
\[
\wh{Q}_{1,0}(Q,p,q)=\frac{1}{4\pi^2}\frac{1}{\ed{p}+\ed{q}}\dint_\ell \frac{d\ell}{|\ell|^2}\big(\wh{Q}(p-\ell,q-\ell)-\sbf{p}\wh{Q}(p-\ell,q-\ell)\sbf{q}\big),
\]
and (\textit{cf} \cite{ptf})
\begin{equation}\label{TFq01}
\wh{Q}_{0,1}(\rho;p,q)=\frac{4\pi}{2^{5/2}\pi^{3/2}}\frac{\wh{\rho}(p-q)}{|p-q|^2}\frac{1}{\ed{p}+\ed{q}}(\sbf{p}\sbf{q}-1).
\end{equation}
The functions $A_{J}^{(\ell_j)_{j=1}^J}$ are defined recursively in \eqref{recur}. We have for instance:

%\begin{equation}
%\left\{\begin{array}{ll}A_1^{\ell_1}\wh{Q}(p,q)&:=\wh{Q}(p-\ell_1,q-\ell_1)-\sbf{p}\wh{Q}(p-\ell_1,q-\ell_1)\sbf{q},\\
%A_{J}^{(\ell_1,\mathbf{L})}\wh{Q}(p,q)&:=A_1^{\ell_1}\big(A_{J-1}^{\mathbf{L}}Q\big)(p,q).
%\end{array}\right.
%\end{equation}

\[
\begin{array}{rl}
A_2^{(\ell_1,\ell_2)}\wh{Q}(p,q)&=A_1^{\ell_1}\big(\wh{Q}(\cdot_p -\ell_2,\cdot_q-\ell_2)-\sbf{\cdot_p}\wh{Q}(\cdot_p -\ell_2,\cdot_q-\ell_2)\sbf{\cdot_q}\big)(p,q)\\
                            &=\Big\{\wh{Q}(p-\ell_1-\ell_2,q-\ell_1-\ell_2)-\sbf{p-\ell_1}\wh{Q}(p-\ell_1-\ell_2,q-\ell_1-\ell_2)\sbf{q-\ell_1}\Big\}\\
                            &\ \ \ \ \ \ \ -\sbf{p}\Big\{\wh{Q}(p-\ell_1-\ell_2,q-\ell_1-\ell_2)-\sbf{p-\ell_1}\wh{Q}(p-\ell_1-\ell_2,q-\ell_1-\ell_2)\sbf{q-\ell_1}\Big\}\sbf{q}.
\end{array}
\]
Writing $L_J:=\sum_{j=1}^J \ell_j$ with $L_0:=0\in\mathbb{R}^3$ we have:
\begin{equation}
\wh{F_{1,0}^{\circ J}}(Q;p,q)=\frac{\alpha^J}{(4\pi^2)^J}\dint_{\ell_1}\cdots \dint_{\ell_J}\frac{d\boldsymbol{\ell}}{\underset{1\le j\le J}{\prod}|\ell_j|^2}\frac{A_J^{(\ell_j)_{j=1}^J}\wh{Q}(p,q)}{\underset{0\le j\le J-1}{\prod}(\ed{p-L_j}+\ed{q-L_j})}.
\end{equation}

In particular the Fourier transform of the density $\rho\big(F_{1,0}^{\circ J}(Q)\big)$ is
\begin{equation}\label{eqg10}
\begin{array}{rl}
\wh{\rho}(F_{1,0}^{\circ J}(Q);k)&=\dfrac{1}{(2\pi)^{3/2}}\dint_u \mathrm{Tr}_{\CC} \wh{F_{1,0}^{\circ J}}(Q;u+\tfrac{k}{2},u-\tfrac{k}{2})du\\
                                                        &=\dfrac{\alpha^J}{(2\pi)^{3/2}(4\pi^2)^J}\underset{u,\ell_1}{\diint}\cdots \underset{\ell_J}{\dint} \mathrm{Tr}_{\CC} \dfrac{dud\boldsymbol{\ell}}{\underset{1\le j\le J}{\prod}|\ell_j|^2}\dfrac{A_J^{(\ell_j)_{j=1}^J}\wh{Q}(u+\tfrac{k}{2},u-\tfrac{k}{2})}{\underset{0\le j\le J-1}{\prod}(\ed{u+\tfrac{k}{2}-L_j}+\ed{u-\tfrac{k}{2}-L_j})}\\
                                                        &=\dfrac{\alpha^J}{(2\pi)^{3/2}(4\pi^2)^J}\underset{u,\ell_1}{\diint}\cdots \underset{\ell_J}{\dint} \dfrac{dud\boldsymbol{\ell}}{\underset{1\le j\le J}{\prod}|\ell_j|^2}\dfrac{\mathrm{Tr}_{\CC}\big\{(1-\sbf{u-\tfrac{k}{2}}\sbf{u+\tfrac{k}{2}})A_{J}^{(\ell_j)_{j=2}^{J-1}}\wh{Q}(u+\tfrac{k}{2},u-\tfrac{k}{2})\big\}}{\underset{0\le j\le J-1}{\prod}(\ed{u+\tfrac{k}{2}-L_j}+\ed{u-\tfrac{k}{2}-L_j})}.
\end{array}
\end{equation}
Above the domain of $\ell_j$ is:
\[
\widetilde{B}_j(r):=\big\{\ell_j,\ \big|u-L_j\pm \tfrac{r}{2}\mathbf{e}\big|<\La\big\},
\]
and the domain of $u$ is $\widetilde{B}_0(r):=\big\{u,\ \big|u\pm \tfrac{r}{2}\mathbf{e}\big|<\La\big\}$. In particular 
\[\text{supp}\, \wh{\rho}(F_{1,0}^{\circ J}(Q))\subset B(0,2\La).\]

\begin{remark}\label{importantremark}
We would like to apply \eqref{eqg10} to the operator $Q_{0,1}(\rho)$. From \eqref{TFq01} we realize that $\wh{Q}_{0,1}(p,q)$ is not a scalar matrix because of the term $\sbf{p}\sbf{q}-\mathrm{Id}$. Yet it is in the algebra spanned by the Dirac matrices $\boldsymbol{\alpha}_1,\boldsymbol{\alpha}_2,\boldsymbol{\alpha}_3,\beta$ as a sum of \textit{even} products of Dirac matrices. The form of $\wh{Q}_{1,0}(Q)$ is similar to $\wh{Q}_{0,1}$: it only adds an \emph{even} number of Dirac matrices to $\wh{Q}$. This is an important remark to be done to prove Theorem \ref{alafurry}.
\end{remark}

For any $J\ge 1$ and $\rho\in\mathcal{C}$, the density $\wh{\rho}(F_{1,0}^{\circ J}(Q_{0,1}[\rho]);k)$ is equal to:
\begin{equation}\label{radial}
\begin{array}{l}
\dfrac{4\pi\alpha^J\wh{\rho}(k)}{(2^{5}\pi^3)^{\tfrac{1}{2}}(2\pi)^{3/2}(4\pi^2)^J}\underset{\underset{0\le j\le J}{\prod} \widetilde{B}_j(r)}{\dint}\dfrac{dud\boldsymbol{\ell}}{|k|^2\underset{1\le j\le J}{\prod}|\ell_j|^2}\dfrac{\mathrm{Tr}_{\CC}\big\{(1-\sbf{u-\tfrac{k}{2}}\sbf{u+\tfrac{k}{2}})A_{J-1}^{(\ell_j)_{j=2}^{J}}(\sbf{u-\tfrac{k}{2}}\sbf{u+\tfrac{k}{2}}-1)\big\}}{\underset{0\le j\le J}{\prod}(\ed{u+\tfrac{k}{2}-L_j}+\ed{u-\tfrac{k}{2}-L_j})}\\
=\wh{\rho}(k)\underset{\underset{0\le j\le J}{\prod} \widetilde{B}_j(r)}{\dint}du d\boldsymbol{\ell}\  S_J(u-L_j\pm\tfrac{k}{2})T_J(u-L_j\pm\tfrac{k}{2})
\end{array}
\end{equation}
where $S_J(u-L_j\pm\tfrac{k}{2})$ is a scalar which is a function of $|u-L_j\pm\tfrac{k}{2}|$ while $T_J(u-L_j\pm\tfrac{k}{2})$ is the trace $\ttr_{\mathbb{C}^4}$ of a sum of products of $\sbf{u-L_j-\tfrac{k}{2}}$. 

We have to deal with $\tfrac{1}{|k|^2}$ and we must show that this integral is well defined. The first problem is easy, the quantity
\[
\dfrac{1}{|k|^2}(\sbf{u-L_J+k/2}\sbf{u-L_J-k/2}-1)(1-\sbf{u-k/2}\sbf{u+k/2})=\dfrac{(\sbf{u-L_J+k/2}\sbf{u-L_J-k/2}-1)}{|k|}\dfrac{(1-\sbf{u-k/2}\sbf{u+k/2})}{|k|}
\]
defines a smooth function by Taylor's formula (for $|k|$ or for $k$ in $\mathbb{R}^3\backslash\{0\}$). Moreover from \eqref{trick1}, we get the estimates:
\[
\begin{array}{rl}
\bigg|\dfrac{\sbf{u-L_J+k/2}\sbf{u-L_J-k/2}-1}{|k|}\dfrac{1-\sbf{u-k/2}\sbf{u+k/2}}{|k|}\bigg|&\le \dfrac{4C_s^2}{E(u-L_J,k/2)}\\
                &\le\frac{4C_s^2}{|u-L_J|E(u,k/2)}.
\end{array}
\]
For any $U$, we have:
\[
\begin{array}{rl}
\dint_\ell\frac{d\ell}{|\ell|^2}\frac{1}{|U-\ell|\ed{U-\ell, k/2}}&\le \dint_\ell\frac{d\ell}{|\ell|^2}\frac{1}{|U-\ell|^2},\\
                           &\le \frac{1}{|U|}\dint\frac{d\ell}{|\ell|^2|\mathbf{e}-\ell|^2}.
\end{array}
\]
Integrating over the $\ell_{j+1}$'s one after the other from from $\ell=J-1$ down to $j=1$ as above with $U=U_j=u-L_j$, there remains but the integral over $u$:
\[
\begin{array}{rl}
\underset{u\in\widetilde{B}_0(k)}{\dint} \dfrac{2C_s^2du}{\ed{u,k/2}}\frac{1}{|u|\ed{u,k/2}}&\times \left\{2\dint\frac{d\ell}{|\ell|^2|\mathbf{e}-\ell|^2}\right\}^J\\
       &\le \left\{2\dint\frac{d\ell}{|\ell|^2|\mathbf{e}-\ell|^2}\right\}^J\underset{u\in\widetilde{B}_0(r)}{\dint}\dfrac{2C_s^2du}{|u|^2\ed{u,k/2}}\\
       &= (K\llo)\times\Big(C_{1,0}'\Big)^{J}.
\end{array}
\]

At last we have:
\begin{equation}
\begin{array}{rl}
\alpha |\wh{\rho}(F_{1,0}^{\circ J}(Q_{0,1}(\rho));k)|&\le \tfrac{\alpha^{J+1}}{(2\pi)^{3/2}(4\pi^2)^J}2^{J+1}C_s^2\left\{\dint\frac{d\ell}{|\ell|^2|\mathbf{e}-\ell|^2}\right\}^J\underset{u\in\widetilde{B}_0(r)}{\dint} \frac{du}{|u|^2\ed{u}}|\wh{\rho}(k)|\\
          &\apprle C_{1,0}\Big(\alpha C_{1,0}'\Big)^J\alpha \llo|\wh{\rho}(k)|.
\end{array}
\end{equation}
As a consequence there holds $\alpha \wh{\rho}(F_{1,0}^{\circ J}(Q_{0,1}(\rho));k)=-g_{\La;J}(k)\wh{\rho}(k)$, and $\sum_{J=0}^{\infty}f_{\La,J}$ is well defined (at least in $L^{\infty}\cap L^2$) as soon as $\alpha$ is sufficiently small. We have
\begin{equation}
\alpha \wh{\tau_{0,1}}(\rho,k)=-\left(\alpha B_\La(k)+\ssum_{J=1}^{+\infty}g_{\La;J}(k)\right)\wh{\rho}(k)=:-\fla(k)\wh{\rho}(k),
\end{equation}
with
\begin{equation}
|\fla(k)|\le \alpha B_\La(k)+\alpha^2\llo K=\mathcal{O}(\alpha\llo).
\end{equation}

\noindent $\mathfrak{2}.$ Let us prove this function is radial. Let $\mathbf{e}_1$ and $\mathbf{e}_2$ in $\mathbb{S}^2$ and $r>0$. We must show that $\fla(r\mathbf{e}_1)=\fla(r\mathbf{e}_2)$. There exists $\mathcal{R}\in \text{SO}_3(\mathbb{R})$ such that $\mathbf{e}_2=\mathcal{R}\mathbf{e}_1$. In \eqref{radial} for $k=r\mathbf{e}_2$, we change variables in the integrals: $v=\mathcal{R}^{-1} u$ and $m_j=\mathcal{R}^{-1}\ell_j$. Writing $M_j=m_1+\cdots+m_j$, we get: $S_J(\mathcal{R}(v-M_J\pm \tfrac{r}{2} \mathbf{e}_1))=S_J(v-M_J\pm \tfrac{r}{2} \mathbf{e}_1)$. We must show the same holds for $T_J$. Let $\mathbf{b}=(b_1,b_2,b_3)$ be the canonical base of $\mathbb{R}^3$. We define
\[
\alpha'_j:=\boldsymbol{\alpha}\cdot \mathcal{R}b_j.
\]
These new matrices satisfy the same relation as the $\alpha$'s:
\[
\begin{array}{l}
\{ \alpha'_j,\alpha'_k\}=2\delta_{jk}\text{\ and\ }\{\alpha'_j,\beta\}=0.
\end{array}
\]
Thus we have $T_J(\mathcal{R}(v-M_J\pm \tfrac{r}{2} \mathbf{e}_1))=T_J(v-M_J\pm \tfrac{r}{2} \mathbf{e}_1)$ and $\fla$ is radial.

From now on we change variables:
\begin{equation}\label{variables}
\left\{
\begin{array}{l}
u_0:=u\text{\ and\ for\ }1\le j\le J,\ u_j:=u-L_j,\ l_{j}=u_{j}-u_{j-1},\\
u_j\in B(|k|):=\big\{ v\in B(0,\La),\ \big|v\pm\frac{|k|}{2}\mathbf{e}\big|<\La\big\}.
\end{array}
\right.
\end{equation}

\noindent $\mathfrak{3}.$ Our purpose is to show that $\fla$ is in $\mathscr{F}(L^1)$ with a (rather) precise bound on $\nlp{1}{\check{f}_\La}$. We already know: $\fla(k)=\alpha B_\La(k)+\mathcal{O}_{L^\infty}(\alpha^2\log(\La))=\mathcal{O}(\alpha\llo).$

As $\fla$ is radial we take a fixed vector $\mathbf{e}\in\mathbb{S}^2$ and study $\fla(k)=\fla(|k|)$ with the help of the integral formulae where $k$ is replaced by $|k|\mathbf{e}$. 

The strategy is to differentiate $\fla$ and prove that its Sobolev norms $\nlp{2}{-\Delta\fla}$ and $\nlp{p}{-\Delta\fla}$ are "small" where $p<2$ is some constant to be chosen later. By Cauchy-Schwartz inequality in Direct space, we obtain an upper bound of $\nlp{1}{\check{f}_\La}$. We will use the \emph{co-area formula} \cite{ev_gar}.

We show that $\check{f}_\La\in L^1$ with $L^1$-norm lesser than $1$ in order to give a meaning to $ \ssum_{\ell=1}^{\infty}(-1)^\ell \big\{\check{f}_\La\big\}^{*\ell}.$

 \begin{remark}\label{sing}
 \begin{enumerate}
\item\ As $\fla$ is radial we have:
\begin{equation}
\begin{array}{rll}
(-\Delta)\fla&=(-\Delta_r)\fla&=-\big(\partial_r^2+\frac{2}{r}\partial_r\big)\fla.
\end{array}
\end{equation}
\item\ For any $u\in\mathbb{R}^3$ and $r\ge 0$ Taylor's formula gives:
\begin{equation}\label{nosing}
\left\{\begin{array}{rl}
(1-\sbf{u+2^{-1}r\mathbf{e}}\sbf{u-2^{-1}r\mathbf{e}})&=r\{\sbf{u}m_1(-\tfrac{r}{2})-m_1(\tfrac{r}{2})\sbf{u}\}\\
\text{with\ }m_1(\tfrac{x}{2})&:=\dint_{t=0}^1d\sbf{u+tx\mathbf{e}/2}\cdot\big(\frac{\mathbf{e}}{2}\big)dt.
\end{array}\right.
\end{equation}
We write $\mathbf{g}(p):=\begin{pmatrix}g_0(p)\\\mathbf{g}_1(p) \end{pmatrix}\in\mathbb{R}^4$ and $\sigma(p):=\tfrac{\mathbf{g}(p)}{\ed{p}}$. 
%souci ci-bas ?

\noindent As we have $\psh{\sigma(u)}{ \text{d}\sigma(u)}=0$, Taylor's Formula at order 2 gives
\begin{equation}\label{nosing2}
\left\{
\begin{array}{l}
\dfrac{1-\psh{\sigma(u+r\tfrac{\mathbf{e}}{2})}{\sigma(u-r\tfrac{\mathbf{e}}{2}})}{r^2}:=\psh{a(u)}{a(u)}+\psh{\sigma(u)}{m_2(r)+m_2(-r)}\\
 \ \ \ \ \ \ \ \ \ \ \ \ +r\psh{a(u)}{m_2(r)-m_2(-r)}+r^2\psh{m_2(r)}{m_2(-r)},\\
a(u):=\text{d}\sigma(u)\cdot\frac{\mathbf{e}}{2}\text{\ and\ }m_2(\tfrac{x}{2}):=\underset{[0,1]^2}{\diint}\text{d}^2\sigma{u+stx\mathbf{e}/2}\cdot\big(\frac{\mathbf{e}}{2},\frac{\mathbf{e}}{2}\big)tdsdt.
\end{array}
\right.
\end{equation}
\item For any $-\tfrac{1}{2}\le x\le \tfrac{1}{2}$:
\begin{equation}\label{inegevidente}
\ed{u+x\mathbf{e}}\ge E(u+x\mathbf{e} )\ge \frac{E(u)}{2}.
\end{equation}
In particular if one takes the modulus of the derivative over $r$ in \eqref{nosing} or \eqref{nosing2} for $0\le r\le 1$, we get the following upper bounds:
\begin{enumerate}
\item  $K/\ed{u}$ for the first derivative,
\item $K/\ed{u}^2$ for the second.
\end{enumerate} 
\end{enumerate}
\end{remark}

%We can decompose $\fla=\alpha B_\La+\gla$ and focus our study on $\gla$ ($\alpha B_\La$ is dealt with in the same way).$\partial _rf(0)=0$ and
\begin{lem}\label{glaco}
The functions $\partial_r\fla$ and $\partial_{r}^2\fla$ are well defined in $\RR$ with support in $\ov{B}(0,2\La)$. Furthermore for $J\in\mathbb{N}$ we have:
\begin{equation}
\left\{\begin{array}{rl | rl}
%|\partial_r f_{\La,J}(p)|&\apprle J\frac{\alpha^{J+1}\llo K^{J+1}\llo}{E(p)}\chi_{|p|<2\La}&|\partial_r \fla(p)|&\apprle  \frac{L}{E(p)}\chi_{|p|<2\La},\\
%|\partial_r^2 f_{\La,J}(p)|&\apprle J\frac{\alpha^{J+1}\llo K^{J+1}\llo}{E(p)^2}\chi_{|p|<2\La}&|\partial_r^2 \gla|&\apprle \frac{\alpha^2\llo }{E(r)^2}\chi_{r<2\La}.
|\partial_r f_{\La,J}(p)|&\apprle J\frac{\alpha^{J+1}\llo K^{J+1}}{E(p)}\chi_{|p|<2\La}&|\partial_r \fla(p)|&\apprle  \frac{L}{E(p)}\chi_{|p|<2\La},\\
|\partial_r^2 f_{\La,J}(p)|&\apprle J\frac{\alpha^{J+1}\llo K^{J+1}}{E(p)^2}\chi_{|p|<2\La}&|\partial_r^2 \gla|&\apprle \frac{\alpha^2\llo }{E(r)^2}\chi_{r<2\La}.%%% dei regarder de plus pres
\end{array}\right.
\end{equation}
%Moreover the function $\partial_r \fla$ is $a$-Hölder for any $0<a<1$ and there exist $C_0,C_1>0$ and $0<\eps<1$ such that:
%\[
%\forall\,p,q\in B(0,\La),\ |p-q|<\eps,\ |\partial_r\fla(p)-\partial_r\fla(q)|\le \frac{C_0|p-q|}{\inf(E(p)^2,E(q)^2)}+\frac{-C_1|p-q|\log |p-q|}{\La^2}.
%\]
 %In particular for such an $a$, there exists $C_{(a)}>0$ such that:
%\[
%\underset{\RR}{\dint}|x|^{1+a}|\check{f}_\La(x)|^2dx\le C_{(a)} L^2.
%\]
\end{lem}
As a consequence:
\begin{lem}\label{ll1}
For $\alpha$ sufficiently small, $\check{g}_\La\in L^1$ and
\begin{equation}
\nlp{1}{\check{g}_\La}\apprle  (\alpha\llo)^2.
\end{equation}
\end{lem}

\begin{remark}
At the very end of the proof of Lemma \ref{glaco}, we refer the reader to the thesis of the author for a (last) technical assumption: proving that $\underset{|x|\to 2\La^{-}}{\lim}\partial_r^2\fla(x)=0$.
\end{remark}

\noindent\textbf{Proof of Lemma \ref{ll1}}

We assume Lemma \ref{glaco}. As $(-\Delta_r)=-(\partial_r^{2}+\tfrac{2}{r}\partial_r)$ we have $\fla\in H^2(\mathbb{R}^3)$ with:
\begin{equation}\label{enfin}
|\Delta \fla(p)|\apprle \tfrac{L}{|p|E(p)}.
\end{equation}

\subparagraph{Proof of $\int_{B(0,1)}|\check{f}_{\La}(x)|dx\apprle L$:} The function $-\Delta\fla$ has a singularity at $r=0$ due to the term $\tfrac{2\partial \fla(r)}{r}$. We split $-\Delta \fla$ w.r.t. $\chi_{|x|\le 1}+\chi_{|x|>1}$. We have
%The $L^2$-norm remains finite since the domain is $\mathbb{R}^3$. More generally, we have
%Although $\partial_r^j\gla:\mathbb{R}_+^*\to\mathbb{R}$ has no singularity at $0$, 
\begin{equation}\label{split_delta}
I_{\La}^{(2)}:=-\Delta \fla\chi_{|x|\le 1}\in L^{p_1}\cap L^3_w,\text{\ and\ }E_{\La}^{(2)}:=-\Delta \fla\chi_{|x|> 1}\in L^{3/2}_w\cap L^{p_2},\ |p_2|>\frac{3}{2},1\le p_1<3.
%-\Delta\fla\in L^{3/2}(\mathbb{R}^3)\cap L^2(\mathbb{R}^3).
%-\Delta\fla\in L^{p}(\mathbb{R}^3),\ p\ge \frac{3}{2}.
\end{equation}
The corresponding norms are respectively $\mathcal{O}(LK(p_1))$ and $\mathcal{O}(LK(p_2))$. We use the Hausdorff-Young inequality and the generalized Young inequality \cite[Vol. II]{ReedSim}. The decomposition \eqref{split_delta} implies the decomposition $\fla=I_{\La}^{(0)}+E_{\La}^{(0)}$ by multiplication by $\tfrac{1}{-\Delta}$.

For $p=1,a=\frac{1}{2},q=2$ and $q'=\tfrac{q}{q-1}$ we have
\begin{align*}
\underset{|x|\le 1}{\dint}\big|\check{I}^{(0)}_{\La}(x)\big|dx&\le \Big(\dint |x|^{aq'}\big|\check{I}^{(0)}_{\La}(x)\big|^{q'}dx\Big)^{1-1/q}\Big(\dint_{|x|\le 1} \frac{dx}{|x|^{aq}}\Big)^{1/q},\\
                  &\apprle \nlp{q}{\,|\nabla|^{a}I_{\La}^{(0)}}\apprle \nlp{q}{\tfrac{1}{|\cdot|^{1+a}}*I_{\La}^{(2)}},\\
                  &\apprle \nlp{p}{I_{\La}^{(0)}}\lVert \frac{1}{|\cdot|^{1+a}}\rVert_{L^{\tfrac{3}{1+a}}_w}\apprle L.
\end{align*}

Similarly let $0<\eps<1$ to be chosen: we have $|\nabla|^{2-\eps}E_{\La}^{(0)}=\tfrac{K(\eps)}{|\cdot|^{3-\eps}}*E_{\La}^{(2)}$. This last function is in $L^2$ provided that $E_{\La}^{(2)}\in L^{\tfrac{6}{3-2\eps}}$. We choose $\eps=3/4$ for instance: this gives
%Let $\tfrac{3}{2}<p<2$ to be chosen and $q=\tfrac{2p}{3p-2}$. We use the generalized Young inequality: $\tfrac{1}{|\cdot|^{-3/q}}\in L^q_w(\mathbb{R}^3)$ and $\tfrac{1}{|\cdot|^{-3/q}}*(-\Delta \fla)\in L^2(\mathbb{R}^3)$. By Plancherel's Theorem this gives in Direct space the following result:
%\[
%\dfrac{|\cdot|^2\check{f}_\La}{|\cdot|^{3-\tfrac{3}{q}}}\in L^2(\mathbb{R}^3),\text{\ that\ is\ }|\cdot|^{\tfrac{3}{q}-1}\check{f}_\La\in L^2(\mathbb{R}^3).
%\]
%We choose $q=\tfrac{7}{5}$ thus:%$\tfrac{3}{p}-\tfrac{3}{2}=4^{-1}$, that is $p=\tfrac{12}{7}$, thus:
\[
\underset{B(0,1)}{\dint}|E_{\La}^{(0)}(x)|dx\le \sqrt{\dint |x|^{5/2} |E_{\La}^{(0)}(x)|^2dx \underset{B(0,1)}{\dint}\frac{dx}{|x|^{5/2}}}\apprle \nlp{4}{E_{\La}^{(2)}}\Big\lvert\Big\lvert \frac{1}{|\cdot|^{9/4}}\Big\rvert\Big\rvert_{L^{4/3}_w}\apprle L.
\]
\subparagraph{Proof of $\int_{|x|\ge 1}|\check{f}_{\La}(x)|dx\apprle L$:} Then it is clear that
\[
\underset{|x|\ge 1}{\dint}|\check{f}_\La(x)|\le \nlp{2}{-\Delta \fla}\sqrt{\underset{|x|\ge 1}{\dint}\frac{dx}{|x|^4}}\apprle L.
\]
\hfill{\footnotesize$\Box$}

\paragraph{Proof of Lemma \ref{glaco}}

The idea of the proof is that each time we differentiate with respect with the radius $r>0$, it leads to an additional term $\tfrac{1}{E(U)}$ in the integrand or a change of the domains and so a better upper bound of the integral.

We will often use the following inequality:
\begin{equation}\label{trickder}
\underset{B(0,\La)}{\dint}\dfrac{dv}{|u-v|^2(\ed{v+\tfrac{k}{2}}+\ed{v-\tfrac{k}{2}})|u+\eps\tfrac{k}{2}|}\le\dfrac{1}{|u+\eps\tfrac{k}{2}|}\dint\frac{dv}{|v|^2|v-\mathbf{e}|^2},
\end{equation}
and for convenience we write:
\begin{equation}
 u^\eps:=u+\eps\frac{k}{2},\ \eps\in\{1,-1\}.
\end{equation}
%(
That the function (and its derivatives) has an extension in $0$ is clear from \eqref{nosing} and \eqref{inegevidente}: differentiating under the integral sign of the Taylor's formula, we get:
\begin{equation}
\Big|d^{J+1}\sbf{u+tr\mathbf{e}/2}\cdot\big((t\mathbf{e}/2)^J ,\frac{\mathbf{e}}{2}\big)\Big|\le K^J \frac{1}{E(u)^{J+1}},\ 0<r,t<1,
\end{equation}

\noindent thus the problem of singularity at $r=0$ drops thanks to \eqref{inegevidente}.

More generally the variable $r$ appears 
\begin{enumerate}
\item  either in the domains $B(r)^{J+1}$,
\item or in a function of $v_j\pm r\tfrac{e}{2}$.
\end{enumerate}
One may write:
\begin{equation}\label{G_J}
\begin{array}{l}
f_{\La,J}(r)=:\underset{B(r)^{J+1}}{\dint}G_J(u_0\pm r\tfrac{\mathbf{e}}{2},\cdots,u_J\pm r\tfrac{\mathbf{e}}{2})d\mathbf{u},\\
G_J=:G^0_J(u_0\pm r\tfrac{\mathbf{e}}{2},\cdots,u_J\pm r\tfrac{\mathbf{e}}{2})\underset{1\le j\le J}{\prod}\dfrac{1}{|u_j-u_{j-1}|^2}.
\end{array}
\end{equation}
It is easy to see that $G^0_J:(\mathbb{R}^3)^{2J+2}\to\mathbb{R}$ is a differentiable function and that each time we take $\partial_{u_j+ r\tfrac{\mathbf{e}}{2}}-\partial_{u_j-r\tfrac{\mathbf{e}}{2}}$ we get a term $K(r^{-1}+\ed{u\pm\tfrac{k}{2}}^{-1})$ for $r>1$ or $K\ed{u}^{-1}$ for $r\le 1$ (see Remark \ref{sing}). This enables us to get upper bounds of the terms of $\partial_r^j f_{\La,J}$ corresponding to derivatives of $G^0_J$.
Indeed for the first derivative: for $\eps,\eps'\in\{ +,-\}$ one has for $1<|k|<2\La$:
\begin{equation}\label{der1}
\left\{
\begin{array}{rll}
\dint \dfrac{du_j}{|u_j-u_{j-1}|^2\ed{u_j+\eps k/2}^2}&\le&\dfrac{1}{|u_{j-1}+\eps k/2|}\underset{\mathbb{R}^3}{\dint}\dfrac{du_j}{|u_j|^2|u_j-\mathbf{e}|^2},\\
\dint \dfrac{du_i}{|u_i-u_{i-1}|^2|u_i+\eps k/2|\ed{u_i, k/2}\ed{u_i+\eps' k/2}}&\apprle& \dfrac{1}{|k|}\big(\frac{1}{|u_{i-1}+ k/2|}+\frac{1}{|u_i-k/2|}\big)\\
   &&\quad\times\underset{(\mathbb{R})^3}{\dint}\dfrac{du_i}{|u_i-\mathbf{e}|^2|u_i|^2}.
\end{array}
\right.
\end{equation}
For the term $(\partial_{u_0+ k/2} -\partial_{u_0-k/2})G_0$ we have:
\begin{equation}\label{der2}
\underset{B(r)}{\dint} \dfrac{du_0}{|u_0-\eps k/2|\ed{u+\eps k/2}}\big(\frac{1}{\ed{u+\eps k/2}^2}+\frac{1}{\ed{u-\eps k/2}^2}\big)\apprle \frac{2}{|k|}\underset{B(0,2\La)}{\dint}\frac{du_0}{\ed{u_0}^2|u_0|}\apprle \dfrac{\llo}{|k|}.
\end{equation}
If $r\le 1$, Remark \ref{sing} enables us to say that
\[
\underset{B(r)^{J+1}}{\dint} \dfrac{|\partial_r G^0_J(u_j\pm r\mathbf{e}/2)|}{\underset{1\le j\le J}{\prod}|u_j-u_{j-1}|^2}\apprle \alpha^{J+1}J\Big(K\dint\dfrac{du}{|u|^2|u-\mathbf{e}|^2}\Big)^J\llo. 
\]

\medskip

\noindent -- In the case of the terms corresponding to $\partial_{v_1}\partial_{v_2} G^0_J$ with $v_a=u_{i(a)}+\eps(a) \tfrac{k}{2}$, the above upper bounds enable us to say that if $i(1)\neq i(2)$ then it suffices to apply twice \eqref{der1},\eqref{der2} and we get an upper bound of the form:
\[
K J^2\big(\chi_{|k|\le 1}+\frac{\chi_{1<|k|<2\La}}{|k|^2}\big) \alpha^{J+1}\Big(K\dint\dfrac{du}{|u|^2|u-\mathbf{e}|^2}\Big)^J\llo,
\]
If $i(1)=i(2)$, then as:
\begin{equation}\label{derr1}
\dint\dfrac{du}{|u-v|^2|u|\ed{u,\tfrac{k}{2}}}\Big(\frac{1}{\ed{u+k/2}^2}+\frac{1}{\ed{u-k/2}^2}\Big)\apprle \dfrac{1}{|k|^2|u|},
\end{equation}
we obtain  an upper bound of the form
\[
K J(\chi_{|k|\le 1}+\frac{\chi_{1<|k|<2\La}}{|k|^2}) \alpha^{J+1}\Big(K\dint\dfrac{du}{|u|^2|u-\mathbf{e}|^2}\Big)^J\llo.
\]
If $i(1)=i(2)=0$, we integrate first over $u_0$, then over $u_1,u_2,\cdots, u_J$ and use \eqref{derr1} with $u=u_0$, $v=u_1$: this gives
\[
\text{for\ }1<r<2\La,\ \Big| \partial_r^2 \dfrac{1-\sbf{u_0+\tfrac{k}{2}} \sbf{u_0-\tfrac{k}{2}}}{r(\ed{u_0+\tfrac{k}{2}}+\ed{u_0-\tfrac{k}{2}})}\Big|\apprle \frac{r^{-2}+\ed{u_0+\tfrac{k}{2}}^{-2}+\ed{u_0-\tfrac{k}{2}}^{-2}}{|u|\ed{u,\tfrac{k}{2}}}.
\]
If $r\le 1$ we use Remark \ref{sing} as before.

\medskip

\noindent  -- There remains to deal with the terms corresponding to differentiation over $r$ in the domain $B(r)^{J+1}$. We rewrite \eqref{G_J} using the \emph{co-area formula}. Indeed, let us write for $\eps\in\{1,-1\}$ and $r\in[0,2\La]$:
\[
B_\eps(r):=\{ p,\ |p+\tfrac{\eps r}{2}\mathbf{e}|< \La,\ \psh{p}{\eps \mathbf{e}}>0\}\text{\ and\ }B(r):=B_1(r)\cup B_{-1}(r)\subset B(0,\La).
\]
In particular $B(\La)=\{p\in B(0,\La),\,\psh{p}{\mathbf{e}}\neq 0\}$. We define the level function:
\[
z:\begin{array}{rll}
B(\La)&\to&[0,2\La]\\
p\in B_\eps(\La)&\mapsto&r\text{\ such\ that\ }\Big|u+\dfrac{r\eps\mathbf{e}}{2} \Big|=\La.
\end{array}
\]

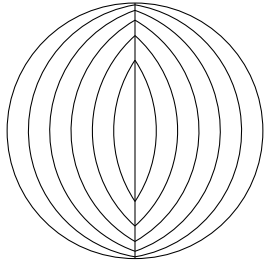
\begin{figure}%[!h]
  \begin{center}
	\begin{tikzpicture}[scale=1.7]
	    %cercle
	    \draw (0,0) circle (1) ;
	    %trait
	    \draw (0,1)--(0,-1) ;
	    %arcs gauches
	    \draw (0,0.986) arc (99.594:260.406:1) ;            % 1/6
	    \draw (0,0.943) arc (109.471:250.529:1) ;		% 2/6
	    \draw (0,0.866) arc (120:240:1) ;			%3/6
	    \draw (0,0.745) arc (131.81:228.19:1) ;		%4/6
	    \draw (0,0.553) arc (146.443:213.557:1) ;		%5/6
	    %arcs droits
	    \draw (0,-0.986) arc (-80.406:80.406:1) ;            % 1/6
	    \draw (0,-0.943) arc (-70.529:70.529:1) ;		% 2/6
	    \draw (0,-0.866) arc (-60:60:1) ;			%3/6
	    \draw (0,-0.745) arc (-48.19:48.19:1) ;		%4/6
	    \draw (0,-0.553) arc (-33.557:33.557:1) ;		%5/6	    
	    
	\end{tikzpicture}

  \end{center}

 \caption{Level sets of the $z$ function}
\end{figure}

We apply the co-area formula with respect to $z$. If $p\in B_{\eps_0}$, we write $\eps(p):=\eps_0$ and
\[
n(p):=\dfrac{p+\eps(p)z(p)\tfrac{\mathbf{e}}{2}}{\big|p+\eps(p)z(p)\tfrac{\mathbf{e}}{2} \big|}=\La^{-1}(p+\eps(p)z(p)\tfrac{\mathbf{e}}{2}).
\]
For $0\le r<2\La$ we write $S(r):=\{p\in B,\,z(p)=r\} $ and $S_\eps(r):=S\cap B_\eps$; each $S_\eps(r)$ is a spherical cap of $S(-\tfrac{r\eps\mathbf{e}}{2},\La)$. The measure of $B(0,\La)\backslash B(\La)$ is zero and the function $z$ is differentiable with 
\[
\nabla z(p)=\frac{-2\eps(p)}{\psh{n(p)}{\mathbf{e}}}n(p).
\]
Thus for any integrable function $F: B(0,\La)\to \mathbb{R}$ and $0\le r<2\La$ we have:
\begin{equation}\label{cor}
\underset{B(r)}{\dint}F(p)\big|\nabla z(p) \big|dp=\dint_{t=r}^{2\La}dt\underset{S(t)}{\dint} F(p)d\mathcal{H}_2(p),
\end{equation}
where $d\mathcal{H}_2(p)$ is the Hausdorff measure on $S(r)$. If we take spherical coordinates with axis $\mathbb{R}\mathbf{e}$ in $S_\eps(r)$ there holds $d\mathcal{H}_2(p)=\La^2\sin(\theta)d\theta d\phi$ in the domain:
\[
M_{\pm}(r)=\{(\theta,\phi)\in(\tfrac{\pi}{2},\tfrac{\pi}{2}\mp \tfrac{\pi}{2})\times (-\pi,\pi),\ \cos(\theta)\ge \frac{r}{2\La}\}.
\]
\begin{notation}
For convenience we write $du$ for both $d\mathcal{H}_2(u)$ (integration over a spherical cap) or $d\mathcal{H}_1(u)$ (integration over a curve).
\end{notation}
%We write $M=M_1(r)\cup M_{-1}(r)$.
%\begin{notation}
%For convenience we write $B_\eps(r):=C(r\mathbf{e})\cap B_\eps$.
%\end{notation}

\noindent-- For each $u_j$ we rewrite the integration over $u_i\in B(r)$.  For each $0\le j\le J$ we need to estimate
%This enables us to get the full derivative $\partial_r f_{\La,J}$.
\[
\underset{B(r)^{j-1}\times S(r)\times B(r)^{J-j}}{\dint}\dfrac{du_0\cdots \wh{du_j}\cdots du_J d\mathcal{H}_2(u_j)}{\underset{1\le j\le J}{\prod}|u_j-u_{j-1}|^2}|G^0_J(u_i\pm \tfrac{k}{2})|.
\]
In $S_\eps(r)$ we take spherical coordinates and write $v=u_{j-1}+\tfrac{\eps r}{2}\mathbf{e}$, if $j=0$ we replace $u_{j-1}$ by $u_2$ and integrate over $u_1,u_2,\cdots u_J$. Using \eqref{der1} we have:
\[
\begin{array}{rl}
\underset{M_\eps(r)}{\dint}\dfrac{\La^2\sin(\theta)d\theta d\phi}{|v-\La n|^2}\dfrac{1}{|\La n|(\ed{\La n}+\ed{\La n-k})}&\le \underset{\mathbb{S}^2}{\dint}\dfrac{\sin(\theta)d\theta d\phi}{|v-\La n|^2}\\
&\le \frac{2\pi}{\La |v|} \log\Big(\frac{\La+|v|}{\La-|v|}\Big).
\end{array}
\]
Then writing $v:=u_{i-1}+\eps \tfrac{k}{2}$ we have:
\[
\begin{array}{l}
\underset{B(r)}{\dint}\dfrac{du_{i}}{|u_i-u_{i-1}|^2 |u_i+\eps \tfrac{k}{2}|}\log\Big(\frac{\La+|u_i+\eps \tfrac{k}{2}|}{\La-|u_i+\eps \tfrac{k}{2}|}\Big)\dfrac{1}{\ed{u_i,k/2}}=\underset{B(r)+\tfrac{\eps k}{2}}{\dint}\dfrac{du_i}{|u_i-v|^2|u_i|\ed{u_i}}\log\Big(\frac{\La+|u_i|}{\La-|u_i|}\Big)\\
\ \ \ \le\underset{B(0,1)}{\dint}\dfrac{du}{|u-v\La^{-1}|^2|u|\ed{\La u}}\log\Big(\dfrac{1+|u|}{1-|u|} \Big)\le2\pi\dint_{r=0}^1 \dfrac{\La dr}{|v|\ed{\La r}} \log\Big(\dfrac{1+r}{1-r} \Big)\log\Big|\dfrac{\La^{-1}|v|+r}{\La^{-1}|v|-r} \Big|\\
\ \ \ \le 2\pi \dint_0^1 \dfrac{\La dr}{|v| \ed{\La r}}\bigg( \log^2\Big(\dfrac{1+r}{1-r} \Big)+\log^2\Big|\dfrac{\La^{-1}|v|+r}{\La^{-1}|v|-r} \Big|\bigg)\apprle  |v|^{-1}.
\end{array}
\]
Finally for sufficiently small $\alpha$, we have
\[
|\partial_r f_{\La,J}(r)|\le KL(\alpha K)^J\big(\chi_{r\le 1}+\tfrac{\chi_{1<r<2\La}}{r}\big).
\]
So by dominated convergence, as $r$ tends to $(2\La)^-$, $\partial_r f_{\La,J}$ tends to $0$ and $g_\La\in H^1(\mathbb{R}^3)$.

%\medskip

%\noindent -- The fact that $\partial_r \fla(0)=0$ follows from the symmetry in $\pm\tfrac{r}{2}\mathbf{e}$.
%\begin{enumerate}
%	\item The term with $\partial_r (\ed{u_j+\tfrac{r}{2}\mathbf{e}}+\ed{u_j-\tfrac{r}{2}\mathbf{e}})^{-1}_{\big|_{r=0}}$ vanishes.
%	\item The terms with $\partial_r (\sbf{u_j+\tfrac{r}{2}\mathbf{e}})_{\big|_{r=0}}$ and $\partial_r (\sbf{u_{J-j}-\tfrac{r}{2}\mathbf{e}})_{\big|_{r=0}}$ cancel each other out. To see this we have to change variables $u'_j:=u_{J-j}$.
%	\item The term with the integration over $B(r=0)^{J-j}\times S(r=0)\times B(r=0)^{j-1}$ vanishes. 
	%\item The terms with the integration over $B(r=0)^{J-j}\times S(r=0)\times B(r=0)^{j-1}$ and the one over $B(r=0)^{j-1}\times S(r=0)\times B(r=0)^{J-j}$ cancel each other out. To see this, we have to change variables: $u_j':=-u_{J-j}$.
%\end{enumerate}

\medskip

\noindent-- Let us deal with the second derivative. There remains the three cases:
\begin{enumerate}
\item One derivative in $B(r)$ and one derivative in the integrand.
\item Two derivatives in two different $B(r)$.
\item Two derivatives in the same $B(r)$.
\end{enumerate}

In fact, we have to deal with the last two cases together because each term alone is not well defined but the sum gives a finite term. Seeing the second derivative as the coefficient of the second term in the Taylor series of $g_{\La,J}(r+\delta r)$, each term is $\underset{\delta r\to 0}{\mathcal{O}}(-\delta r \log(\delta r))$ but the sum is $\mathcal{O}(\delta r)$ due to some cancellation.

\noindent 1.

\noindent 1.1. One derivative in $u_{i_1}\pm\tfrac{r}{2}\mathbf{e}$ and one in the domain of $u_{i_2}$ with $i_1\neq i_2$. Up to integrating over $u_j$ from $j=0$ to $j=J$, we can suppose that $i_1<i_2$. We split $S(r)$ between $S_+(r)$ and $S_-(r)$. In $S_\eps(r)$, we use \eqref{trickder} (and \eqref{trick1} at the beginning), this gives:
\[
\begin{array}{l}
 \underset{S_\eps(r)}{\dint}\dfrac{du_{i_2}}{|u_{i_2-1}-u_{i_2}|^2\ed{u_{i_2}+\eps \tfrac{k}{2}}|u_{i_2}+\eps\tfrac{k}{2}|}\le \underset{\mathbb{S}^2}{\dint}\dfrac{du_{i_2}}{\La^2|\tfrac{|u_{i_2-1}^\eps|}{\La}-u_{i_2}|^2}\\
 \ \ \ \apprle \dfrac{1}{\La|u_{i_2-1}^\eps|}\log\Big(\dfrac{1+\tfrac{|u_{i_2-1}^\eps|}{\La}}{\La-|u_{i_2-1}^\eps|}\Big).
\end{array}
 \]
We take spherical coordinates with respect to $-\eps\tfrac{k}{2}$: for any $v\in B=B_+\cup B_-$ we have
\[
 \begin{array}{rll}
  \underset{B(0,\La)}{\dint}\dfrac{du}{|u_{i_2-1}-v^\eps|^2|u_{i_2-1}|\ed{u_{i_2-1}}}\log\Big(\dfrac{1+\tfrac{|u_{i_2-1}|}{\La}}{1-\tfrac{|u_{i_2-1}|}{\La}}\Big)&\apprle& \dfrac{1}{|v|}\dint_0^1dz\log\Big(\dfrac{1+z}{1-z}\Big)\log\Big(\dfrac{\tfrac{|v^\eps|}{\La}+z}{\tfrac{|v^\eps|}{\La}-z}\Big)\\
     &\apprle & \dfrac{1}{|v^\eps|}\dint_0^2\log\Big(\dfrac{1+z}{1-z}\Big)^2,\\
  \underset{B(0,\La)}{\dint}\dfrac{du}{|u_{i_2-1}|\ed{u_{i_2-1}^2}}\log\Big(\dfrac{1+\tfrac{|u_{i_2-1}|}{\La}}{1-\tfrac{|u_{i_2-1}|}{\La}}\Big)&\apprle&\dint_0^\La\dfrac{dz}{E(z)}\log\Big(\dfrac{1+\tfrac{z|}{\La}}{1-\tfrac{z}{\La}}\Big)\\
  &\apprle& 1+\La^{-1}.
 \end{array}
\]
Then we use the same method as for the first derivative: when integrating over $u_{i_1}$, we use $(\ed{u_{i_1}+\tfrac{k}{2}}+\ed{u_{i_1}-\tfrac{k}{2}})^{-1}\le \ed{\tfrac{k}{2}}^{-1}$. In this first subcase, we get an upper bound of the form:
\[
 \dfrac{J^2(K\alpha)^{J+1}\llo}{\La E(k)}.
\]

\noindent 1.2. One derivative in $u_i\pm \tfrac{r}{2}\mathbf{e}$ and one in the domain of $u_i$. Splitting the integration over $S_+(r)$ and $S_-(r)$, and using \eqref{trickder}, we have to estimate
\begin{equation}\label{d2domint}
 \underset{S_\eps(r)}{\dint}\dfrac{du_i}{|u_i-v|^2|u_i+\eps\tfrac{k}{2}|\ed{u_i,k/2}\ed{u_i+\eps'\tfrac{k}{2}}}\le  \underset{S_\eps(r)}{\dint}\dfrac{du_i}{|u_i-v|^2|u_i+\eps\tfrac{k}{2}|\ed{u_i,k/2}\ed{u_i-\eps\tfrac{k}{2}}}.
\end{equation}
Above $v$ is either $u_{i+1}$ or $u_{i-1}$ depending on the order of integration (from $u_{J}$ to $u_0$ or from $u_0$ to $u_{J}$ if the derivatives act on $u_0+\tfrac{k}{2}$). Moreover $\eps,\eps'\in\{1,-1\}$ and the term with $\eps'$ comes from the derivative in the integrand. By using \eqref{trickder} several times (starting with \eqref{trick1}) we get the term $|u_i+\eps\tfrac{k}{2}|=|u_i^\eps|$ in \eqref{d2domint}.

In \eqref{d2domint}, we use spherical coordinates to obtain the following upper bound:
\begin{equation}\label{tech_int_0}
 \underset{\mathbb{S}^2}{\dint}\dfrac{\La^2du}{\La^2}\dfrac{1}{|\La u-v^\eps|^2\ed{\La u-r\mathbf{e}}}\le 2 \underset{\mathbb{S}^2}{\dint}\frac{du}{\big|\La u- |v^\eps|\mathbf{e}\big|^2E(u-r\mathbf{e})}.
\end{equation}

Let us assume for the moment that this integral is lesser than:
\begin{equation}\label{tech_int}
 \dfrac{K}{\La^2|v^\eps|}\Big(1-\chi_{|v^\eps|>(2-\sqrt{3})\La} \log\big(1-\frac{|v^\eps|}{\La}\big)\Big).
\end{equation}
In the process of integrating over the $u_i$'s, we have to integrate over $v$ with this upper bound. Taking spherical coordinates with respect to $-\tfrac{\eps r}{2}\mathbf{e}$, we have: 
\[
\left\{
\begin{array}{l}
 \underset{B(0,\La)}{\dint}\frac{dv}{|v'-v|^2\ed{v}|v|}\apprle \dfrac{1}{|v'|}\dint\dfrac{dv}{|v|^2|v-\mathbf{e}|^2}\\
 \underset{B(0,\La)}{\dint}\frac{dv}{|v|\ed{v}^2}\apprle \llo.
 \end{array}
 \right.
\]
Moreover, writing $A_\La:=A(0,(2-\sqrt{3})\La,\La)$ the annulus, we have:
\[
\left\{
\begin{array}{rll}
 \underset{A_\La}{\dint}\frac{\log(1-\tfrac{|v|}{\La})dv}{|v'-v|^2|v|\ed{v}(\La^2+|v|^2)}&\apprle&\dfrac{1}{\La^2|v'|}\dint_{2-\sqrt{3}}^1\dfrac{-\log(1-z)dz}{z(1+z^2)}\log\Big|\frac{\tfrac{|v'|}{\La}+z}{\tfrac{|v'|}{\La}-z}\Big|\\
      &\apprle&\dfrac{1}{\La^2|v'|},\\
  \underset{A_\La}{\dint}\frac{\log(1-\tfrac{|v|}{\La})dv}{|v|\ed{v}^2(\La^2+|v|^2)}&\apprle&\dfrac{1}{\La^2}\dint_{2-\sqrt{3}}^1\dfrac{-\log(1-z)dz}{z^3}.
 \end{array}
 \right.
\]

\noindent\textit{Proof of \eqref{tech_int_0}$\le$\eqref{tech_int}} We write 
\[
x:=|v^\eps|,\ A:=\La^2+x^2,\ B:=\sqrt{1+\La^2+r^2},\ a:=\dfrac{2x\La}{x^2+\La^2}\text{\ and\ }b:=\dfrac{2\La r}{1+\La^2+r^2}.
\]
The upper bound \eqref{tech_int_0} is equal to
\begin{equation}\label{2nd_form}
\dfrac{4\pi}{AB}\dint_{-1}^1\frac{dy}{(1-ay)\sqrt{1-by}}=\dfrac{4\pi}{ABa}\underset{0}{\overset{\tfrac{2b}{\sqrt{1+b}+\sqrt{1-b}}}{\dint}}\dfrac{dz}{z^2+2z\sqrt{1-b}+b(\tfrac{1}{a}-1)}.
\end{equation}
If $a\le \tfrac{1}{2}$, then this integral is $\mathcal{O}\left( \dfrac{1}{AB}\dint_{-1}^1\dfrac{dy}{\sqrt{1-by}}\right)=\mathcal{O}\left( \dfrac{1}{\La^2E(v^\eps)}\right).$
%\[
%\mathcal{O}\left( \dfrac{1}{AB}\dint_{-1}^1\dfrac{dy}{\sqrt{1-by}}\right)=\mathcal{O}\left( \dfrac{1}{\La^2E(v^\eps)}\right).
%\]

\medskip

 Similarly, if $b\le\tfrac{1}{2}$, we get: $\mathcal{O}\left(\dfrac{1}{AB}\dint_{-1}^1\dfrac{dy}{1-ay} \right)=\mathcal{O}\left( \dfrac{1}{\La^2E(v^\eps)}\right).$
%\[
% \mathcal{O}\left(\dfrac{1}{AB}\dint_{-1}^1\dfrac{dy}{1-ay} \right)=\mathcal{O}\left( \dfrac{1}{\La^2E(v^\eps)}\right).
%\]

\noindent If $\tfrac{1}{2}<a,b\le 1$, we consider formula \eqref{2nd_form}.

\noindent We have $z^2\ge 2z\sqrt{1-b}$ for $z\ge 2\sqrt{1-b}$ and $2\sqrt{1-b}< \frac{2b}{\sqrt{1+b}+\sqrt{1-b}}$ \emph{iff} $b> \tfrac{4}{5}$. 

\medskip

For $\tfrac{1}{2}<b\le \tfrac{4}{5},a> \tfrac{1}{2}$ we get the upper bound:
\[
\begin{array}{l}
\dfrac{20\pi}{AB}\dint_{-1}^1\dfrac{dy}{1-ay}\apprle \dfrac{1}{\La^2 |v^\eps|}.
\end{array}
\]
For $b> \tfrac{4}{5},a> \tfrac{1}{2}$, we have the upper bound
\begin{equation}\label{int_chiant}
\dfrac{4\pi}{AaB}\left(\underset{0}{\overset{2\sqrt{1-b}}{\dint}}\dfrac{dz}{2z\sqrt{1-b}+b(\tfrac{1}{a}-1)}+\underset{2\sqrt{1-b}}{\overset{\tfrac{2b}{\sqrt{1+b}+\sqrt{1-b}}}{\dint}}\dfrac{dz}{z^2+b(\tfrac{1}{a}-1)}.\right)
\end{equation}
The first integral of \eqref{int_chiant} gives (without $4\pi/(AB)$)
\[
 \dfrac{1}{2a\sqrt{1-b}}\log\left(1+\dfrac{4(1-b)}{b(\tfrac{1}{a}-1)}\right)\le\dfrac{1}{\sqrt{1-b}}\log\left(1+5\dfrac{1-b}{1-a}\right).
\]
If $1-b\le 1-a$, then this gives $\mathcal{O}((1-b)^{-1/2})$, else this gives $\mathcal{O}(\tfrac{\log(1-a)}{\sqrt{1-b}})$.

The second integral of \eqref{int_chiant} gives (without $4\pi/(AaB)$ and writing $X:=(a^{-1}-1)^{-1}$):
\[
\begin{array}{rl}
 \sqrt{\dfrac{X}{b}}\dint_{2\sqrt{\tfrac{1-b}{b}}X}^{\tfrac{2\sqrt{bX}}{\sqrt{1-b}+\sqrt{1+b}}}\dfrac{dz}{z^2+1}&\apprle\dint_{2\sqrt{1-b}}^{2}\dfrac{Xz^2}{1+Xz^2}\dfrac{dz}{z^2}\\
                       &\apprle \dfrac{1}{\sqrt{1-b}}=\dfrac{\sqrt{1+\La^2+r^2}}{\sqrt{1+(\La-r)^2}}.
 \end{array}
\]
We have: $\dfrac{\log(1-a)}{AB\sqrt{1-b}}=2\dfrac{\log\big|\frac{\sqrt{\La^2+x^2}}{\La-x} \big|}{(\La^2+x^2)\sqrt{1+(\La-x)^2}}\apprle\dfrac{1+\log(1-\tfrac{|v^\eps|}{\La})}{(\La^2+|v^\eps|^2)\sqrt{1+(\La-|v^\eps|)^2}}.$
%\[
% \dfrac{\log(1-a)}{AB\sqrt{1-b}}=2\dfrac{\log\big|\frac{\sqrt{\La^2+x^2}}{\La-x} \big|}{(\La^2+x^2)\sqrt{1+(\La-x)^2}}\apprle\dfrac{1+\log(1-\tfrac{|v^\eps|}{\La})}{(\La^2+|v^\eps|^2)\sqrt{1+(\La-|v^\eps|)^2}}.
%\]

Let us emphasize that the condition $a>2^{-1}$ is equivalent to $\frac{|v^\eps|}{\La}\ge 2-\sqrt{3}.$

Bringing all those estimates together ends the proof of \eqref{tech_int_0}$\le$\eqref{tech_int}.
%\[
% \frac{|v^\eps|}{\La}\ge 2-\sqrt{3}.
%\]

\noindent 2.

\noindent 2.1. One derivative in the domain of $u_j$ and one in the domain of $u_{i}$ with $i-j\ge 2$. We integrate over $u_{j'}$ from $j'=0$ to $j'=j$ and from $j'=J$ to $j'=i$ using the method for the first derivative. The integration over $u$ with $u$ either $u_j$ or $u_i$ (resp. with $v$ either $u_{j+1}$ or $u_{i-1}$) gives:
\begin{equation}
\begin{array}{rl}
\ssum_{\eps\in\{1,-1\}}\underset{S_\eps(r)}{\dint}\frac{du}{|u-v|^2}\frac{1}{|u+\eps\dfrac{k}{2}|\ed{u+\eps \dfrac{k}{2}}}\apprle \dfrac{1}{\La^2}& \ssum_\eps\dint_{\tfrac{r}{2\La}}^{1}\frac{dy}{\La^2+|v^\eps|^2-2\La |v^\eps|y}\\
 &\apprle \ssum_\eps \dfrac{1}{\La|v^\eps|}\log\left(\dfrac{\La+|v^\eps|}{\La-|v^\eps|}\right).
\end{array}
\end{equation}
If $j+2\le i$, then by integrating over $u_{j+1}$ we have:
\[
\begin{array}{l}
\underset{B(r)}{\dint}\dfrac{du_{j+1}}{|u_{j+1}-u_{j+2}|^2}\dfrac{1}{|u_{j+1}^\eps|(\ed{u_{j+1}^+}+\ed{u_{j+1}^-})}\log\left(\dfrac{\La+|u_{j+1}^\eps|}{\La-|u_{j+1}^\eps|}\right)\\
 \ \ \ \ \apprle \dfrac{1}{\La}\underset{B(0,1)}{\dint}\frac{du}{|u|^2|u-\La^{-1}u_{j+2}^\eps|^2}\log\left(\dfrac{1+|u|}{1-|u|}\right)\\
     \ \ \ \ \apprle \dint_0^1\frac{dr}{|u_{j+2}^\eps|}\log\left(\dfrac{1+r}{1-r}\right)\log\left(\dfrac{r+\tfrac{|u_{j+2}^\eps|}{\La}}{r-\tfrac{|v^\eps|}{\La}}\right)\apprle \dfrac{1}{|u_{j+2}^\eps|},
\end{array}
\]
and we conclude as before. Else $j+1=i$ and we have:
\[
\begin{array}{rl}
\underset{B(r)}{\dint}\dfrac{du_{j+1}}{|u_{j+1}^\eps|^2}\dfrac{1}{\ed{u_{j+1}^+}+\ed{u_{j+1}^-}}\log\left(\dfrac{\La+|u_{j+1}^\eps|}{\La-|u_{j+1}^\eps|}\right)^2&\apprle \dfrac{1}{\La}\dint_{z=0}^\La\frac{dz}{\ed{z}}\log\left(\dfrac{1+\tfrac{z}{\La}}{1-\tfrac{z}{\La}}\right)^2(1-\tfrac{r}{2\La})\\
&\apprle (1-\tfrac{r}{2\La})(\llo+1).
\end{array}
\]

\noindent 2.2. One derivative in the domain of $u_j$ and one in the domain of $u_{j+1}$. We only look at the corresponding coefficient in the Taylor series of $g_{\La,J}(r+\delta r)$ with $r'=r+\delta r$. Indeed, let us treat for instance
\[
\begin{array}{l}
\underset{(u_{j},u_{j+1})\in B(r')\times S(r)}{\diint}\frac{du_jdu_{j+1}}{|u_j-u_{j+1}|^2}\dfrac{|\psh{n(u_{j+1})}{\mathbf{e}}|}{2}\underset{B(r)^{J-1}}{\dint}\dfrac{G_J^0(u_\ell\pm\tfrac{k}{2})}{\prod_{a\neq j+1} |u_a-u_{a+1}|^2}\\
\ \ \ \ \ \ =:\underset{(u_{j},u_{j+1})\in B(r')\times S(r)}{\diint}\frac{du_jdu_{j+1}}{|u_j-u_{j+1}|^2}\dfrac{|\psh{n(u_{j+1})}{\mathbf{e}}|}{2} G_{J,j}(u_j,u_{j+1}).
\end{array}
\]
We substract the integral of the same function but over $(u_j,u_{j+1}, \mathbf{u}')$ in

\noindent $ B(r)\times S(r)\times B(r)^{J-1}$ where $\mathbf{u}'=(u_0,\cdots, \wh{u}_j,\wh{u}_{j+1},\cdots)$ and use the co-area formula. This gives
\begin{equation}\label{cancel1}
\dint_{r+\delta r}^{r} dt \underset{S(t)}{\dint}\underset{S(r)}{\dint}\dfrac{du_j du_{j+1}}{|u_j -u_{j+1}|^2}\dfrac{|\psh{n(u_{j})}{\mathbf{e}}|}{2} \dfrac{|\psh{n(u_{j+1})}{\mathbf{e}}|}{2} G_{J,j}(u_j,u_{j+1}).
\end{equation}
We deal with $G_{J,j}(u_j,u_{j+1})$ as in the case 2.1. Let us say for instance $0<\delta r\ll 1$, then for any $(u_{j+1},t)\in S(r)\times (r,r')$  we have:
\[
\text{dist}(u_{j+1},S(t))\ge \La\big| \sqrt{1+\tfrac{\psh{n_u}{\mathbf{e}}}{\La}\delta r+(\tfrac{(t-r)}{2})^2}-1\big|=\underset{\delta r\to 0}{\mathcal{O}}(\La |(t-r)\psh{n_u}{\mathbf{e}}|).
\]
By the Theorem of projection onto a closed convex $\RR$, we have 
\[
|u_{j+1}-u_j|^2\ge |u_{j+1}-\Pi_{S(t)}u_{j+1}|^2+|\Pi_{S(t)}u_{j+1}-u_j|^2.
\] 
If $r'<r$, then we consider instead the projection of $u_j\in S(r)$ onto $B(t)$. Anyway the quantity in \eqref{cancel1} is
\[
\begin{array}{l}
\underset{\delta r\to 0}{\mathcal{O}}\left(\dfrac{(\alpha K)^J}{\La^2}\dint_{r}^{r+\delta r} dt \underset{\mathbb{S}^2}{\dint}da |\psh{a}{\mathbf{e}}|\log\Big(1+\dfrac{1}{|t-r|^2|\psh{a}{\mathbf{e}}|^2}\Big)=\dfrac{(\alpha K)^J}{\La^2}\delta r(1-\log(\delta r))\right).
 \end{array}
\]
The corresponding term is not Lipschitz because of the term $-\delta_r \log(\delta_r)$.

\medskip

\noindent 3. Let us write the expansion of
\begin{equation}\label{cancel21}
\underset{B(r')}{\dint}\dfrac{|\psh{n_{u_j}}{\mathbf{e}}|du_j}{2}\underset{B(r)^J}{\dint}du_0\cdots \wh{du_j}\cdots du_J G_J(u_\ell\pm\tfrac{k}{2})=:\underset{B(r')}{\dint} \widetilde{G}_{J,j}(u_j)du_j.
\end{equation}
We substract $\underset{B(r)}{\dint} \widetilde{G}_{J,j}(u_j)du_j$ in \eqref{cancel21} and get
\begin{equation}\label{cancel22}
\dint_{r+\delta r}^r dt\underset{S(t)}{\dint} du_j \widetilde{G}_{J,j}(u_j).
\end{equation}%[
We split \eqref{cancel22} between integration over $S_+(t)$ and $S_-(t)$. For any $t\in (r,r']$, we write $s:=t-r$ and:%)
\begin{equation}\label{phit}
\Phi_t :\begin{array}{rll}S(t)&\to & S(r)\\ u\in S_\eps(t) & \mapsto & v(u):=u+z_t(u) n_u\in S_\eps(r)\text{\ where\ }|z(u)|=\underset{\delta r\to 0}{\mathcal{O}}(\delta r)\end{array}.
\end{equation}
From now on we assume $v\in S(r)$ and $u\in S(t)$ and write $\ov{n}_u$ instead of $n_u$ to emphasize this is a function of $u\in S(t)$ and not of $v\in S(r)$. The function $z_t:S(t)\to \mathbb{R}$ satisfies the equation
\begin{equation}\label{z21}
\Big|u+z_t(u)\ov{n}_u+\eps \frac{r}{2}\mathbf{e}\Big|^2=\La^2\text{\ that\ is\ }z_t\big(1+\frac{z_t}{2\La}-\tfrac{\eps s\psh{\ov{n}_u}{\mathbf{e}}}{2\La}\big)=\frac{\eps s\psh{\ov{n}_u}{\mathbf{e}}}{2}-\frac{s^2}{8\La}.
\end{equation}
Changing variables in the integration over $S(t)$ we have:
\[
\underset{S(t)}{\dint} du_j \widetilde{G}_{J,j}(u_j)=\underset{\Phi_t(S(t))}{\dint} dv \widetilde{G}_{J,j}(\Phi_t^{-1}(v))\text{J}(\Phi_t; \Phi_t^{-1}(v))^{-1}dv.
\]

\noindent -- We need to compute $\Phi_t^{-1}(v)$ and $\text{J}(\Phi_t; \Phi_t^{-1}(v))$. First we have:
\[
\ov{n}_u=\dfrac{v-z_t\ov{n}_u+\eps \tfrac{r+s}{2}\mathbf{e}}{\La}=n_v+\eps\frac{s}{2\La}\mathbf{e}-\frac{z_t }{\La}\ov{n}_u,
\]
thus
\begin{equation}\label{nunv}
\ov{n}_u=\dfrac{1}{1+\tfrac{z_t}{\La}}(n_v+\eps\tfrac{s}{2\La}\mathbf{e}),
\end{equation}
and
\begin{equation}\label{nvnu}
n_v=\big(1+\dfrac{z_t}{\La}\big)\ov{n}_u-\dfrac{\eps s}{2\La}\mathbf{e}.
\end{equation}
Using the formula \eqref{nunv} in \eqref{z21}, we obtain the following equation satisfied by $z_t$:
\begin{equation}\label{z22}
z_t\Big(1+\dfrac{z_t}{\La}-\dfrac{\eps s}{2\La(1+\tfrac{z_t}{\La})}\big( \psh{n_v}{\mathbf{e}}+\frac{\eps s}{2\La}\big)\Big)=\frac{\eps s}{2(1+\tfrac{z}{\La})}\Big( \psh{n_v}{\mathbf{e}}+\frac{\eps s}{2\La}\Big)-\dfrac{s^2}{8\La}.
\end{equation}
In particular there holds:
\begin{equation}\label{ztdr}
z_t(u)=\dfrac{\eps s}{2\La}\psh{n_v}{\mathbf{e}}+\underset{\delta r\to 0}{\mathcal{O}}((\delta r)^2).
\end{equation}
We differentiate $z_t$ in \eqref{z21} and get:
\begin{equation}\label{dz21}
\text{d}z_t(u):\begin{array}{rll} \text{T}_uS_\eps (t)&\to & \mathbb{R}\\
   h&\mapsto & \dfrac{\eps s}{2\La}\dfrac{\psh{h}{\mathbf{e}}\big(1+\frac{z_t}{\La}\big)}{1+\tfrac{z_t}{\La}-\tfrac{\eps s}{2\La}\psh{\ov{n}_u}{\mathbf{e}}}.
\end{array}
\end{equation}
Thus differentiating in \eqref{phit} and using \eqref{nunv} in \eqref{dz21} give
\begin{equation}
\text{d}\Phi_t(u):\begin{array}{rll} 
    \text{T}_uS_\eps (t)&\to          & \text{T}_vS_\eps (r)\\
   h                                &\mapsto &\big(1+\frac{z_t}{\La}\big)h+ \dfrac{\eps s}{2\La}\dfrac{\psh{h}{\mathbf{e}}\big(1+\frac{z_t}{\La}\big)}{1+\tfrac{z_t}{\La}-\tfrac{\eps s}{2\La(1+\tfrac{z_t}{\La})}(\psh{n_v}{\mathbf{e}} +\tfrac{\eps s}{2\La})}\dfrac{n_v+\frac{s}{2\La}\mathbf{e}}{1+\frac{z}{\La}}   
\end{array}
\end{equation}
Let $(a,b)$ be an orthonormal basis of $\text{T}_uS_\eps (t)$ with $b\times \ov{n}_u=a$, then we have:
\[
\begin{array}{rll}
\text{J}(\Phi_t;u)&=&\psh{\text{d}\Phi_t(u)a\times \text{d}\Phi_t(u) b}{n_v}\\
    &=&\psh{\big(\big[1+\frac{z_t}{\La}\big] a+\ov{n}_u \text{d} z_t(u)a \big)\times \big(\big[1+\frac{z_t}{\La}\big] b+\ov{n}_u \text{d} z_t(u)b \big) \times}{n_v}\\
    &=&\big(1+\frac{z_t}{\La}\big)^2\psh{\ov{n}_u}{n_v}-\big(1+\frac{z_t}{\La}\big)\big[\psh{a}{n_v} \text{d} z_t(u)a+\psh{b}{n_v} \text{d} z_t(u)b\big]\\
    &=&\big(1+\frac{z_t}{\La}\big)(1+\tfrac{\eps s}{2\La}\psh{n_v}{\mathbf{e}})+\dfrac{\eps s}{2\La}\big(1+\frac{z_t}{\La}\big)\big( \psh{a}{\mathbf{e}} \text{d} z_t(u)a+\psh{b}{\mathbf{e}} \text{d} z_t(u)b\big)\\
    &=&\big(1+\frac{z_t}{\La}\big)(1+\tfrac{\eps s}{2\La}\psh{n_v}{\mathbf{e}})+\frac{\eps s}{2\La}\Big(1- \dfrac{(\psh{n_v}{\mathbf{e}}+\tfrac{\eps s}{2\La})^2}{(1+\tfrac{z_t}{\La})^2}\Big)\times\\
    & &\ \ \ \ \dfrac{1+\tfrac{z_t}{\La}}{1+\tfrac{z_t}{\La}-\tfrac{\eps s}{2\La(1+\tfrac{z_t}{\La})}(\psh{n_v}{\mathbf{e}} +\tfrac{\eps s}{2\La})}\\
    &=&1+\dfrac{\eps s}{2\La}\Big[1-\frac{(\psh{n_v}{\mathbf{e}}+\tfrac{\eps s}{2\La})^2}{(1+\tfrac{z_t}{\La})^2}+\psh{n_v}{\mathbf{e}}\big(\frac{1}{\La}+1\big)\Big]+\underset{\delta r\to 0}{\mathcal{O}}((\delta r)^2).
\end{array}
\]
\noindent -- As $u=v-z_t\ov{n}_u=v+\tfrac{\eps s}{2}\psh{n_v}{\mathbf{e}}n_{v_j}+\underset{\delta r\to 0}{\mathcal{O}}((\delta r)^2)$, we get:
\begin{equation}\label{dd21}
\begin{array}{rl}
\underset{S_\eps(t)}{\dint} du_j \widetilde{G}_{J,j}(u_j)r&=\underset{\Phi_t(S_\eps(t))}{\dint} dv_j \widetilde{G}_{J,j}(v_j+\tfrac{\eps s}{2}\psh{n_{v_j}}{ \mathbf{e}}n_{v_j} +\underset{\delta r\to 0}{\mathcal{O}}((\delta r)^2))\times\\
&\ \ \ \ \ \Big(1-\dfrac{\eps s}{2\La}\Big[1-\frac{(\psh{n_v}{\mathbf{e}}+\tfrac{\eps s}{2\La})^2}{(1+\tfrac{z_t}{\La})^2}+\psh{n_v}{\mathbf{e}}\big(\frac{1}{\La}+1\big)\Big]+\underset{\delta r\to 0}{\mathcal{O}}((\delta r)^2)\Big) dv.
\end{array}
\end{equation}

We have $\Phi_t(S_\eps(t))\neq S(r)$. In spherical coordinates $(r,\theta,\phi)$ with respect to $-\eps\tfrac{r}{2}\mathbf{e}$ and positive vertical half-line $\RR_+\eps \mathbf{e}$ we have
\begin{equation}\label{domain}
\Phi_t(S_\eps(t))\simeq\Big\{(\La,\theta,\phi),\ \dfrac{rs}{2\La\sqrt{1-\tfrac{rs}{2\La^2}+\tfrac{s^2}{4\La^2}}}=\cos(\theta_t)\le \cos(\theta)\le 1\Big\},
\end{equation}
and $\cos(\theta_t)=\tfrac{r}{2\La}-\tfrac{r^2}{8\La^2}s+\underset{\delta r\to 0}{\mathcal{O}}((\delta r)^2)$.

At this point, we need to differentiate $\widetilde{G}_{J,j}$: we have
\[
\widetilde{G}_{J,j}(u_j)=\dfrac{|\psh{n_{u_j}}{\mathbf{e}}|}{2}\underset{B(r)^J}{\dint}du_0\cdots \wh{du_j}\cdots du_J \dfrac{G^0_J(u_\ell\pm\tfrac{k}{2})}{\underset{1\le i\le J}{\prod} |u_i-u_{i-1}|^2}.
\]
We change variables as follows: $v_i:=u_i-u_j$, this enables us to remove $u_j$ from the term $|u_{j}-u_{j\pm 1}|^{-2}$. Writing $B_\eps(r,u_j):=\{v:\ |v+u_j+\eps\tfrac{k}{2}|<\La\}$, $B(r;u_j)=B_+(r;u_j)\cup B_-(r;u_j)$ and $S_\eps(r,u_j):=\partial B_\eps(r,u_j)$, we have
\begin{equation}\label{stuj}
\widetilde{G}_{J,j}(u_j)=\dfrac{|\psh{n_{u_j}}{\mathbf{e}}|}{2}\underset{B(r;u_j)^J}{\dint}dv_0\cdots \wh{dv_j}\cdots dv_J\dfrac{G^0_J(v_\ell+u_j\pm\tfrac{k}{2})}{\underset{1\le i\le J}{\prod} |v_i-v_{i+1}|^2},
\end{equation}
with the convention $v_j=0$. We differentiate the formula \eqref{stuj}: $u_j$ appears in the integrand and in the domains $B(r;u_j)$. We deal with the terms corresponding to differentiation of the integrand as before. Then we have for any integrable function $\text{F}$ and small displacement $\delta u\in\RR$:
\begin{equation}\label{tresfin}
\underset{B_\eps(r,u_j+\delta u)}{\dint} \text{F}(v)dv-\underset{B_\eps(r,u_j)}{\dint}\text{F}(v)dv=\underset{S_\eps(r;u_j)}{\dint} \text{F}(v)(\psh{n(v-u_j)}{\delta u}+\underset{\delta u\to 0}{\mathcal{O}}(|\delta u|^2))dv,
\end{equation}
where $n(v-u_j)$ is the outward normal of $S_\eps(r,u_j)$ at $v$. Substituting in \eqref{dd21}, as in the case 2.2. we get terms which are $\underset{\delta u\to 0}{\mathcal{O}}\Big(|\delta u|(1-\log |\delta u|)\Big)$. Writing $u_j=u$ we have
\begin{equation}\label{compar}
\begin{array}{l}
\widetilde{G}_{J,j}(v-\tfrac{\eps s}{2}\psh{n_{v}}{\mathbf{e}}n_{v} +\underset{\delta r\to 0}{\mathcal{O}}((\delta r)^2))=\underset{\delta r\to 0}{\mathcal{O}}(-(\delta r)^2\log(\delta r))\\
+\widetilde{G}_{J,j}(v)-\frac{\eps s\psh{n_{v}}{\mathbf{e}}}{2}\underset{i\neq j}{\ssum}\underset{{\small\begin{array}{l} v_i\in S(r;v)\\
\mathbf{v}'_i\in  B(r,v)^{J}\end{array}}}{\dint}\dfrac{du_id\mathbf{v}'_i}{\underset{1\le \ell\le J}{\prod}|v_\ell-v_{\ell-1}|^2}G^0_J(v_\ell\pm\tfrac{r}{2}\mathbf{e})\psh{n(v_i-u_j)}{n_{v}}.
\end{array}
\end{equation}

We write $\mathcal{C}(r):=S_+(r)\cap S_-(r)$ (this is a curve): the integration of $\widetilde{G}_{J,j}(v_j)$ over 

\noindent $S_\eps(r)\Delta \Phi_{\delta r}(S_\eps(\delta r))$ gives rise to a term:
\[
-\dfrac{2r^2}{8\La^2}\underset{u_j\in \mathcal{C}(r),(u_i)_{i\neq j}\in B(r)^J}{\dint}\dfrac{r}{4\La}du_0\cdots du_J G_{J}(u_\ell\pm\tfrac{k}{2})+\underset{\delta r\to 0}{\mathcal{O}}((\delta r)^2).
\]
Thus we get a term of order% $\underset{\delta r\to 0}{\mathcal{O}}((\delta r)^2)=\underset{\delta r\to 0}{o}(\delta r)$.
\[
-\dfrac{2r^2}{8\La^2}\underset{u_j\in \mathcal{C}(r),(u_i)_{i\neq j}\in B(r)^J}{\dint}\dfrac{r}{4\La}du_0\cdots du_J G_{J}(u_\ell\pm\tfrac{k}{2})=\mathcal{O}\Big(\dfrac{(\alpha K)^{J+1}}{\La^2}\Big).
\]
By integrating the term $\widetilde{G}_{J,j}(v)\times(\text{J}(\Phi_t; u_j)^{-1}-1)$, we get a well defined number in the limit $\delta r\to 0$. Furthermore this term is
\[
\mathcal{O}\Big(\frac{1}{\La}\underset{u_j\in S(r)}{\dint}\underset{(u_0,\cdots, \wh{u_j},\cdots,u_J)\in B(r)^J}{\dint\cdots \dint} du_0\cdots du_J |G_J(u_\ell\pm\tfrac{k}{2})|\Big)=\mathcal{O}\Big( \dfrac{(\alpha K)^{J+1}\llo}{\La^2}\chi_{r<2\La}\Big).
\]

\noindent -- To conclude, we consider $\widetilde{G}_{J,j}(\Phi_t^{-1}(v))-\widetilde{G}_{J,j}(v)$ to deal with the problem of case 2.2.

\medskip

Up to a term $-\delta^2 \log(\delta r)=\underset{\delta r\to 0}{o}(\delta r)$, we can take $S(r)$ instead of $\Phi_t(S_\eps(t))$ and $1$ instead of the full jacobian $\text{J}(\Phi_t;u)$. 
We have $\eps \psh{n_v}{\mathbf{e}}=|\psh{n_v}{\mathbf{e}}|$. In \eqref{compar} we take back the previous variables $u_i=v+v_j$, this gives

\[
\delta r \underset{v\in S_\eps(r)}{\dint}\ssum_{i\neq j}\underset{(u_i,\mathbf{u}')\in S(r)\times B(r)^{J-1}}{\dint}du_0\cdots du_J\dfrac{|\psh{n_v}{\mathbf{e}}|}{2}\left(-\dfrac{|\psh{n_v}{\mathbf{e}}|\psh{n_v}{n_{u_i}}}{2}\right)G_J(u_\ell\pm\tfrac{k}{2}).
\]
When we sum this term with that of \eqref{cancel1}, for each $i\neq j$ we have
\[
\begin{array}{rll}
\Big||\psh{n_{u_i}}{\mathbf{e}}|-|\psh{n_v}{\mathbf{e}}|\psh{n_v}{n_{u_i}}\Big|&=&\Big|\eps(u_i)\psh{n_{u_i}}{\mathbf{e}}-\eps(v)\psh{n_v}{\mathbf{e}}\times \eps(v)\eps(u_i)\psh{n_v}{n_{u_i}} \Big|,\\
&\le& \min(\sqrt{2}|n_{u_i}-n_v|,2).
\end{array}
\]

Thus there is no more logarithmic divergence: for $u=u_j$ and $v=u_{j-1}$ or $v=u_{j+1}$, we use the same method as that for \eqref{cancel1} and get
\[
\underset{S(r)\times S(r)}{\diint}\dfrac{|n_u-n_v||\psh{n_u}{\mathbf{e}}|dudv}{|u-v|^2}\dfrac{1}{\ed{\La}^2\La^2}.%\apprle \underset{\mathbb{S}^2}{\dint}\dfrac{du}{\max(\La|u-\mathbf{e}|,\La^2|u-\mathbf{e}|^2)}.
\]
We split the domain in $4$: $S_\eps(r)\times S_{\eps'}(r)$: the case $\eps=\eps'$ gives finite number. Indeed if we use spherical coordinates with respect to $-\tfrac{\eps r}{2}\mathbf{e}$, we have $|n_u-n_v|\le \tfrac{|u-v|}{\La}$, and the integral is
\[
 \mathcal{O}\left(\underset{\mathbb{S}^2}{\dint}\dfrac{du}{\La^2|u-\mathbf{e}|}\right)=\mathcal{O}\left(\dfrac{1}{\La^2}\right).
\]
The integration over $S_+(r)\times S_-(r)$ is also finite. To see this we proceed as follows.

For convenience we write $x:=\frac{r}{2\La},\theta_1^0=\text{arccos}(x),\theta_{-1}^0=\text{arccos}(-x)$ and $s(\cdot)$ (resp. $c(\cdot)$) for $\sin$ (resp. $\cos$). We take spherical coordinates with respect to $-\eps\tfrac{r}{2}\mathbf{e}$ for any $S_\eps(r)$ and obtain:
\[
\begin{array}{l}
\dfrac{2\pi}{\La^2}\underset{(\theta_1,\theta_{-1},\phi)\in(0,\theta_1^0)\times(-\pi,\theta_{-1}^0)\times(-\pi,\pi)}{\diiintdens}\dfrac{s(\theta_1)s(\theta_{-1})d\theta_1d\theta_{-1}d\phi}{(c(\theta_1)-c(\theta_{-1})-2x)^2+s(\theta_{-1})^2s_{\phi}^2+(s(\theta_1)-s(\theta_{-1})c_\phi)^2}\\
\ \ \apprle\dfrac{1}{\La^2}\underset{(\theta_1,\theta_{-1},\phi)\in(0,\theta_1^0)\times(-\pi,\theta_{-1}^0)\times(-\pi,\pi)}{\diiintdens}\dfrac{s(\theta_1)s(\theta_{-1})d\theta_1d\theta_{-1}d\phi}{(c(\theta_1)-c(\theta_{-1})-2x)^2+c(\theta_{-1})^2\phi^2}=:\dfrac{A}{\La^2}.
\end{array}
\]
We write $\theta_\eps=\theta_\eps^0-\eps \phi_\eps$: there holds
\[
\begin{array}{rll}
 \eps c(\theta_\eps)-x&=&x(c(\phi_\eps)-1)+\sqrt{1-x^2}s(\phi_\eps),\\
x(c(\phi_\eps)-1)+\sqrt{1-x^2}s(\phi_\eps)&\ge &\phi_\eps\Big(\dfrac{2}{\pi}\sqrt{1-x^2}-x\dfrac{\phi_\eps}{2}\Big),\\
 &\ge&\dfrac{2\phi_\eps}{\pi}\Big(\sqrt{1-x^2}-\dfrac{\pi}{4}x\arccos(x)\Big)\ge \dfrac{2\phi_\eps\sqrt{1-x}}{\pi}\Big(1-\dfrac{x\pi}{4}\Big)\\
 &\ge&\dfrac{2\phi_\eps\sqrt{1-x}}{\pi}\Big(1-\dfrac{\pi}{4}\Big)\ge \dfrac{\sqrt{1-x^2}\phi_\eps}{\pi}\Big(1-\dfrac{\pi}{4}\Big).
 \end{array}
\]
Thus we have
\[
 \begin{array}{rll}
  A&\apprle&\underset{\phi_1,\phi_{-1}\in(0,\theta_1^0)}{\diint}\dfrac{\sin(\theta_1)d\phi_1 d\phi_{-1}}{\sqrt{1-x^2}\sqrt{\phi_1^2+\phi_{-1}^2}}\\
   &\apprle&\underset{\phi_1\in(0,\theta_1^0)}{\dint}\dfrac{d\phi_1}{\sqrt{1-x^2}}\log\Big(1+\dfrac{\arccos(x)}{\phi_1}\Big)\\
   &\apprle&\underset{\phi\in(0,1)}{\dint}\log(1+\phi^{-1})d\phi.
 \end{array}
\]

\subparagraph{Conclusion} 
We obtain at last the following upper bound for the terms of cases 2. and 3.:
\[
J^2\dfrac{(\alpha K)^{J+1}\llo}{\La^2}.
\]
It is possible to show that the function $\partial_{r}^2\fla(x)$ tends to zero as $|x|$ tends to $2\La$, this is proved in the thesis of the author (to appear in 2014).

\hfill{\footnotesize$\Box$}

\paragraph{Alternative $\Fla$}

In the proof of Theorem \ref{existence}, one is lead to consider a pertubative self-consistent equation with $\D$ replaced by $\D+\frac{2}{\la}\frac{\D}{|\D|}$. In particular we need Lemma \ref{witfla} below for the proof of Lemma \ref{der_des_der}. We can write
\[
\D+\frac{2}{\la}\frac{\D}{|\D|}=\beta \widetilde{w}_0(-i\nabla)+\boldsymbol{\alpha}\cdot \frac{-i\nabla}{|-i\nabla|}\widetilde{w}_1(-i\nabla).
\]
The formulae are the same with $g_0,g_1$ replaced by $\widetilde{w}_0,\widetilde{w}_1$, estimates of the same kind hold. The alternative functions are marked with a tilde: $\widetilde{B}_\La$ and $\widetilde{g}_\La$.

We can easily estimate $\int_{|x|\ge R} |\mathscr{F}^{-1}(\wit{F}_\La)(x)|dx$ for $R\ge 1$: writing $\mathfrak{f}_\La:=\mathscr{F}^{-1}(\wit{F}_\La)$ we have the following Lemma:

\begin{lem}\label{witfla}
For $\la,\La\gg 1$ we have:
\begin{equation}
\int_{|x|\ge R} |\mathfrak{f}_\La(x)|dx\le \nlp{2}{-\Delta \wit{F}_\La}\sqrt{4\pi R^{-1}}=\mathcal{O}(L R^{-1/2}).
\end{equation}
%In particular, with $R^{-1}=\alpha\ll 1$ and $L\le L_0$ it is lesser than $K L\alpha^{1/2}$.
\end{lem}

\end{appendix}

\noindent\textbf{Acknowledgment} The author wishes to thank  \'E. S\'er\'e for useful discussions. This work was partially supported by the Grant ANR-10-BLAN 0101 of the French Ministry of Research.

\begin{footnotesize}
\bibliographystyle{plain}
\bibliography{bibliothese.bib}
\end{footnotesize}

\end{document}